\RequirePackage{fix-cm}
\documentclass[natbib,smallextended]{svjour3}       
\smartqed  
\usepackage{graphicx}
\usepackage{amsmath}
\usepackage{epigraph}
 \journalname{Space Science Reviews}
\newcommand{\aap}{{Astron. Astrophys.}}

\usepackage{aas_macros}

\usepackage[backref,breaklinks,colorlinks,citecolor=blue]{hyperref}

\begin{document}

\title{A Hitch-hiker's Guide to Stochastic Differential Equations}
\subtitle{Solution Methods for Energetic Particle Transport in Space
  Physics and Astrophysics}

\titlerunning{A Hitch-hiker's Guide to Stochastic Differential Equations}        

\author{R. Du Toit Strauss \and Frederic Effenberger}

\institute{R. Du Toit Strauss \at
              Center for Space Research, North-West University, Potchefstroom, 2531, South Africa \\
              National Institute for Theoretical Physics (NITheP), Gauteng, South Africa \\
              \email{dutoit.strauss@nwu.ac.za}           
              \and F. Effenberger \at
              Department of Physics and KIPAC, Stanford University, Stanford, CA 94305, USA \\
              \email{feffen@stanford.edu} }

\date{Received: date / Accepted: date}

\maketitle

\begin{abstract}
  In this review, an overview of the recent history of stochastic
  differential equations (SDEs) in application to particle transport
  problems in space physics and astrophysics is given. The aim is to
  present a helpful working guide to the literature and at the same
  time introduce key principles of the SDE approach via ``toy
  models''. Using these examples, we hope to provide an easy way for
  newcomers to the field to use such methods in their own
  research. Aspects covered are the solar modulation of cosmic rays,
  diffusive shock acceleration, galactic cosmic ray propagation and
  solar energetic particle transport. We believe that the SDE method,
  due to its simplicity and computational efficiency on modern
  computer architectures, will be of significant relevance in
  energetic particle studies in the years to come.

  \keywords{Cosmic Rays \and Energetic Particles \and Stochastic
    Differential Equations \and Numerical Methods \and Diffusive Shock
    Acceleration \and Transport Equations }
\end{abstract}

\vspace{-0.3cm}
\tableofcontents
\vspace{-0.1cm}

\newpage
\section{Introduction}
\epigraph{\emph{Would it save you a lot of time if I just gave up and
    went mad now?}}{\citet[][h2g2]{h2g2}}
\label{Sec:introduction}

Studying the transport of non-thermal particles through turbulent
plasmas is ubiquitous in space and astrophysics. Examples include,
amongst others, the propagation of cosmic rays through the galaxy and
the heliosphere. The transport of these particles is usually
described by a Fokker-Planck type diffusion equation that must, for
most applications, be solved numerically due to the complexity of both
the plasma geometry and the associated transport parameters. In recent
years, stochastic differential equations (SDEs) have been used more
frequently to numerically solve a variety of problems in space- and
astrophysics. It appears that SDEs have, to some extent, replaced or
at least complemented more traditional (i.e. largely finite difference
(FD) type) numerical schemes to solve multi-dimensional partial
differential equations (PDEs) which involve diffusive processes. As
such, we have decided to compile this review, which consists of two
parts or aspects: (i) Simple 1D ``toy models" are constructed,
discussed in detail, and used to illustrate how SDEs are solved
numerically for a variety of different boundary conditions, how
source/sink terms are handled numerically and how to deal with some
numerical pitfalls that might exist (for example, how to select an
appropriate time step). (ii) Directly after each numerical section, a
review is given on some of the most relevant and contemporary
SDE-based models that are currently available and implement the
previously discussed numerical techniques for the solution of
real-world problems\footnote{As with any review paper, some references
  will inadvertently fall through the cracks and we would like to
  apologize for this in advance.}. The primary aim of this paper is
therefore to introduce SDE-type numerical techniques to novices in the
field and to give a guide to the reader in applying these methods in
his/her own work.

The scope of applications considered here can loosely be defined as
energetic particle transport in space and astrophysical plasmas. The
common modeling ground for corresponding transport processes is a
Fokker-Planck type equation which involves both diffusive
(i.e. stochastic or random) and convective (i.e. deterministic)
processes\footnote{The terms convection and advection are usually used
  interchangeably without any consensus on their potentially different
  meaning (e.g. an active vs. passive process). We do not attempt to
  use one over the other rigorously in this text.}. The limits of this
approach can be understood in the kinetic behaviour of the particles:
at low energies and sufficiently high densities the particles may
thermalize, making a kinetic treatment superfluous. At very high
energies and low densities, the particles may not scatter anymore with
the background plasma fluctuations on the relevant length and time
scales, and thus a completely deterministic treatment of the problem
is possible, e.g. by solving the Newton-Lorentz equations of motion
for individual particles \citep[see also the discussions
in][]{schlikeiser2002,Shalchi-2009}. We do not consider these two
limits here and suppose that a Fokker-Planck like transport equation
is always the proper starting point \citep[see, e.g.,][for a classical
derivation of the Fokker-Planck equation in the context of cosmic ray
transport]{Schlickeiser-1989}.

We begin by describing very briefly the mathematical background of the
problem, before we consider different aspects of the modeling approach
and practical computational issues in conjunction with selected
applications. These are, in the order of the subsequent chapters: the
solar modulation of cosmic rays, diffusive shock acceleration,
galactic cosmic ray propagation and solar energetic particle
transport.

And above all, we emphasize: \emph{Don't Panic}.

\subsection{Mathematical background}
Several monographs deal with the mathematical formalism related to
SDEs and their application in a variety of scientific fields
\citep[e.g.][]{Kloeden2,Kloeden,Oksendal,Gardiner}. For those
completely new to the subject, we recommend the excellent introductory
book by \citet{Lemons}. Also, the classical review by Chandrasekhar on
stochastic problems \citep{Chandrasekhar-1943} needs to be
mentioned. The discussion in this section is based on these
sources. We only present the most relevant aspects for the purposes of
this paper, without any attempt for strict mathematical rigor or
derivations.

For our purposes, it is sufficient to define an SDE as any
equation that can be cast into the general form
\begin{equation}
\label{Eq:SDE_define}
\frac{d x(t)}{d t} = a(x,t) + b(x,t) \zeta (t)
\end{equation}
in the 1D case where $a(x,t)$ and $b(x,t)$ are continuous functions,
while $\zeta (t)$ represents a rapidly varying stochastic function,
also referred to as a noise term. In SDE nomenclature, the first term
on the right-hand side of Eq. \ref{Eq:SDE_define} is referred to as
the {\it drift} or deterministic term, while the second is referred to
as the {\it diffusion} term (not to be confused with the physical
drift and diffusive processes that will be considered later
on). Moreover, only SDEs of the $\mathrm{It\bar{o}}$ type are
considered here, where Eq. \ref{Eq:SDE_define} can be rewritten as
\begin{equation}
\label{Eq:ito_sde}
dx(t) = a(x,t)dt + b(x,t) dW(t)
\end{equation}
with $W(t)$ representing the Wiener process; a time stationary
stochastic L\'evy process where the time increments have a Normal
distribution with a mean of zero (i.e. a Gaussian distribution) and a
variance of $d t$; $dW(t) = W(t+dt) - W(t) \sim \mathcal{N}(0,d
t)$.
In fact, the Wiener process can be understood as the integral of the
stochastic function present in Eq. \ref{Eq:SDE_define}, i.e. in its
differential form we have $dW(t) = \zeta(t) dt$. See especially the
introduction by \citet{Gardiner}.

Eq.~\ref{Eq:ito_sde} can be integrated formally as
\begin{equation}
\label{Eq:SDE_int}
x(t) = x_0 + \int_0^t a(x,t')dt' + \int_0^t b(x,t') dW(t')
\end{equation}
where the first integral is a normal (Riemann or Lebesgue) integral,
while the second is an $\mathrm{It\bar{o}}$-type \footnote{Various
  different types of SDEs exist, such as the widely used Stratonovich
  stochastic formulation. Here we choose to be as brief as possible
  and refer the interested reader to e.g. \citet{Gardiner1983} for a
  comprehensive review of these different formulations. It must
  necessary be kept in mind that, due to the different temporal
  discretizations used in these SDE formulations, the numerical
  methods described here may not be applicable to solve SDE that are
  not of the $\mathrm{It\bar{o}}$-type; this is certainly true for the
  Stratonovich formulation.} stochastic integral. Analytical solutions
for SDEs are, however, only available for very few limited scenarios
\citep[for an example, see the discussion of the Ornstein-Uhlenbeck
process in][]{Lemons}, and as such, Eq.~\ref{Eq:SDE_int} is usually
integrated numerically. In the most common approach, the SDEs are
integrated by using the Euler-Maruyama numerical scheme
\citep[][]{maruyama}, where a finite time step $\Delta t$, is chosen
and Eq.~\ref{Eq:ito_sde} is solved iteratively
\begin{equation}
\label{SDE_numer}
x(t + \Delta t) = x(t) + a(x,t)\Delta t + b(x,t)\Delta W(\Delta t)
\end{equation}
from an initial position $x = x_0$ at $t=0$ and continued until either
a boundary is reached at $x = x_e$ at $t=t_e$, or until a temporal
integration limit is reached at $t = T$.  Higher-order schemes (in
time) are available \citep[see, e.g., the book by][]{Kloeden}, but
often the simple Euler time stepping is sufficient, particularly in
diffusion-dominated cases. For a comparison of different higher-order
numerical schemes, see also \citet{anna2015}.

By using
$\mathcal{N}(0,dt) \stackrel{d}{=} \mathcal{N}(0,1) \sqrt{dt}$ (where
the symbol $\stackrel{d}{=}$ indicates that the random processes
follow the same distribution) it follows that the Wiener process can
be discretized as
\begin{equation}
\Delta W(\Delta t) = \sqrt{\Delta t} \cdot  \Lambda(t),
\end{equation}
where $\Lambda(t)$ is a simulated Gaussian distributed pseudo-random
number (PRN). The temporal evolution of $x$ forms a trajectory through
phase space, which is generally referred to as the trajectory of a
{\it pseudo-particle} (this is a widely used, but somewhat
inappropriate term, that has become part of the nomenclature of the
field; the actually meaning of the term is discussed in
Sec. \ref{subsec:liouville} where, for the case of charged particle
propagation, we show that this {\it pseudo-particle} actually
represents the temporal evolution of an ensemble of real particles, or
equivalently, the evolution of a phase-space density
element). Integrating Eq. \ref{SDE_numer} for a single pseudo-particle
has no significance, as integration must be carried out over a large
number of possible trajectories (i.e. over a large number of possible
Wiener processes; see Eq. \ref{Eq:SDE_int}), so that
Eq. \ref{SDE_numer} is usually integrated $N \gg 1$ times, each time
using a different series of simulated Wiener processes, starting the
integration process with different seeds to the pseudo random number
generator (PRNG); this process is sometimes referred to as tracing $N$
pseudo-particles. Numerically, the independence of the
  pseudo-random numbers (PRNs; referring to the computational
  technique of simulating random numbers via deterministic algorithms)
  is a very important condition to fulfil in the numerical model, as
  some PRNGs are not independent and have a quite short cycling period
  (meaning the same set of PRNs are repeated after a certain time). A
  ``high level'' PRNG is therefore needed, e.g. the Mersenne Twister
  PRNG \citep[][]{twister}. When SDEs are solved on parallel computing
  architectures, it is important that the PRNs, simulated on different
  compute cores, are also independent \citep[e.g.][]{flippie}.

We can generalize Eq.~\ref{Eq:ito_sde} to an $n$-dimensional set of
SDEs, so the general formulation becomes
\begin{equation}
\label{Eq:bigsde}
dx_i = a_i (x_i,s)ds + \sum_{j=1}^{n} b_{ij}(x_i,s)  dW_i(s),
\end{equation}
where $\vec{a}$ is an $n$-dimensional vector and $\mathbf{b}$ is a
$n\times n$ matrix. In general, this system of SDEs can be thought of
as being integrated either backward or forward in time (more on the
relative merits of each approach in the next section). When time
backward integration is performed, Eq.~\ref{Eq:bigsde} is {\it
  equivalent} to the following Fokker-Planck equation (also referred
to as the time backward Kolmogorov equation)
\begin{equation}
\label{Eq:backward}
-\frac{\partial \rho(x_i,s)}{\partial s} =  \sum_{i=1}^n  a_i(x_i,s) \frac{\partial \rho(x_i,s)}{\partial x_i}  + \frac{1}{2} \sum_{i=1}^n \sum_{j=1}^n   C_{ij}(x_i,s) \frac{\partial^2 \rho(x_i,s)}{\partial x_i \partial x_j} ,
\end{equation}
where $\rho(x_i,s)$ is a conditional probability density, depending on
all $x_i$, $s$ and a final condition for $\rho$ at time $T$. Note that
we introduced $s$ as a new time-marching coordinate to indicate that
it can be different from actual (forward moving) time $t$ (the
equivalence between time forward and time backward integration is
discussed in more detail in Sec. \ref{Sec:CRs}).

If time forward integration is performed instead, the corresponding
Fokker-Planck equation (also referred to as the time forward
Kolmogorov equation in order to distinguish it from
Eq. \ref{Eq:backward}) is given by
\begin{equation}
\label{Eq:forward}
\frac{\partial \rho(x_i,t)}{\partial t} = - \sum_{i=1}^n
\frac{\partial }{\partial x_i}\left[\tilde{a}_i(x_i,t)\rho(x_i,t)\right]   + \frac{1}{2}
\sum_{i=1}^n \sum_{j=1}^n  \frac{\partial^2 }{\partial x_i \partial
  x_j} \left[\tilde{C}_{ij}(x_i,t)  \rho(x_i,t)\right] .
\end{equation}
Note that the main difference between both equations is the implicit
vs. explicit formulation in the coefficients, which may then differ
for the different formulations. For most cases,
$\vec{a} \neq \tilde{\vec{a}}$ and $\vec{C} \neq \tilde{\vec{C}}$.  We
point to \citet{koppetal2012} and \citet{Bobik_etal_2016} for a
detailed discussion on how the time forward and backward SDE
formulation is related to different Fokker-Planck equations. Moreover,
we note that the ``full" Fokker-Planck equation can also contain
source and linear terms, while are not included in
Eqs. \ref{Eq:backward} and \ref{Eq:forward}. These additional terms
are however included in Secs. \ref{Sec:galactic_transport} and
\ref{Sec:importance_sampling}, respectively.

The diffusion tensor $C_{ij}$ is given by
\begin{equation}
C_{ij}(x_i,s) \equiv  \left( b_{ij}(x_i,s) \cdot b_{ji}(x_i,s)  \right).
\end{equation}
The quantity $b_{ij}$ is sometimes referred to as the volatility
matrix (especially in mathematical texts), to illustrate that it is
related to, but is not the same as the diffusion matrix (or tensor). A
suitable Fokker-Planck like PDE can therefore be cast into the form of
Eq. \ref{Eq:backward}, the quantities $\vec{a}$ and $\mathbf{b}$
obtained, and the equivalent SDE formulation emerges
naturally. \footnote{As $b_{ij}$ is basically the square root of
  $C_{ij}$, calculating $b_{ij}$ for some scenarios can be very tricky
  in higher dimensions, but is always possible as $C_{ij}$ is a
  positive definite tensor \citep[][]{Gardiner1983}, and usually also
  symmetric \citep[][]{koppetal2012}. It is also interesting to note
  that $b_{ij}$ is not unique but different choices of $b_{ij}$ lead
  to the same solution as they are all mathematically equivalent.} The
$n$-dimensional PDE is thereby transformed into a set of $n-1$
two-dimensional SDEs, with the latter usually much easier to solve
numerically.  In the remainder of this paper, we will be concerned
with such solution methods and corresponding applications. It should
be mentioned at this point that linear loss or gain and source terms
can, of course, be of importance in many circumstances when
Fokker-Planck type models are used. We will return to some of the
issues related to the inclusion of such terms in the corresponding
following sections.

\subsection{Why should one use the time-backward method?}
\begin{figure}[!ht]
    \centering
    \includegraphics[width=1\textwidth]{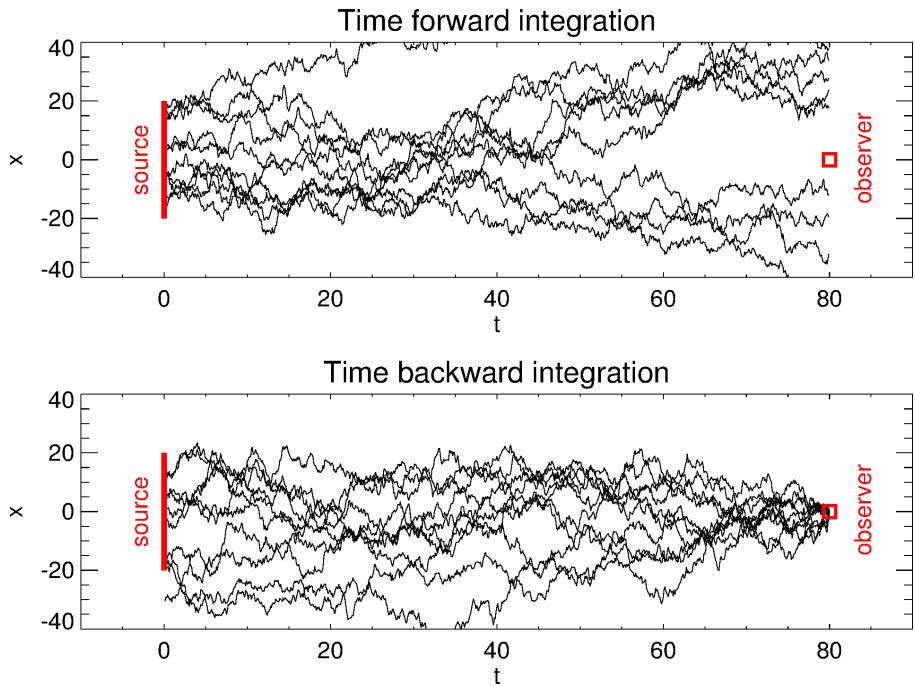}
    \caption{Illustrating time forward (top panel) and time backward
      (bottom panel) integration of SDEs in a 1D scenario.}
    \label{fig1_1}
\end{figure}
In energetic particle transport applications of SDEs, the time
backward integration approach is mostly followed. In
Fig. \ref{fig1_1}, it is demonstrated why this is the case: The top
panel shows, for illustration purposes, 10 solutions of a 1D SDE, as a
function of time, starting at a ``source" region (the thick red line
at $t=0$). In most astrophysical and heliospheric applications, we are
however only interested in calculating the solution of the PDE at a
collection of phase-space points, which may be, for example, along the
trajectory of a spacecraft or an energy spectrum at a given
position. For the top panel of the figure, we are therefore interested
in the intensity at an ``observational point" (indicated by the red
square at $t=80$). It is clear that for this set-up, none of the SDE
solutions pass through the observational point, and hence, they do not
contribute to the intensity at the observational point -- these
pseudo-particle trajectories were thus unnecessarily solved and simply
wasted computational resources. A much more efficient set-up is
illustrated in the bottom panel of the figure: Integration is started
at the observational point (where one wants the solution) and traced
(solved/integrated) backward in time until model boundaries are
reached \citep[for a comparison between the time forward and backward
approaches, see][]{Bobik_etal_2016}. For many applications, the latter
set-up is more efficient.\footnote{Of course, the relative efficiency
  of both approaches depends on the relative sizes of the source and
  boundary surfaces compared to the size of the effective observer. As
  suggested by \citet{milsteinetal_2004}, some solutions are also hard
  to evaluate in the time backward scenario.} In this review, we will
thus focus almost solely on the backward integration method.

\subsection{Advantages of the SDE method}
There are numerous advantages to adopt an SDE-based numerical
scheme above more traditional finite difference methods. These
advantages include: (i) The SDE numerical scheme is unconditionally
stable, although this does not necessarily imply numerical accuracy.
(ii) Related to this, the method can handle large gradients,
something that finite difference models struggle with, but care has to
be taken in the choice of the time step (see
Section~\ref{Sec:timestep}). (iii) The SDE approach does not calculate
solutions on a specific grid, but at a number of discrete phase-space
positions. There is thus no need to store unnecessary information,
saving computational memory and making computations in higher
dimensions possible. (iv) Each solution (i.e. each pseudo-particle) is
completely independent, making it possible to perform calculations on
parallel computational environments, significantly speeding up
calculations. Calculations can also performed in parallel on graphics
processing units \citep[GPUs,][]{flippie}. (v) The SDE approach allows
one to visualize more of the physics (processes) included in the
transport equation under investigation. This will be illustrated
further in this work.

\newpage
\section{Handling Boundary Conditions: Solar Modulation of Cosmic Rays}
\label{Sec:CRs}

\epigraph{\emph{For a moment, nothing happened. Then, after a second
    or so, nothing continued to happen.}}{h2g2}

We start by considering a simple one-dimensional diffusion equation
\begin{equation}
\label{Eq:original_PDE}
\frac{\partial f(x,t)}{\partial t} = \kappa \frac{\partial^2 f}{\partial x^2}
\end{equation}
where $\kappa = 1/2$ is assumed to be constant. This equation is
solved on the interval $x\in [x_{\min}, x_{\max}] = [-1,1]$. Referring
to Eqs. \ref{Eq:ito_sde} and \ref{Eq:backward}, the time-backward SDE,
being equivalent to Eq. \ref{Eq:original_PDE}, is
\begin{equation}
dx = b  dW
\end{equation}
with $b=1$. For this one-dimensional situation, $\vec{b}$, which is in
the general case a tensor, reduces to a scalar. Similarly for the
general tensor $\vec{C}$, where a comparison between
Eqs. \ref{Eq:backward} and \ref{Eq:original_PDE} show that, for this
specific set-up, $C=2 \kappa$.

In general, the solution at any point $x$, at any time $T$, can
obtained by solving the convolution, \citep[see e.g.][]{peietal2010}
\begin{equation}
f(x,T) = \int_0^T \int_x G(x',t) f_b(x',t)   dx' dt
\label{Eq:general_convolution}
\end{equation}
where $f_b(x',t)$ denotes the boundary value for either Dirichlet-type
or initial boundary conditions. Here Eq. $G(x,t)$ is a Green's
function of the considered PDE \citep[e.g.][]{webbgleeson1977}, where
the normalization condition
\begin{equation}
\int_{0}^{\infty}\int_{-\infty}^{\infty}G(x',t)dx'dt= 1.
\end{equation}
holds.

For the scenarios considered in this section, $f_b(x,t)$ has the general form
\begin{equation}
f_b(x,t) = \underbrace{g(t)\delta(x+1)}_\text{left boundary}
+ \underbrace{h(t)\delta(x-1)}_\text{right boundary} 
+ \underbrace{k(x)\delta(t)}_\text{initial condition}
\end{equation}
where $g(t)$ and $h(t)$ are boundary conditions specified at $x=\pm1$ respectively and $k(x)$ is an initial condition. Eq. \ref{Eq:general_convolution} then becomes
\begin{eqnarray}
f(x,T) &=& \underbrace{\int_0^T g(t) G(x=-1,t)  dt}_\text{left boundary contribution} \nonumber\\
         &+&  \underbrace{\int_0^T h(t) G(x=1,t)  dt}_\text{right boundary contribution} \nonumber\\
         &+& \underbrace{\int_{-1}^{1} k(x) G(x,t=0) dx}_\text{initial condition contribution}.
\end{eqnarray}
The normalization condition of $G(x,t)$ of course still applies

\begin{equation}
\underbrace{\int_0^{\infty} G(x=-1,t)  dt}_\text{left boundary} +  \underbrace{\int_0^{\infty}  G(x=1,t)  dt}_\text{right boundary} + \underbrace{\int_{-1}^{1}  G(x,t=0) dx}_\text{initial condition} = 1
\end{equation}

For the results presented here, time backward integration is
performed, and the solution is only calculated at $x=0$ by means of
the SDE model. Although, in this section, we calculate the solution
only at a single $x$ point, extension of the scheme for different
values of $x$ is simple and will be discussed in the following
sections. The SDE results are compared to the solution from a finite
difference numerical scheme in order to validate the numerical
approach applied here.\footnote{Although for some of the simple
  problems that are considered here, some (semi) analytic solutions exist, we
  decided not to include them here, to keep the focus on the numerical
  approach.}

Backward time $s$, is related to forward time $t$, via

\begin{equation}
s = T - t
\end{equation}

where $T$ is our temporal boundary. Also note that, for example,
$G(x,t=0) = G(x,s=T)$.

As we want the solution of the SDE at $x=0$, $N$ pseudo-particles are
released from $x=0$ at $s=0$ (this implies
$G(x,s)=\delta(x)\delta(s)$) and their stochastic trajectories are
traced until the temporal boundary $s=T$ is reached. The trajectory of
a small number ($N=5$) of such particles are shown in panel (g) of
Fig. \ref{fig1}, where they are integrated from $s=0$ to $s=0.5$. At
this point, $G(x,t)$ is calculated by binning the position of the
pseudo-particles in either a spatial and/or temporal grid. For
example, to apply the initial condition, the spatial grid is divided
into $M$ equally spaced grid cells and the number of pseudo-particles
in each cell, $N_j(x_j,s)$, where $x_j$ refers to the mid-point of
each spatial cell, at $s=T$ is recorded, or the pseudo-particles
interact with the spatial boundaries, dividing by the total number of
pseudo-particles injected initially, one may calculate
\begin{equation}
G(x_j,t=0)dx \approx \frac{N_j}{N} .
\label{Eq:calculate_G_1}
\end{equation}
From here it is easy to calculate
\begin{equation}
f'(x,T) = \int_{-1}^{1} k(x) G(x,t=0)  dx \approx \frac{1}{N} \sum_{j=1}^{M} k(x_j) N_j 
\end{equation}
where $x_j$ is the spatial position of the $j$-th bin. $f'(x,T)$ is
the contribution of the initial condition to the value of $f(x,T)$. A
similar approach is then also applied to the boundaries, although
integration for this scenario is performed over time and not space. In
a later section it will become clear how this binning process works
and that, for some applications, the binning process is redundant so
that the solution can be obtained directly from the exit position of
the pseudo-particles.

\subsection{Dirichlet Boundary Conditions}
\label{subsec:general_dirichlet}

\begin{figure}[!ht]
    \centering
    \includegraphics[width=1\textwidth]{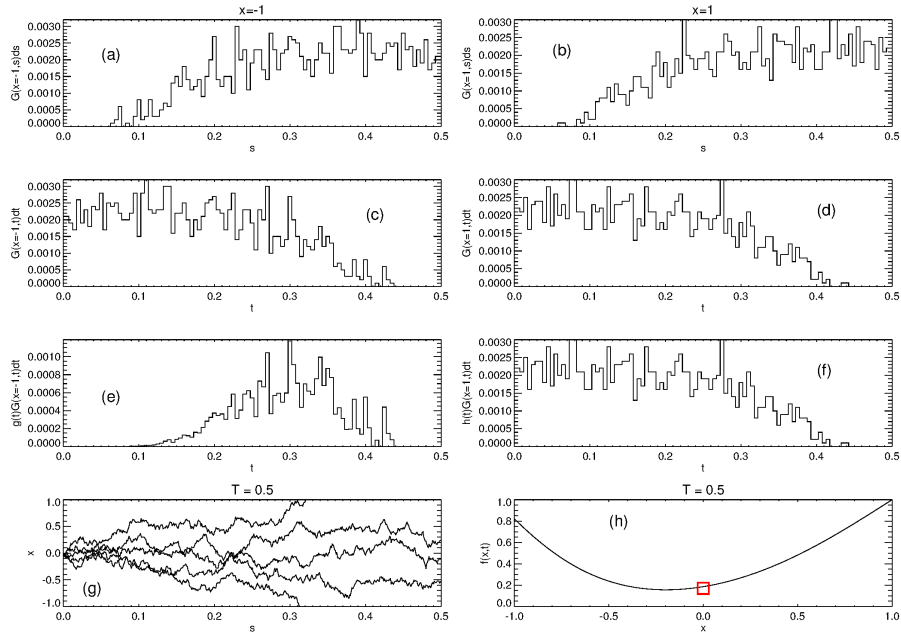}
    \caption{Calculating the solution of the PDE, given by Eq. \ref{Eq:original_PDE}, by means of a SDE
      numerical method for $(x=0,t=0.5)$. Panels (a) and (b) show the
      calculated probability distributions ($G(x=-1,s)$ and
      $G(x=1,s)$) for either the left or right boundaries, binned in
      terms of backward time. These are then converted to forward
      time and shown in panels (c) and (d), while the convolutions
      with the boundary conditions are shown in panels (e) and
      (f). The solution of the SDE approach is shown in panel (h) as
      the red box while the solid line is the solution of a FD scheme. For
      illustrative purposes, panel (g) shows the trajectory of 5
      pseudo-particles.  }
    \label{fig1}
\end{figure}

\begin{figure}[!ht]
    \centering
    \includegraphics[width=1\textwidth]{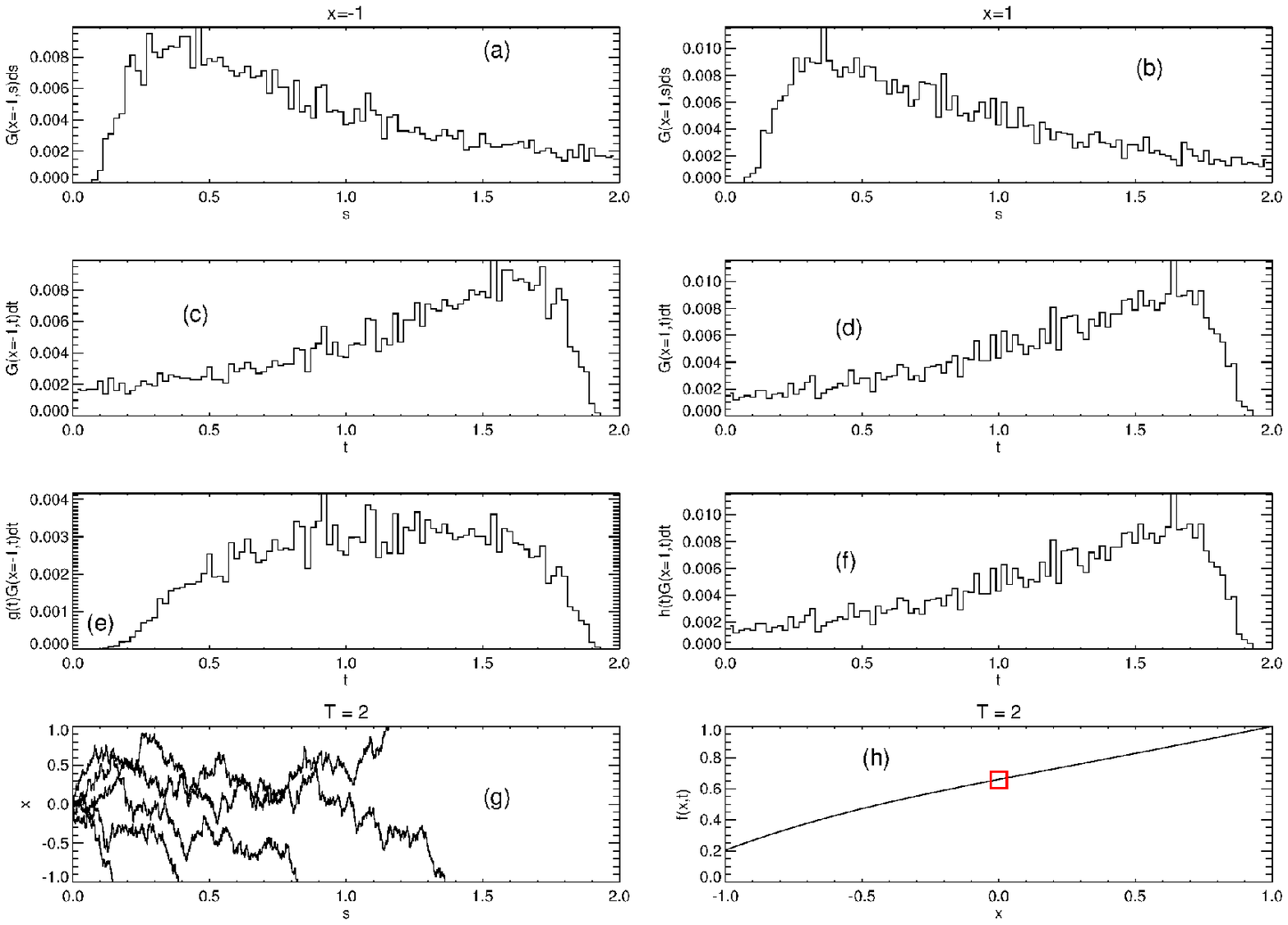}
    \caption{Similar to Fig. \ref{fig1}, but for $T=2$.}
    \label{fig2}
\end{figure}

We first consider Eq. \ref{Eq:original_PDE} for the case of
$f(x,t=0)=0$ (an empty system as an initial condition, i.e. $k(x)=0$)
and the following boundary conditions
\begin{eqnarray}
g(t) &=& \frac{5}{t}\exp \left[- \frac{1}{t} - \frac{t}{1}  \right] \\
h(t) &=& 1.
\end{eqnarray}

The complete methodology to calculate the solution at two different
times, $f(x=0,t=0.5)$ and $f(x=0,t=2)$, is illustrated in
Figs. \ref{fig1} and \ref{fig2}. In Fig. \ref{fig1}, the different
panels show the calculated $G(x,s)ds$ for $x=-1$ (panel a) and $x=1$
(panel b) for $T=0.5$. They are thus the binned (in terms of
backward-time) normalized number of pseudo-particles reaching either
the left or right boundary. These quantities are then converted to
forward-time and shown as $G(x=-1,t)dt$ (panel c) and $G(x=1,t)dt$
(panel d). Note that for the discretized SDE formulation, all
particles that exit the computational region are assumed to interact
with the spatial boundaries, therefore all particles with e.g.
$x \leq -1$ are counted in the calculation of $G(x=-1,t)dt$. The
convolutions $g(t)G(x=-1,t)dt$ (panel e) and $h(t)G(x=1,t)dt$ (panel
f) are shown for illustrative purposes, while the solution of the SDE
scheme at $(x=0,t=0.5)$ is shown as the red box in panel (h). In this
panel, the solid line shows the solution of the FD scheme. Panel (g) shows the trajectory of 5
pseudo-particles for illustrative purposes, all released from $x=0$
(the position at which we want the solution). The inversion
$G(x,s)ds \rightarrow G(x,t)dt = G(x,T-s)ds$ is not really needed in
calculations as the convolutions may be done directly in terms of
backward time, e.g.
\begin{equation}
\int_0^T h(t)G(x,t)dt = \int_T^0 h(T-s)G(x,s)ds,
\end{equation}
but is shown here for completeness sake. 

Interesting for the SDE scenario is that the steady-state solution is
well defined (in contrast to FD scheme where convergence to a steady
state must be checked continuously) for the time backward SDE
formulation -- it is simply when all the pseudo-particles have exited
the computational domain, i.e.
\begin{equation}
t\rightarrow \infty \Rightarrow G(-1<x<1,t)dt \rightarrow 0.
\end{equation}
Fig. \ref{fig2} is similar to Fig. \ref{fig1}, but for a later time $T=2$.

\subsection{Initial Conditions}

\begin{figure}[!ht]
    \centering
    \includegraphics[width=1\textwidth]{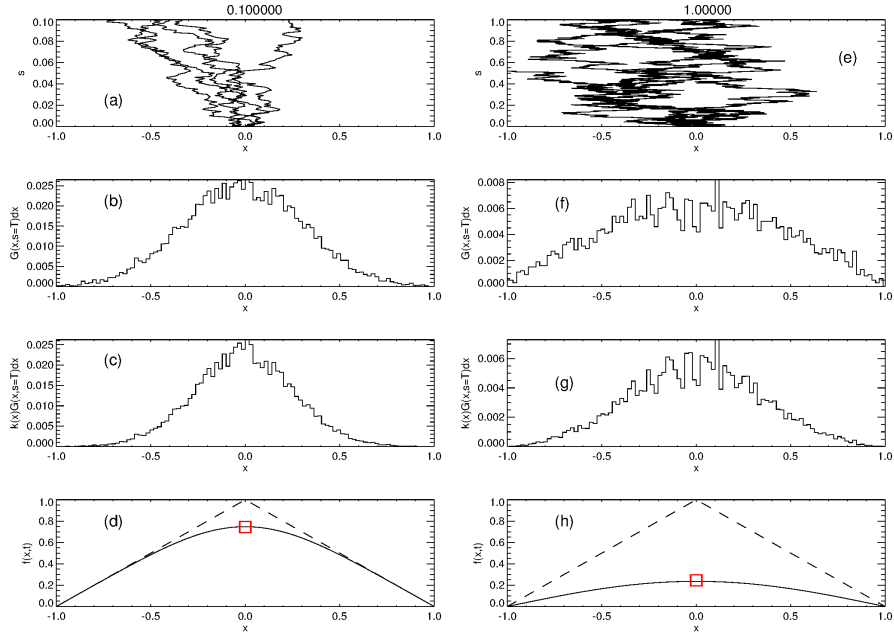}
    \caption{The calculation of $f(x=0,t)$ for $T=0.1$ (left panels)
      and $T=1$ (right panels). In the bottom panels, the red box
      shows the solution of the SDE approach, the solid line that of a
      FD scheme and the dashed line indicates the initial condition
      specified at $t=0$.}
    \label{fig3}
\end{figure}

We now turn to the case of a prescribed initial condition. We choose
$k(x) = - |x| + 1$ along with absorbing boundary conditions at the
spatial boundaries, i.e. $g(t)=h(t)=0$.

Fig. \ref{fig3} shows the calculation of $f(x=0,t)$ for $T=0.1$ (left
panels) and $T=1$ (right panels). Panels (a) and (e) again shows
illustrative pseudo-particle trajectories, while panels (b) and (f)
show $G(x,s=0.1)dx$ and $G(x,s=1)dx$. This time the pseudo-particles
are binned in terms of spatial position at $s=T (t=0)$. The
corresponding convolutions with the initial condition are shown in
panels (c) and (g), while panels (d) and (h) again compare the final
solution of the SDE approach to that of the FD scheme. The dashed line
indicates the initial solution. For the SDE approach, the solution at
$T=0 \Rightarrow s,t=0$ is readily obtained: Keeping in mind that
$G(x,t=0)dx = G(x,s=0)dx = \delta(x=0)\delta(t=0)$, the solution is
simply $f(x,t=0)=k(x)$ as required.

\subsection{Von Neumann Boundary Conditions}

\begin{figure}[!ht]
    \centering
    \includegraphics[width=1\textwidth]{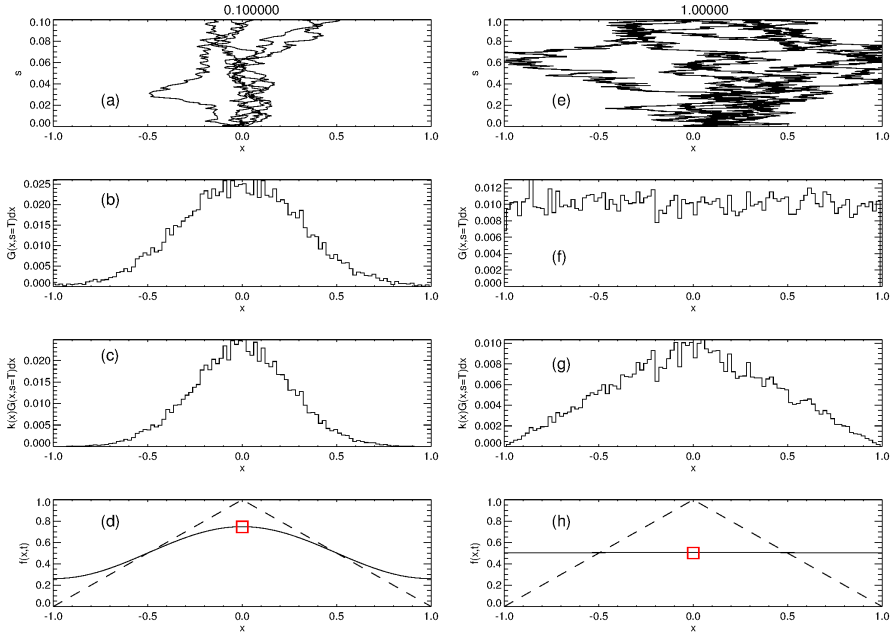}
    \caption{Similar to Fig. \ref{fig3}, except that reflecting
      boundary conditions are assumed at $x=\pm1$.}
    \label{fig4}
\end{figure}

The solutions presented here are similar to that of the previous
section, except that reflecting conditions are assumed, where

\begin{equation}
\left. \frac{\partial f(x,t)}{\partial x} \right|_{x=\pm1} = 0.
\end{equation}

Reflecting conditions are implemented in the SDE formulation as

\begin{eqnarray}
x < x_{\min} &:& x \rightarrow -2x_\mathrm{min} -x = -2 -x\\
x > x_{\max} &:& x \rightarrow 2x_\mathrm{max} - x = 2-x
\end{eqnarray}

similarly to hard sphere reflection off a planar surface. Note that
for this situation, $G(x=-1,t)=G(x=1,t)=0$, so that all
pseudo-particles stay in the computational region
$x\in(x_\mathrm{min},x_\mathrm{max})=(-1,1)$. The calculation of
$f(x=0,t)$ for this scenario is shown in Fig. \ref{fig4}. The figure
is similar to Fig. \ref{fig3}, except with the choice of reflecting
boundaries.

Here, the steady-state is less well defined for the SDE approach. As
all pseudo-particles must stay in the computational region, the steady
state is defined as a limit on the computational time $T\gg 1$. In
this limit, $G(x,t)dt$ should become independent of time, and in this
particular case also of $x$. This means the pseudo-particle are
distributed evenly across the spatial computational region. It is also
easy to write down the steady state solution: For reflecting
boundaries and the limit $T\rightarrow \infty$, we require
\begin{equation}
G(x,t\rightarrow\infty)dx \rightarrow G_{\infty}dx
\end{equation}
where $G_{\infty}$ is simply a constant (i.e. the spatial bins are equally
populated by pseudo-particles). Because of the normalization
condition, we have that
\begin{equation}
\int_{-1}^{1} G(x,t)dx = G_{\infty} \int_{-1}^{1} dx = 1 \Rightarrow G_{\infty} =
\frac{1}{\int_{-1}^{1} dx} = \frac{1}{2} .
\end{equation}
The solution of $f(x,t\rightarrow\infty)$ is then simply
\begin{equation}
f(x,t \rightarrow \infty)= \int_{-1}^{1}k(x) G_{\infty} dx = \frac{1}{2} \int_{-1}^{1} k(x)dx
\end{equation}
which is the average of $k(x)$.

\subsection{Periodic Boundary Conditions}
\label{Sec:periodic_boundaries}

In the SDE approach, periodic boundary conditions are specified as \citep[see e.g.][]{straussetal2011}
\begin{eqnarray}
x<x_{\min} &:& x \rightarrow x_{\max} - x_{\min} + x \\
x> x_{\max} &:& x \rightarrow x_{\min} - x_{\max} + x,
\end{eqnarray}
in other words, when a pseudo-particle exists the left boundary,
$x_{\min}$, it simply re-appears at the right boundary,
$x_{\max}$. This formulation will be illustrated and become clear in
the following section, see especially
Eq. \ref{Eq:spherical_renormalization}, where the solution of SDEs in
spherical spatial coordinates is considered.

\subsection{Cosmic ray modulation studies}
\label{Sec:cosmic_rays}

Galactic cosmic rays (GCRs) are highly energetic and fully ionized
charged particles that originate outside of the heliosphere with
energies of up to $10^{12}$ GeV \citep[e.g.][]{hillas2006}. The most
likely GCR accelerators are believed to be supernovae throughout the
galaxy \citep[e.g.][]{moskalenkoetal2002}, although other
astrophysical systems, e.g. pulsars, are good acceleration candidates
\citep[e.g.][]{buschingetaL2008}. These particles propagate through
the galaxy (see Section \ref{Sec:galactic_transport}) and can then
reach the edge of the heliosphere -- the region of interstellar space
influenced by the Sun's plasma outflow. Once these particles enter the
heliosphere, their intensities are modulated by the turbulent
heliospheric magnetic field and embedded irregularities or turbulent fluctuations. This is a
time- and energy-dependent process, where the amount of modulation
generally increases with decreasing energy \citep[see the recent
review of][]{potgieter2013}.

In order to model the heliospheric modulation of GCRs, i.e. their
transport from the heliopause (HP; that is, the outer boundary of the
heliosphere) to Earth, the intensity of the different GCR species
directly at the HP must be known and prescribed as a Dirichlet-type
boundary condition in any numerical modulation model. To achieve this,
results from galactic propagation models
\citep[e.g.][]{moskalenkoetal2002,strongetal2011}, are used to
estimate the intensity of the GCRs at the HP; simply assumed to be the
local interstellar spectrum (LIS). This LIS is then modulated as the
GCRs propagate towards the inner heliosphere. The recent crossing of
the HP by the Voyager 1 spacecraft
\citep[e.g.][]{Stone_etal_2013,gurnett_etal_2013} has shed some light
on the LIS, at least for some GCR species in certain energy regimes,
so that this quantity cannot be treated as a free parameter any more
\citep[see also the work of
e.g.][]{vos_potgieter_2015,Corti_et_al_2016}.

\subsubsection{Transport equation and relevant numerics}

The transport of GCRs inside the heliosphere is governed by the
\citet{parker1965} transport equation (TPE),
\begin{equation}
\label{Eq:TPE_parker}
\frac{\partial f}{\partial t} = -\left(\vec{u} + \vec{v}_d \right)\cdot\nabla f + \nabla \cdot \left( \mathbf{K}\cdot \nabla f \right) + \frac{1}{3} \left( \nabla \cdot \vec{u}  \right) \frac{\partial f}{\partial \ln P}  
\end{equation}
given in terms of the gyro- and isotropic CR distribution function
$f(r,\theta,\phi,P,t)$ (in spherical spatial coordinates) which is
related to the GCR differential intensity by
$j(r,\theta,\phi,P,t)=P^2 f(r,\theta,\phi,P,t)$, where $P$ is particle
rigidity (defined as $P=pc/q$ with $p$ particle momentum, $c$ the
speed of light and $q$ particle charge). A re-derivation of the TPE
was given by \citet{webbgleeson1979}, while the TPE in the above form
can also be derived by averaging the Fokker-Planck equation over pitch
angle \citep[e.g.][]{schlikeiser2002,stawicki2003}. The processes
included in the TPE and described by the associated terms, are from
left to right: Temporal changes, convection with the plasma (solar
wind) velocity $\vec{u}$ due to the embedded nature of the
heliospheric magnetic field (HMF), usually taken to be Parkerian in
its simplest formulation \citep[][]{parker1958}, particle drifts,
diffusion and adiabatic energy changes due to an expansion or
compression of the background plasma. In the supersonic solar wind
$\nabla \cdot \vec{u} >0$, and hence, GCRs generally loose energy is
this region \citep[e.g.][]{parker1966} Additional processes,
e.g. momentum diffusion (Fermi II acceleration), can also be included
into the TPE by including the corresponding terms, while diffusive
shock (Fermi I) acceleration is already included in the adiabatic term
in its present form \citep[e.g.][]{strauss_etal_2010}. The components
of the 3D diffusion tensor, $\mathbf{K}$ are related to the underlying
HMF turbulence evolution \citep[as discussed by,
e.g.,][]{oughtonetal2011,zanketal2012,usmanovetal2014} and different
scattering theories coupling this to pitch-angle particle scattering
\citep[e.g.][]{jokipii1966,Matthaeusetal2003}. The ab initio approach
to cosmic ray modulation \citep[e.g.][]{engelbrechtburger2013} aims to
unite these two aspects in a comprehensive modulation model. Although
the diffusion tensor can therefore be specified in a HMF aligned
coordinate system, it must be transformed into the global coordinate
system in which Eq. \ref{Eq:TPE_parker} will be solved
\citep[e.g.][]{effenbergeretal2012}. Usually the latter is chosen to
be in spherical spatial coordinates. Due to the large scale HMF, GCRs
will undergo a combination of gradient, curvature and current sheet
drift, the latter resulting because of the switch in HMF polarity
across the heliospheric current sheet (HCS). Drifts were neglected in
modulation models until \citet{jokipiietal1977} pointed out that
drifts may indeed play a dominant role in GCR modulation and can
explain the observed HMF polarity dependent CR observations
\citep[e.g.][]{jokipiikopriva1979,potgietermoraal1985,burgerpotgieter1989}. See
also the comprehensive review by \citet{potgieter2013}.

As pointed out in Section \ref{Sec:introduction}, the 4D transport
equation given by Eq. \ref{Eq:TPE_parker} can be cast into a set of 3,
2D SDEs, with the latter much easier to handle numerically. The
relevant set of SDEs for GCR modulation is
\begin{eqnarray}
dr &=& a_r \cdot ds + b_{rr}\cdot dW_r + b_{r\theta}\cdot dW_{\theta} +  b_{r\phi}\cdot dW_{\phi} \nonumber  \\
d\theta &=& a_{\theta} \cdot ds  + b_{\theta\theta}\cdot dW_{\theta} +  b_{\theta\phi}\cdot dW_{\phi}  \nonumber \\
d\phi &=& a_{\phi} \cdot ds  +  b_{\phi\phi}\cdot dW_{\phi}  \nonumber \\
dp &=& a_{p} \cdot ds ,
\end{eqnarray}
where the coefficients $a_i$ and $b_{ij}$ depend on the coefficients
of the convective and diffusive terms in the TPE. We do not state the
explicit form of these here to keep the presentation compact, but they
are given for example in \citet{straussetal2011}. We encourage the
interested reader to derive and/or look up these coefficients in
different coordinate systems and implement them in their own SDE
framework. Also note that this is the simplest possible formulation
for CR transport when a Parker field is assumed. For a more general
scenario, see e.g. \citet{peietal2010}.

The multidimensional Wiener process used in the previous equation is
given simply by
\begin{equation}
d\vec{W} = [dW_r,dW_{\theta},dW_{\phi}].
\end{equation}	
If momentum (energy) diffusion is neglected, $\mathbf{b}$ can be
reduced from a $4 \times 4$ matrix to a $3 \times 3$ matrix (see also
the discussion in \citet{koppetal2012}). To calculate $\mathbf{b}$,
the square root of the diffusion tensor must be calculated. As
discussed by, e.g., \citet{johnsonetal2002}, the magnitude and form of
$\mathbf{b}$ may not be unique, but all of these $\mathbf{b}$'s will
be mathematically equivalent.

The GCR boundary outer condition (the so-called LIS specified at the
HP) can be handled by applying
Eq.\ref{Eq:general_convolution}. Neglecting initial conditions, this
can be rewritten as
\begin{equation}
\label{Eq:flux_general}
f(\vec{x}_0,t_0) = \int_0^{t_0} \int_{\vec{x}\in \Omega_b} f_b(\vec{x},t) \rho (\vec{x},t) d\Omega dt,
\end{equation}
where $f_b(\vec{x},t)$ represents the boundary condition (i.e. the
LIS), $\rho(\vec{x},t)$ the {\it conditional} probability density,
$f(\vec{x}_0,t_0)$ the phase space position where the intensity is
calculated, i.e. the {\it observational} point, $\Omega_b$ the
boundary of the integration domain and where
$\vec{x} =\{r, \theta, \phi, p \}$ in phase-space coordinates.

Many studies focus on stationary (steady state) solutions of the TPE,
where
$t \rightarrow \infty \Rightarrow \rho(\vec{x},t): \rightarrow
\rho(\vec{x})$, reducing Eq. \ref{Eq:flux_general} to
\begin{equation}
\label{Eq:flux_steadystate}
f(\vec{x}_0) =  \int_{\vec{x}\in \Omega_b} f_b(\vec{x}) \rho (\vec{x}) d\Omega  .
\end{equation}
For GCRs, the boundary values are only momentum, or energy, dependent
so that only integration over momentum space is performed,
i.e. $d\Omega : \rightarrow dp$,
\begin{equation}
\label{Eq:flux_energyboundary}
f(\vec{x}_0) = \left. \int_{p}^{\infty} f_b(p') \rho (\vec{x}) dp' \right|_{\vec{x} \in \Omega_b} ,
\end{equation}
which is essentially a convolution of the boundary condition and the
probability distribution. Here, we have assumed that only energy
losses are present, i.e.\ the pseudo-particles can only reach the
boundary with a higher momentum than they started with. If, however,
particle acceleration is also included in the model, pseudo-particles
can exit the system with lower energy than they started with (keeping
in mind this is the time-backward solution) and the momentum
integration boundaries must be changed from $[p,\infty]$ to
$[0,\infty]$.  Note that Eq. \ref{Eq:flux_energyboundary} can be
expanded when two disjointed boundaries are present, such that
$\Omega_b' \cap \Omega_b=0$, to
\begin{equation}
\label{Eq:two_boundaries}
f(\vec{x}_0) = \left. \int_{p}^{\infty} f_b(p') \rho (\vec{x}) dp' \right|_{\vec{x} \in \Omega_b}  + \left. \int_{p}^{\infty} f_b'(p') \rho (\vec{x}) dp' \right|_{\vec{x} \in \Omega_b'},
\end{equation}
which is the case if, for example, the Jovian magnetosphere, which is
a strong source of low energy electrons
\citep[e.g.][]{chenette1980,moses1987}, is included into a modulation
model as a second CR species.

Two approaches can be used to incorporate these boundary
conditions. Firstly, the set of SDEs can be solved to provide
$\rho(\vec{x})$, whereafter Eq. \ref{Eq:flux_energyboundary} is used
to obtain the CR flux -- this is the most general solution method
discussed in Section \ref{subsec:general_dirichlet}. A second, and
numerically easier approach is to calculate the weighted value of
$f(\vec{x}_0)$ for each pseudo-particle individually, before
normalizing this to the correct magnitude at the end of the
integration process. For a single pseudo-particle, reaching a momentum
dependent boundary with $p_i=p_i^e$ (where $i$ labels the
pseudo-particles),
\begin{equation}
\label{Eq:deltading}
\rho_i(\vec{x}) = \left. \delta (p_i - p_i^e) \right|_{\vec{x} \in \Omega_b},
\end{equation}
because of the normalization condition. This implies that
\begin{equation}
f_i(\vec{x}_0) = f_b(p_i^e),
\end{equation}
so that each pseudo-particle traced to the boundary makes some
(weighted; see Eq. \ref{Eq:die_aap_}) contribution to the total
solution. Repeating for $N \gg 1$ pseudo-particles, the numerical
solution of the TPE can be calculated as
\begin{equation}
\label{Eq:die_aap_}
f(\vec{x}_0)  = \frac{1}{N} \sum_{i=1}^N f_b (p_i^e),
\end{equation}
which follows from Eq. \ref{Eq:flux_energyboundary}.

Although the SDE approach does not have to prescribe boundary
conditions at the computational domains of the angular coordinates,
the angular position of the pseudo-particles needs to be renormalized
so that $\theta \in [0,\pi]$ and $\phi \in [0,2\pi]$. This is done by
specifying the following conditions:
\begin{eqnarray}
\label{Eq:spherical_renormalization}
\phi < 0 &:& \phi \rightarrow \phi + 2\pi \nonumber \\
\phi > 2 \pi &:& \phi \rightarrow \phi - 2\pi \nonumber \\
\theta < 0 &:& \theta \rightarrow |\theta|; \phi \rightarrow \phi \pm \pi \nonumber \\
\theta > \pi &:& \theta \rightarrow 2\pi - \theta; \phi \rightarrow \phi \pm \pi
\end{eqnarray}
which occur when the pseudo-particle either propagates around the
ecliptic plane or cross the solar poles. Although this is really a
coordinate re-normalization, it is very similar to using periodic
boundary conditions (see also Section \ref{Sec:periodic_boundaries}).

\subsubsection{Selected results}
\label{Ssec:Moduilation_results}

\begin{figure}[!ht]
    \centering
    \includegraphics[width=0.5\textwidth]{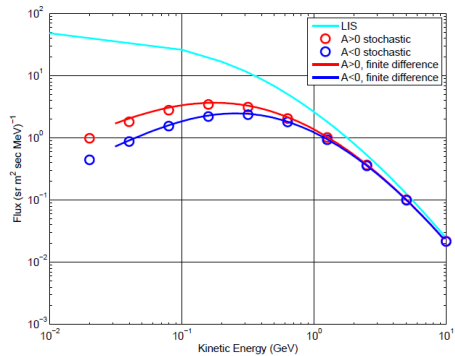}
    \caption{An example of the exhaustive benchmarking studies
      presented in earlier SDE-model papers. Although, in retrospect,
      these benchmarking studies seem superfluous (the mathematical
      equivalence is already established), they were needed in order
      to illustrate the validity of SDE approach to the heliospheric
      community. This example is taken from \citet{peietal2010}.}
    \label{GCR_fig3}
\end{figure}

The first\footnote{Earlier, \citet{jokipii_levy_1997} used the first-
  and second-order moments of the TPE to construct a random walk model
  for CRs. However, as this model did not implement a Wiener process,
  but rather a uniform distribution of random numbers, it is not an
  SDE model in a strict sense.} SDE-based GCR modulation model was
presented by \citet{yamadaetal1998} for the spatially 1D
scenario. These authors applied their model to the case of GCR proton
modulation and illustrated the validity of the SDE approach by
benchmarking the SDE approach to a traditional FD scheme -- the
comparison was, of course, a success. The SDE approach, however, only
gained widespread interest after \citet{zhang1999} showed the
applicability of this approach in 3D, including particle drift in a
flat (i.e. a tilt angle of $0^{\circ}$) HCS set-up. In this paper,
Zhang illustrated that the SDE approach can also lead to additional
insight into the modulation process by visually illustrating the
modulation process; we will touch on these illustrative results again
later on. Later models applied essentially the same SDE formulation,
applied to GCR protons, electrons and Jovian electrons, but most also
included extensive benchmarking studies
\citep[e.g.][]{gervasietal1999,bobik_etal_2008,
  peietal2010,straussetal2011}. An example of such a benchmarking
study is shown in Fig. \ref{GCR_fig3} and taken from
\citet{peietal2010}. Here, the authors compare GCR proton results from
a 3D SDE model to that of a FD model, including drifts. These
exhaustive benchmarking studies gave confidence in the SDE approach,
which led to the heliospheric community quickly embracing this ``new"
generation of numerical modulation models.

\begin{figure}[!ht]
    \centering
    \includegraphics[width=1\textwidth]{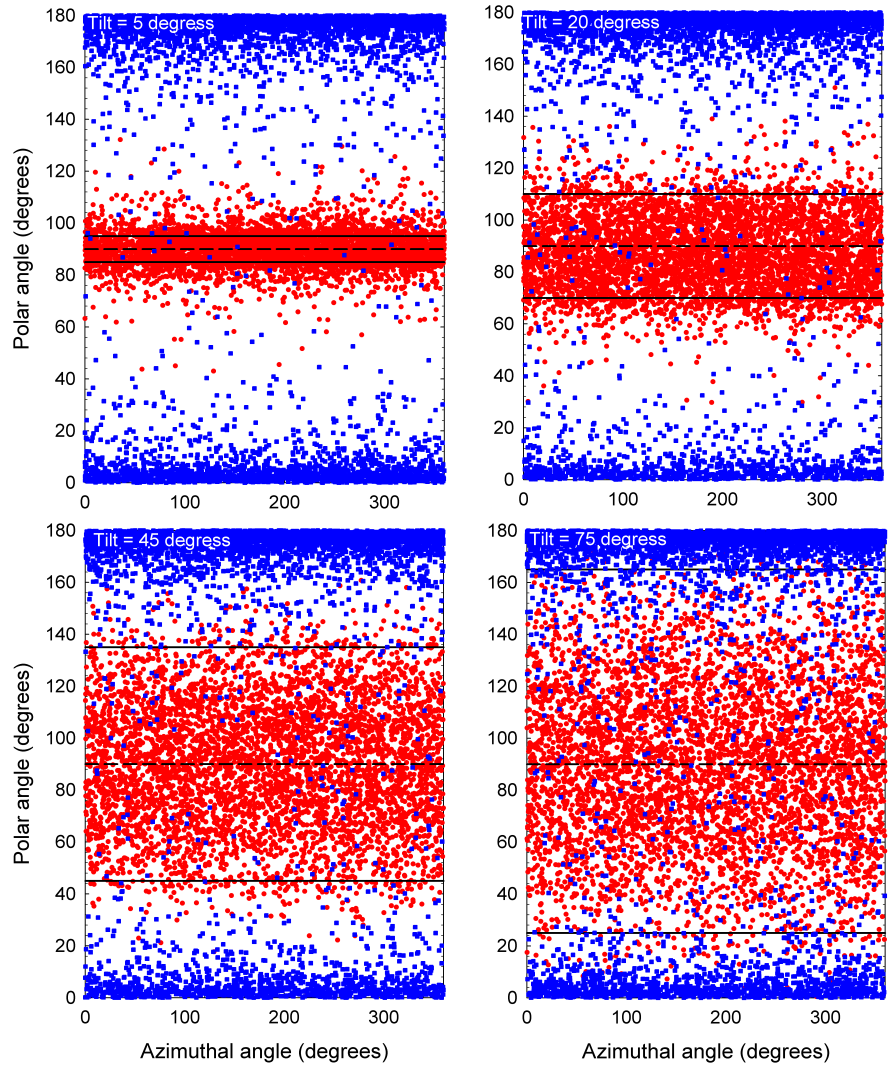}
    \caption{Illustrating drift effects in the heliosphere by plotting
      the exit position of protons in the so-called $A>0$ (blue) and
      $A<0$ (red) drift cycles for various tilt angles. The solid and
      dashed lines show the latitudinal extend of the wavy HCS, with
      the different tilt angles indicated on each panel. The figure is
      taken from \citet{straussetal2012}. With permission of Springer.}
    \label{GCR_fig4}
\end{figure}

The next step in modulation models was the implementation of a wavy
HCS. This is, however, by no means an easy task to accomplish in
3D. The first model to achieve this was given by \citet{miyake2005};
this work was summarized in a conference proceeding but never
published in detail. A more detailed methodology (but only in 2D) was
given by \citet{ososkin}, and for 3D models by \citet{pei2011} and
\cite{straussetal2012}. The inclusion of a wavy HCS also gave the
opportunity to effectively illustrate the drift process, and an
example of such an illustration is shown in Fig. \ref{GCR_fig4}. Here
the exit positions (that is, the position, in terms of polar and
azimuthal angles, at which a pseudo-particle exits the heliosphere,
i.e. reaches the edge of the computational domain) at a spherical heliopause for
GCR proton pseudo-particles are shown. This representation needs some
careful interpretation: The pseudo-particles are released at Earth and
exits the heliosphere at the shown positions. In the time-forward
scenario, it can thus be concluded that GCR, that enter the
heliosphere at these positions, will end up at Earth. It does, however,
not mean that GCRs only enter the heliosphere at these positions (the
GCR flux is assumed isotropic and uniform outside the
heliosphere). The simulations presented in Fig. \ref{GCR_fig4} are
performed for the $A>0$ (blue points) and $A<0$ (red points) drift
cycles and varying values of the HCS tilt angle. In the $A>0$ drift
cycle, GCR protons generally drift from the polar regions to reach
Earth, whereas, in the $A<0$ cycle, they mainly drift along the HCS
\citep[for details see also][]{straussetal2012c}. See also the modelling of \citet{Raath-etal-2016,Raath-etal-2015}. The effectiveness of
SDE-type models in modeling and reproducing drift effects, have
recently led to SDE models being increasingly applied to model charge
sign dependent modulation effects
\citep[e.g.,][]{Della_torre_etal_2012,bobik_etal_2012,maccione_2013},
that is, simultaneously modeling and reproducing particle and
anti-particle ratios in order to gauge the effectiveness of the drifts
in different polarity cycles.

From the onset, \citet{zhang1999} illustrated that, in the SDE
formulation, it is possible to calculated the propagation (resident)
time and energy losses suffered by GCRs
directly. \citet{straussetal2011b} showed that these quantities,
calculated numerically via a SDE model, compare well with earlier
analytical models. \citet{florinksipogorelov2009}, for example,
calculated the propagation time for particles in different regions of
the heliosphere and illustrated that they spend a long time in the
heliosheath (the region between the termination shock and the
heliopause); a region where CRs suffer little or no energy
losses. Similarly, \citet{bobik_et_al_2011} included and illustrated
the effect of additional deterministic energy loss processes. Although
such calculations make the modulation process clearer and more
understandable, it is not yet conclusively established how these
quantities are related to CR intensities or observations \citep[see
e.g.][for an attempt at this]{straussetal2013a}. It must always be
kept in mind that {\it pseudo-particles} are not real particles! More
recent work by, e.g., \citet{adrian}, suggests more complicated
weighting procedures for the propagation time in order for this
quantity to be more representative of reality.

\begin{figure}[!ht]
    \centering
    \includegraphics[width=1\textwidth]{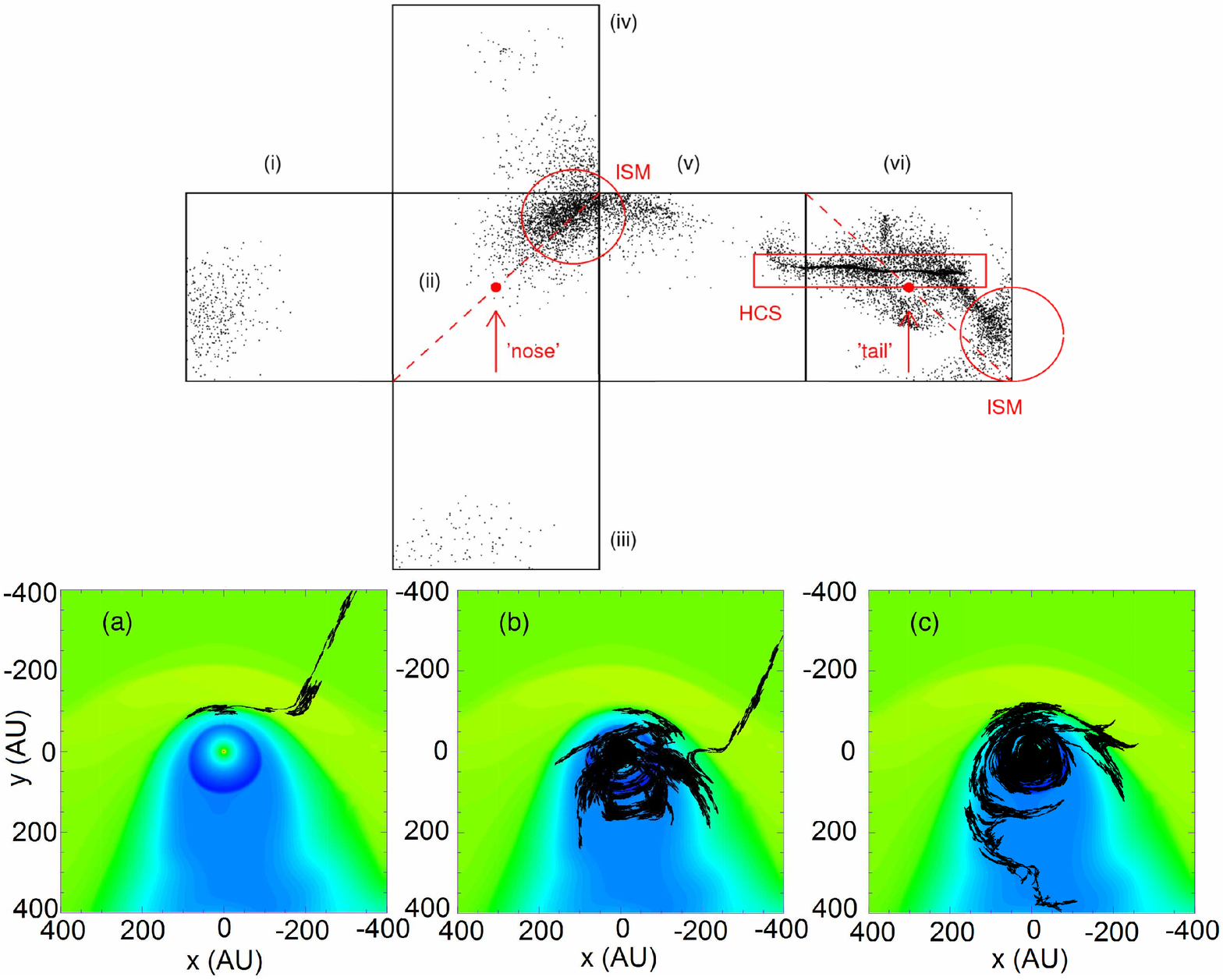}
    \caption{A solution of a hybrid CR modulation model, taken from
      \citet{straussetal2013b}. The bottom panels show the trajectory
      of 3 pseudo-particles in an MHD simulated heliospheric
      environment (the contour plots show the plasma density). The top
      panel shows the exit positions of pseudo-particles, initially
      released from Earth, at the boundaries of the model. For this
      scenario, the model boundary is a cube, which is ``unfolded" in
      the top panel. The large red circles indicate the position of
      pseudo-particles that have entered the heliosphere by
      propagating along the assumed interstellar magnetic field
      (labelled ISM), while the red square shows particles which
      drifted along the heliospheric current sheet (labelled HCS). The
      ``nose" and ``tail" of the heliosphere are indicated on panels
      (ii) and (vi) respectively. }
    \label{GCR_fig1}
\end{figure}

Due to the stability of the SDE numerical scheme, coupled with the low
memory requirements when solving SDEs in higher dimensions, it has
also become possible to create hybrid CR modulation models. These are
models that use the input from magneto-hydrodynamic (MHD) models of
the heliosphere geometry and plasma flow fields, and solve the TPE
using essentially a test (pseudo-) particle approach (i.e. coupling
the TPE to an MHD simulated background plasma). Such models can be
considered more realistic in the sense that they capture the complex
and asymmetrical structure of the heliosphere. The first of these
models was presented by \citet{balletal2005}, where after the
approach was followed by \citet{florinksipogorelov2009},
\citet{straussetal2013b}, \citet{luoetal2013},
\citet{guo_florinski_2014} and most recently by
\citet{guo_florinski_2014_b}. These models were mostly applied to GCR
transport in the large-scale heliosphere (with the focus on the
heliosheath; see also Section \ref{subsec:time_step_review}), while
\citet{guo_florinski_2014_b} modeled the effect of co-rotating
interaction regions (CIRs) on GCR intensities in the inner
heliosphere. An example solution of such a hybrid modulation model is
shown in Fig. \ref{GCR_fig1}.

Although the numerical approach to integrating SDEs is, per
definition, time dependent (it can be considered an explicit numerical
scheme), most of the above-mentioned work considered the steady-state
solution, i.e. all pseudo-particles are integrated until they reach a
boundary. Some work that has focused on time-dependent (meaning here
that the transport coefficients are time-dependent and no stationary
solution exists) solutions is, however, available
\citep[e.g.][]{yamada_etal_1999,luo_etal_2011,guo_florinski_2014_b,polish2014}.

\subsubsection{Pseudo-particle interpretation and an illustration of Liouville's theorem}
\label{subsec:liouville}

Because the SDE model solves for $f(\vec{x},t)$, a pseudo-particle
represents a realization of $f(\vec{x},t)$, so that a pseudo-particle
is defined as an ensemble of CR particles, where the ensemble is
constructed such that it results in a gyro- and isotropic collection
of particles. A more accurate term for a pseudo-particle is that of a
{\it phase space density element}. However, the nomenclature of a
pseudo-particle is ingrained into the field of SDE based modulation
models, and as such, will be used throughout this paper. Many
discussions have shown that it is sometimes mistakenly assumed that a
pseudo-particle represents an actual CR particle, or the guiding
centre of such a particle. This is of course not the case, as a single
GCR particle would follow the Lorentz-Newton equations of motion. The way in which particle
source and/or loss terms are treated in the SDE approach (see Sections
\ref{Sec:galactic_transport} and \ref{Sec:importance_sampling}) also
shows that pseudo-particles are in fact not physical particles; `pseudo'
may be assumed to refer to the fact that they are simply mathematical
realizations of particle distribution functions \citep[see also the
comments by][]{koppetal2012}.

The trajectory of a pseudo-particle also has an often overlooked and
very important physical significance. Along the trajectory of a single
pseudo-particle (where $\vec{x}'$ labels the set of phase space
coordinates along this trajectory) 
\begin{equation}
\rho (\mathcal{\vec{x}'}) = 1,
\end{equation}
which is an extension of Eq. \ref{Eq:deltading}, and occurs because of
the normalization condition of $\rho(\vec{x},t)$ as shown in
Eq. \ref{Eq:deltading}. Because
$f(\vec{x},t) \propto \rho(\vec{x},t)$, it follows that
\begin{equation}
f({\vec{x}'}) = \mathrm{constant},
\end{equation}
so that
\begin{equation}
\label{Eq:liouvilles}
\frac{D f({\vec{x}'})}{Dt} = 0,
\end{equation}
which is a Lagrangian derivative along
$\vec{x}'$. Eq. \ref{Eq:liouvilles} is simply a re-statement of
Liouville's theorem: Along the trajectory of a pseudo-particle through
phase space, the CR distribution function remains constant. The
trajectory of a pseudo-particle therefore also shows a graphical
illustration of Liouville's theorem. For cases, described later on,
where sources and/or sinks of particles are considered, the
phase-space distribution along the path of a pseudo-particle is no
longer constant, and hence, the connection with Liouville's theorem
cannot be made.

\newpage
\section{Choosing the Correct Time Step: Diffusive Shock Acceleration and the Transport of Cosmic Rays Beyond the Heliopause}
\epigraph{\emph{``Ford!" he said, "there's an infinite number of monkeys outside who want to talk to us about this script for Hamlet they've worked out."}}{h2g2}
\label{Sec:timestep}

\begin{figure}[!ht]
    \centering
    \includegraphics[width=1\textwidth]{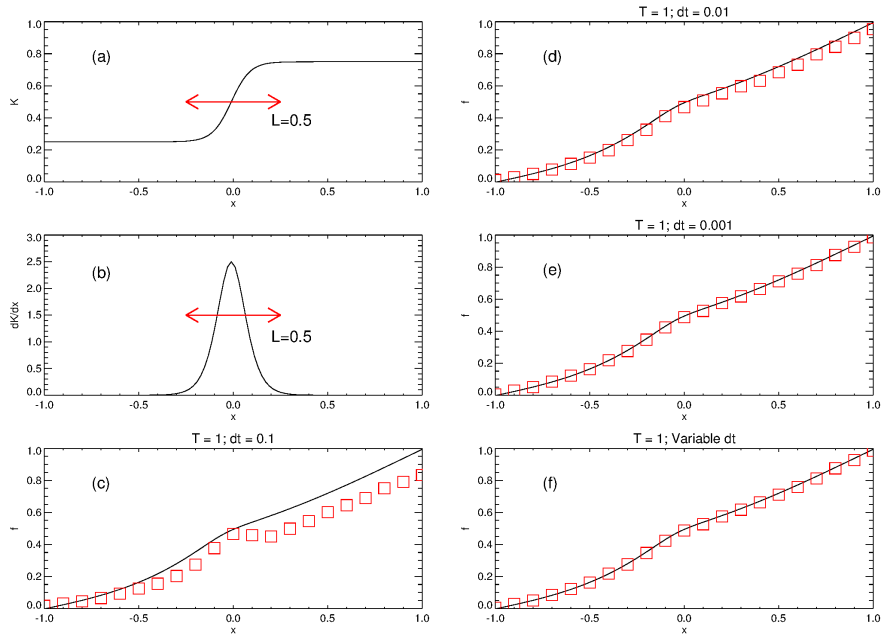}
    \caption{Comparing the SDE (open squares) and finite difference
      (solid line) methods for different choices of $\Delta t$. Panel
      (a) shows $\kappa(x)$, panel (b) its derivative and panels (c) -
      (e) the SDE results for $\Delta t = 0.1$, $\Delta t = 0.01$ and
      $\Delta t = 0.001$ respectively. The SDE solutions in panel (f)
      adopt a variable time step.}
    \label{fig3_1}
\end{figure}

In this section, we again consider the 1D PDE
\begin{equation}
\frac{\partial f(x,t)}{\partial t} = \frac{\partial}{\partial x} \left( \kappa(x) \frac{\partial f(x,t)}{\partial x}  \right)
\end{equation}
where $\kappa$ is no longer constant, but assumed to be
\begin{equation}
\kappa(x) = 0.5 + 0.25 \tanh \left( 10x \right).
\end{equation}
$\kappa(x)$, and its derivative, are shown in panels (a) and (b) of
Fig. \ref{fig3_1}. Note that $\kappa(x)$ changes by a factor of three
over a length scale of ${L}=0.5$; the meaning of which will
become clear during this section.

The corresponding SDE formulation is
\begin{equation}
dx = \frac{d \kappa(x)}{dx} dt + b(x)  dW
\end{equation}
where $b(x) = \sqrt{2\kappa(x)}$.

We focus on the effect of changing the numerical step size
$\Delta t$ in the SDE solver. For the simulation
used here, no initial condition is assumed, with $f(x=-1,t)=0$ and
$f(x=1,t)=1$ and the results are compared to those of the FD code at
time $T=1$.

Unlike finite differences, the SDE method does not have a prescribed
restriction on the time step; i.e. no equivalent SDE CFL \citep[Courant-Friedrichs-Lewy, see][]{CFL} formulation
is applicable. This has advantages, namely that the SDE method is
therefore unconditionally stable. However, as we will show, the SDE
solution does not converge to the correct solution if $\Delta t$ is
chosen to be too large.

Panel (c) of Fig. \ref{fig3_1} shows the results when $\Delta t = 0.1$
is used in the SDE scheme (red squares) as compared to the FD scheme
(solid line) which derives its value of $\Delta t$ from the
appropriate CFL condition. This choice of $\Delta t$ is clearly too
large and the comparison shows that the SDE results are clearly
incorrect. In panel (d) a reasonable value of $\Delta t = 0.01$ is
adopted. However, the results of the SDE model, especially for $x>0$,
is clearly still incorrect. Only for $\Delta t = 0.001$ (panel e) do
both numerical methods converge.

A question is however how to correctly choose the value of $\Delta t$?
Our methodology to do this consistently starts with evaluating the
first- and second-order moments of the SDE
\begin{eqnarray}
\langle \Delta x \rangle &=& \frac{d\kappa(x)}{dx} \Delta t\\
\langle \Delta x^2 \rangle &=& 2\kappa(x) \Delta t = b^2(x) \Delta t.
\end{eqnarray}

In the SDE approach, it can be assumed that the pseudo-particles, on
average, move a distance
$l^2=\max \left\{ \langle \Delta x \rangle^2, \langle \Delta x
  ^2\rangle \right\} $
in a time step $\Delta t$. We can therefore use the numerical spatial
step size to constrain the time step, i.e. by specifying a required
average jump length $l$ we can determine what $\Delta t$ should
be. For the diffusion equation we are currently considering, the
transport parameters change over a spatial region of
${L}$. Assuming we want the pseudo-particles to sample this
transition, we require that $l \ll {L}$. The time step is thus
constrained by the expression
\begin{equation}
\label{Eq:varying_delta_t}
\Delta t(x)   =\min \left\{  \frac{l}{\left| \frac{d\kappa(x)}{dx}\right| }   ,  \frac{l^2}{b^2(x)}    \right\}.
\end{equation}
where $l$ is specified and must be therefore shorter than the shortest
spatial structure present in the computational
domain. \citet{krullsachterberg1994} state that a necessary
requirement for an SDE model to be accurate is given
$\langle \Delta x \rangle^2 \ll \langle \Delta x^2 \rangle$. Using our
specific example, this imposes the following condition on the
timestep, $\Delta t \ll 2 \kappa / (d \kappa /dx)^2$, stating that the
stochastic integration must be dominated by stochastic processes and
not by deterministic drift terms. Depending on the integration scheme
used, the deterministic terms may be traced, however, quite accurately
\citep{Kloeden}.

For our present example, the smallest length scale of interest is that
over which $\kappa$ changes and we choose $l =0.1 {L}=0.05$.  The
results of the SDE scheme using this variable $\Delta t$ are shown in
panel (f) of Fig. \ref{fig3_1}. It can be seen that for this choice of
$\Delta t$ a good agreement exists between the SDE and FD schemes.

It is thus clear that, in order to correctly solve an SDE numerically,
the modeled pseudo-particles must sample all features in the
computational region. This can be done very efficiently, as
illustrated in this section, by adopting a variable $\Delta t$,
constrained by the first- and second-order moments of the SDE and
prescribing the shortest length scale that must be sampled,
$l$. Computationally, this is also much cheaper than using an
artificially short time step over the entire spatial grid; $\Delta t$
is only reduced where e.g. $\kappa$ changes significantly but could be
larger elsewhere.

\subsection{Literature Review}
\label{subsec:time_step_review}

In the previous section we have illustrated how crucial the choice of an appropriate time step is in a SDE model; the general rule of thump being that the pseudo-particles should sample the smallest region of interest in the computational domain. In the following section we discuss two scenarios for which this is especially important.

A further scenario was already alluded to in Section \ref{Sec:cosmic_rays}, namely that of drifts along the HCS. As shown by, e.g., \citet{burgeretal1985}, CRs will experience the HCS, and start drifting along, when they are within 2 Larmor radii ($r_L$) from it. Therefore, based on the discussion in the previous section, we need to choose $\Delta t$ such that $l \ll 2r_L$ in order to capture the full effect of the HCS on CR modulation. Furthermore, because $r_L$ changes with both particle energy and spatial position (due to the changing HMF magnitude), is would be logical to implement a varying $\Delta t$ in such a numerical model.

\subsubsection{Diffusive Shock Acceleration}

\begin{figure}[!ht]
    \centering
    \includegraphics[width=0.45\textwidth]{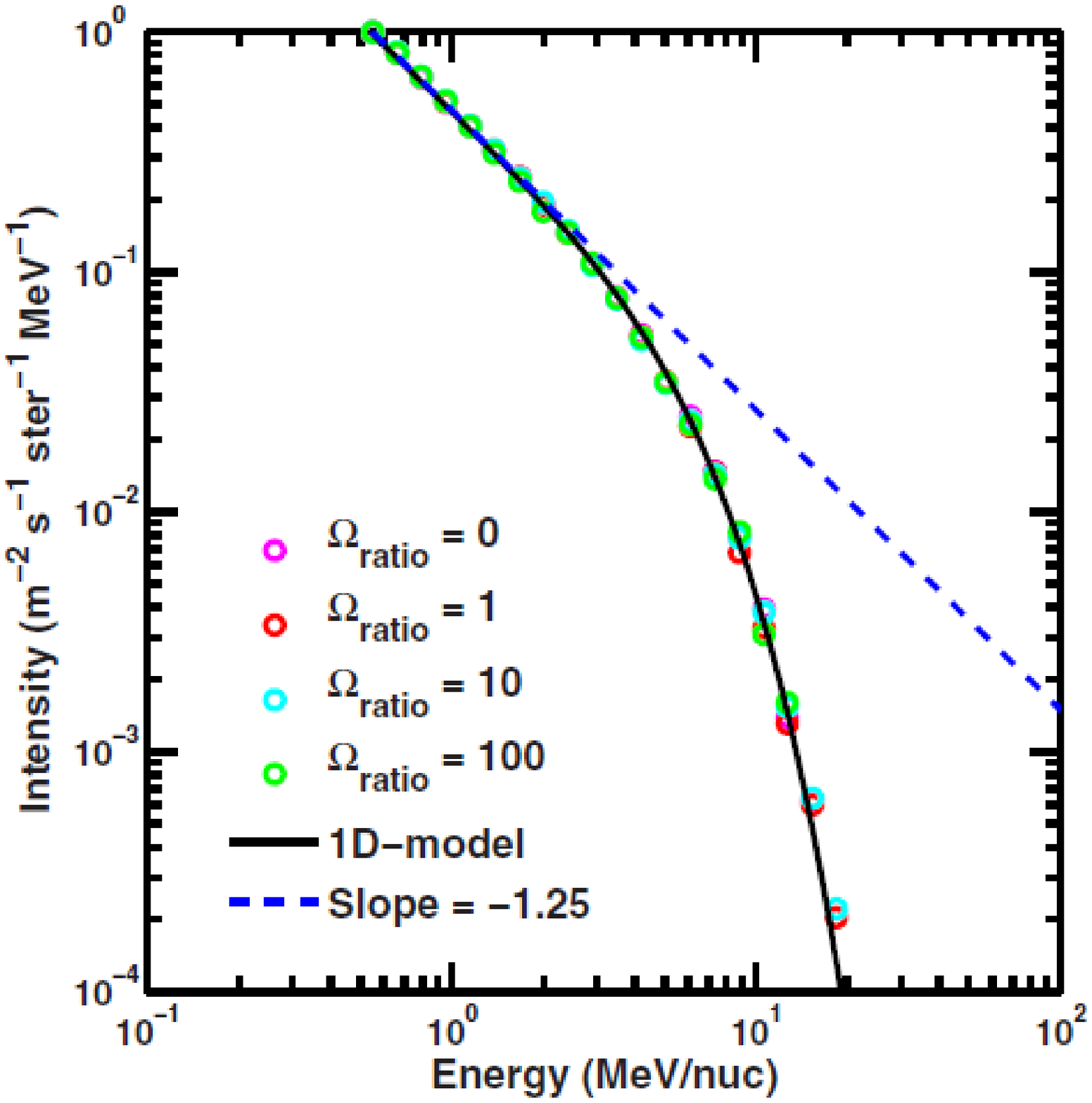}
        \includegraphics[width=0.45\textwidth]{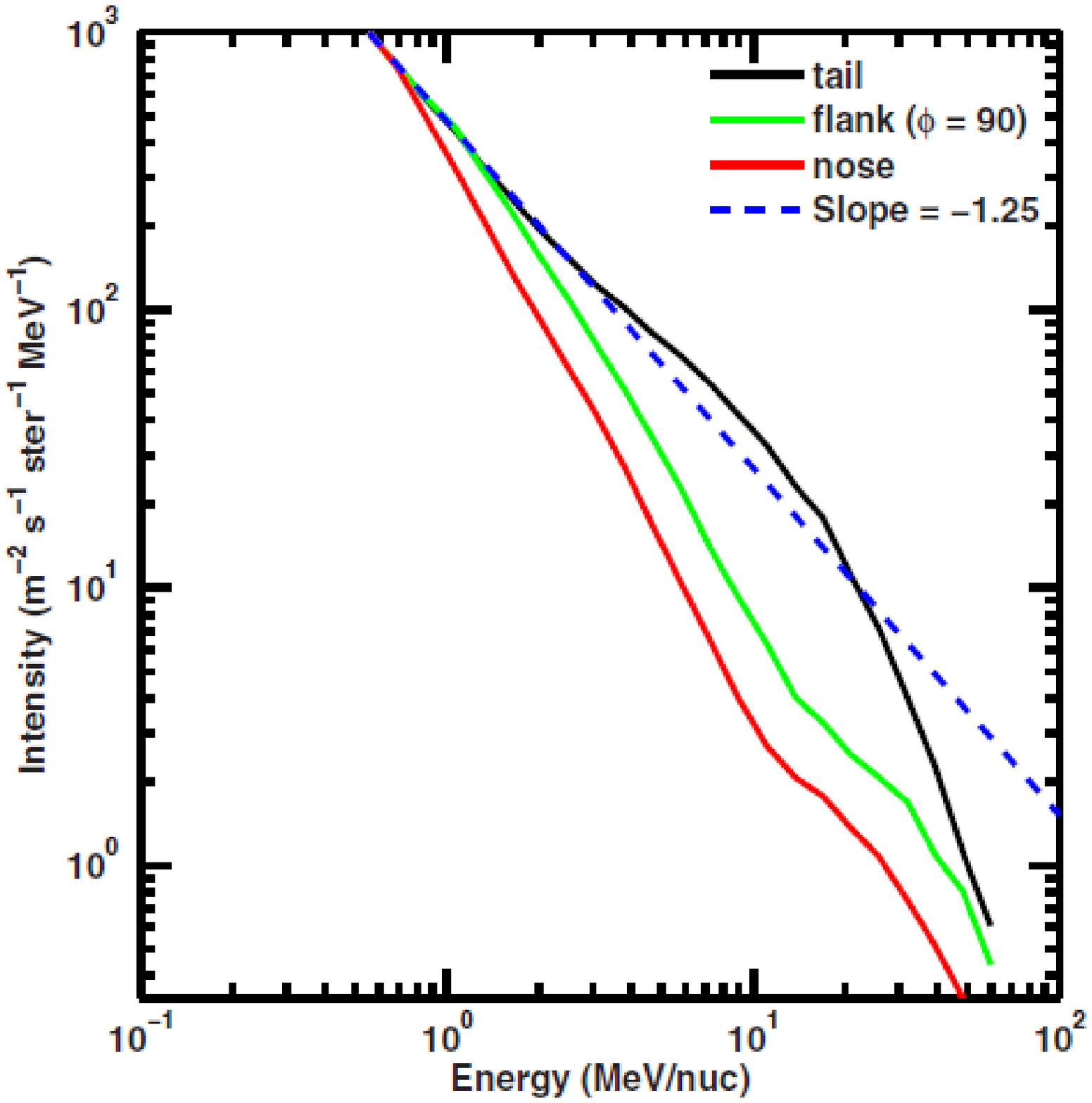}
        \caption{Simulating DSA with an SDE model, as applied to ACR
          acceleration at the TS \citep[taken
          from][]{senanayake_florinski_2013}. The left panel shows a
          comparison between the SDE results for spherically symmetric
          scenario (the symbols) and that of a 1D FD model (the solid
          line) as compared to the analytical result for an infinite
          planar shock (dashed line). The right panel shows the
          resulting ACR spectrum at the TS when a more realistic
          heliospheric geometry is incorporated, leading to preferred
          acceleration at some regions along the TS. \copyright\
          AAS. Reproduced with permission.}
    \label{DSA_fig1}
\end{figure}

Without knowing its full implications, \citet{fermi1949} proposed the
basis for shock acceleration: CR particles are scattered by forward
and backward moving waves, making ``head-on" (resulting in a kinetic
energy gain) and ``trailing" (resulting in a loss of kinetic energy)
collisions. If the intensity of the forward and backward moving waves
are roughly equal, the process can lead to diffusion of the original
CR distribution function in energy space. The process is referred to
as momentum diffusion, or Fermi II acceleration -- referring to the
fact that the energy gain per collision cycle is
$\sim \left( v/c \right)^2$, where $v$ is particle speed and $c$ is
the speed of light. In the presence of a shock wave (where the
supersonic flow slows to subsonic speeds), and assuming the CRs have
mobility across the shock, the process is much more efficient -- CRs
make more head-on collisions when moving upstream of the shock than
trailing collisions when moving downstream -- and the process is
referred to Fermi I acceleration (the energy gain per collision cycle
is $\sim v/c$), or more frequently, diffusive shock acceleration (DSA)
\footnote{A shock is, however, not a prerequisite for Fermi I
  acceleration to occur. See, e.g., the SDE modeling by
  \citet{armstrong_etal_2012} and the references therein, where a
  beam of CRs can be accelerated via Fermi I acceleration when waves,
  travelling in a preferred direction, are present.}. DSA is
considered the dominant process responsible for particle
acceleration at supernovae blast waves (thus creating the GCR
component) and, more, locally at the heliospheric termination shock
(TS, thus creating the anomalous cosmic ray (ACR) component), while
travelling interplanetary shocks can also energize CRs. DSA can be
modeled in the Parker formalism, by noting that the plasma flow
divergence term, $\nabla \cdot \vec{u}$, governing energy changes,
becomes large and negative at a shock, and if treated correctly, can
be used to simulate DSA on the distribution function level. For
details, see the comprehensive review by \citet{drury1983}, or, more recently, by \citet{Fichtner2001}.

In order to numerically model DSA, Parker's equation needs to be
solved, accounting for the large value of $\nabla \cdot \vec{u}$
directly at the shock. Mainly for numerical reasons, but also
motivated by physical arguments \citep[see e.g.][]{leroux_etal_1996},
the shock is taken to have a finite width (denoted by
$L_{\mathrm{shock}}$). In the SDE formulation, the pseudo-particles
should therefore sample this acceleration region $L_{\mathrm{shock}}$
continuously and not only interact with the shock sporadically. In the
terminology of the previous section, the numerical step size should be
such that $l \ll L_{\mathrm{shock}}$. A thorough discussion regarding
this, and SDE model solutions including DSA, are given by
\citet{achterberg_krulls_1992}, \citet{krullsachterberg1994},
\citet{park_petrosian_1996}, \citet{Marcowith-Kirk-1999} and more recently by
\citet{schterberg_schure_2011}. The technique of using a variable
$\Delta t$, as illustrated in the preceding section, is thus ideally
suited for simulating DSA at quasi-discontinuous shocks via SDE models
\footnote{\citet{zhang_2000} explores a technique for simulating DSA,
  at a discontinuous TS by means of SDEs using what is termed to be
  ``skew Brownian motion", where the spatial coordinates are rescaled
  in order to remove the discontinuity.}.

The acceleration of ACRs at the TS was recently modeled, utilizing a
SDE approach, by \citet{senanayake_florinski_2013}, with some of their
results shown in Fig. \ref{DSA_fig1}. The acceleration of ACRs proved
to be a contentious research topic after Voyager 1 did not observed
the expected power law spectrum at the TS
\citep[e.g.][]{stoneetal2005,deckeretal2005}, but rather a modulated
form. Although several explanations followed, including the inclusion
of momentum diffusion into the transport model
\citep[e.g.][]{strauss_etal_2010}, this modulated form is generally
attributed to more efficient acceleration of ACRs at certain positions
along the TS, also evident from the results shown in
Fig. \ref{DSA_fig1}.

Recently, modeling DSA for anisotropic particle distributions,
i.e. adopting the more general \citet{skilling_1971} transport
equation, has attracted some attention, with
\citet{Zuo_etal_2011,zuo_etaL_2013} and \citet{Kartavykh-etal-2016} applying a pitch-angle dependent
SDE model.

\subsubsection{Outer Heliosheath Modulation}

\begin{figure}[!ht]
    \centering
    \includegraphics[width=1\textwidth]{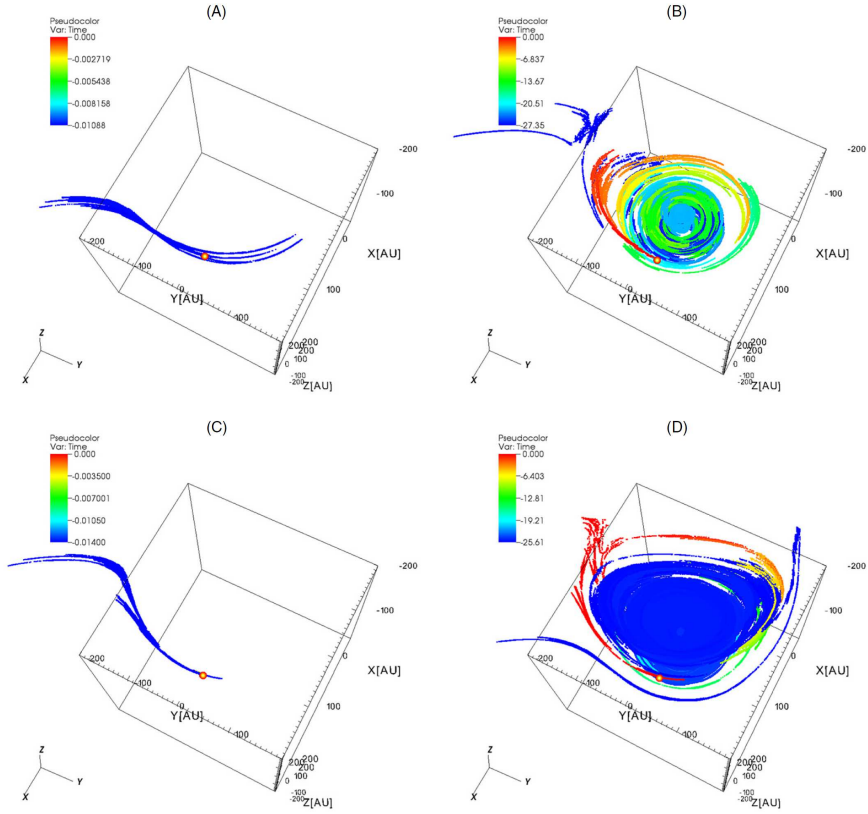}
    \caption{An example of pseudo-particle trajectories in a hybrid CR
      modulation model, taken from
      \citet{guo_florinski_2014}. Although the diffusion coefficients
      change by orders of magnitude across the HP, a variable
      $\Delta t$ forces the pseudo-particles to sample different
      regions of the computational regime with the same spatial
      precision. Here, the authors choose $l$ to be smaller than the
      grid size of the MHD model. This is essential to capture the
      full effect of CR modulation across the HP.  \copyright\
          AAS. Reproduced with permission.}
    \label{HP_fig1}
\end{figure}

Anticipating the Voyager crossing of the HP, \citet{schereretal2011}
questioned the use of a Dirichlet boundary condition for GCR
modulation at the HP \footnote{In an earlier paper by
  \citet{jokipii2001}, the use of a Dirichlet boundary condition was
  motivated by analytical considerations.}. Because GCRs diffuse
through both the heliosphere and the local interstellar medium, it is
quite conceivable that GCRs might ``leak" into the heliosphere faster
than they are replenished from the galaxy, leading to a positive
gradient outside the HP. If this is true, the LIS for GCRs (i.e. the
Dirichlet boundary condition) must be places at some distance beyond
the HP (the region between the HP and the bow-shock if referred to as
the outer heliosheath; the region of interstellar plasma still
influenced by the presence of the heliosphere). A study by
\citet{schereretal2011}, using SDEs, and a follow-up study by
\citet{herbstetal2012} found that this might be the case, depending on
the assumed parameters. \citet{kota_jokipii_2014}, using a
semi-analytical model, however, suggested that GCR modulation beyond
the HP is not possible.

The models mentioned in the previous paragraph are simplified, to such
an extent that they may not be able to capture the complex nature of
GCR transport beyond the HP. To mitigate this,
\citet{straussetal2013b} applied a hybrid GCR modulation model to the
problem (see also the discussion in Section
\ref{Ssec:Moduilation_results}). From a modeling point-of-view, the
transport of GCRs across the HP presents an intriguing problem: The
diffusion coefficients across the HP change by up to a factor of
$10^5$, due to the different turbulent processes believed to occur in
these distinct regions (the HP can be considered a tangential
discontinuity so that the two plasmas -- the heliospheric plasma on the
one side and the interstellar medium on the other -- cannot mix and
interact). This large, fairly abrupt change in $\mathbf{K}$ leads to a
large $\nabla \cdot \mathbf{K}$ drift type term that, if handled
incorrectly, can lead to incorrect results. For this reason,
\citet{straussetal2013b} employed a variable $\Delta t$, in a similar
fashion to Eq. \ref{Eq:varying_delta_t}, mimicking a SDE-type CFL
condition. The pseudo-particles can, therefore, not have unreasonable
high propagation speeds and are able to sample the HP region
sufficiently. Results from this paper corroborated that of
\citet{schereretal2011}: Modulation beyond the HP is possible, but the
level of modulation is heavily parameter-dependent. Two more hybrid
GCR modulation models that addressed this possibility followed;
firstly by \citet{guo_florinski_2014} (with some of their results
highlighted in Fig. \ref{HP_fig1}) and recently by \citet{zhang2015} and
\citet{luo_etaL_2016}. Although the latter confirmed the results of
\citet{straussetal2013b}, the former rejected these. The question of
whether modulation of GCRs beyond the HP occurs, and therefore also
whether a Dirichlet boundary condition at the HP is appropriate,
remains unanswered.

Directly after crossing the HP, Voyager 1 measured anisotropic CR
distributions in the interstellar medium
\citep[][]{krimigis_etal_2013}. This necessitates the use of a
pitch-angle dependent CR transport equation
\citep[e.g.][]{skilling_1971}, rather than the Parker equation, valid
only for isotropic CR distributions. Due to the strong gradients
emphasized earlier in this section, SDE models were again applied to
the problem by \citet{florinski_etaL_2013} and
\citet{strauss_fichtner_2015}. Explaining these anisotropies remains an
ongoing topic of study.

\newpage
\section{Handling Source/Sink Terms: Galactic Cosmic Ray Propagation}
\epigraph{\emph{We demand rigidly defined areas of doubt and uncertainty!}}{h2g2}
\label{Sec:galactic_transport}

\begin{figure}[!ht]
    \centering
    \includegraphics[width=1\textwidth]{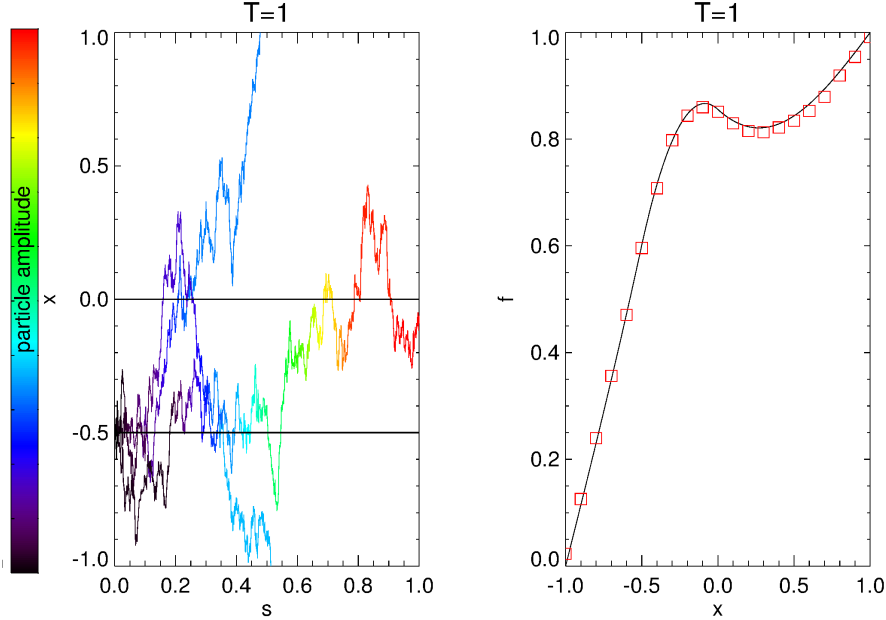}
    \caption{The left panel shows traces of pseudo-particles starting
      at $x=-0.5$, colored according to their amplitude as given by Eq. \ref{Eq:amplitude}: Initially
      $a=0$ (as indicated by black), but $a$ increases when the
      pseudo-particles interact with the source region. The right
      panel again compares the results of the SDE and FD models, using
      $10 000$ particles at each $x$ point in the SDE approach.  }
    \label{fig4_1}
\end{figure}

\begin{figure}[!ht]
    \centering
    \includegraphics[width=1\textwidth]{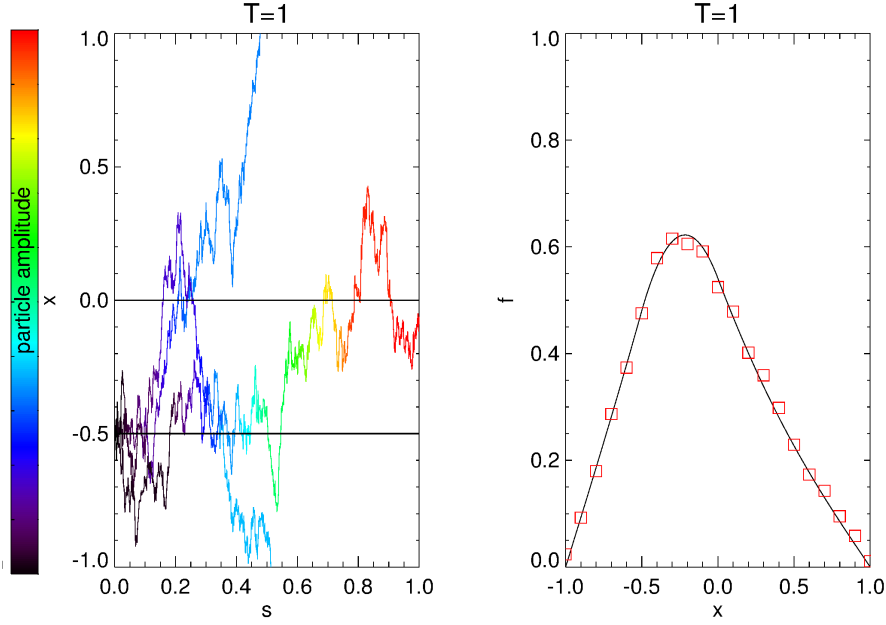}
    \caption{Similar to Fig. \ref{fig4_1}, but using the boundary conditions $f(x=-1,t)=f(x=1,t)=0$.  }
    \label{fig4_2}
\end{figure}

We consider again the 1D diffusion equation,

\begin{equation}
\label{Eq:with_sources}
\frac{\partial f(x,t)}{\partial t} =  \kappa \frac{\partial^2 f}{\partial x^2} + Q(x)
\end{equation}
with $\kappa = 0.5$ constant, but with the addition of a source term $Q(x)$, given by
\begin{equation}
Q(x) = \left\{ \begin{array}{l}
2  : x \in [-0.5,0]  \\
0  : \text{otherwise}.  \end{array} \right.
\end{equation}

The equivalent SDE formulation is once again
\begin{equation}
dx =  b  dW,
\end{equation}
but, with the addition of sources, a particle amplitude $a$ needs to
be calculated along the integration trajectory for each
pseudo-particle. This quantity is initially zero for each
pseudo-particle $a_i^{s=0}=0$ (where $i$ refers to a single
particle). The value of $a$ then changes during integration as
\begin{equation}
\label{Eq:amplitude}
a_i(s + ds) = a_i(s) + Q(x)ds.
\end{equation}
The value of $a$ can be positive (as in our example where a source is
considered), but can also be negative (if sinks are considered, i.e
$Q<0$).

To calculate $f$, Eq. \ref{Eq:general_convolution} now
needs to be generalized to account for the varying particle amplitude
\begin{equation}
f(x,T) = \underbrace{\int_0^T \int_x G(x',t) f_b(x',t)   dx dt}_\text{contribution from boundary conditions} + \underbrace{\mathcal{A}(x,T)}_\text{source contribution},
\label{Eq:solution_sources}
\end{equation}
where 
\begin{equation}
\mathcal{A}(x,T) =  \frac{1}{N} \sum_{i=1}^{N} a_{i}(s=T)
\end{equation}
in its discretized form and where $N$ is the total number of
pseudo-particles solved at a given $x$.

The particle amplitude can be considered as a ``correction term" when
sources/sinks are considered. This can be understood when re-writing
Eq. \ref{Eq:with_sources} as

\begin{equation}
\frac{\partial f(x,t)}{\partial t} =  \kappa \frac{\partial^2 f}{\partial x^2} + \frac{da}{dt}
\end{equation}

with the auxiliary equation

\begin{equation}
\label{Eq:aux_eq}
\frac{d a}{d t} = Q(x) \Rightarrow \Delta a = Q(x) \Delta t
\end{equation}

which must be solved simultaneously. The ``standard" SDE methodology
is followed to solve the diffusion part, while $a$ (the particle
amplitude) is updated continuously along the pseudo-particle's
trajectory by solving Eq. \ref{Eq:aux_eq}, via Eq. \ref{Eq:amplitude},
and adding the averaged correction term, $\mathcal{A}$, in
Eq. \ref{Eq:solution_sources}.

The SDE solutions are presented in Fig. \ref{fig4_1} using the
boundary conditions $f(x=-1,t)=0$ and $f(x=1,t)=1$. The right panel of
the figure shows a comparison between the SDE solutions (red squares)
and a FD scheme (solid line), showing good agreement. The left panel
shows the trajectory of 3 pseudo-particles originating from $x=-0.5$,
where the curves are colored according to their particle amplitude
(black corresponds to a zero value and red to a high positive
value). It is clear that $a_i$ changes when the source region is
encountered.

Fig. \ref{fig4_2} is similar to Fig. \ref{fig4_1}, but this time using
the boundary conditions $f(x=-1,t)=f(x=1,t)=0$. For this case, the
only contribution to $f$ is from the $\mathcal{A}$ term in
Eq. \ref{Eq:solution_sources}.

\subsection{Galactic cosmic ray propagation studies}
The transport of cosmic rays (CRs) in our galaxy is a topic of
considerable interest, both from a fundamental point of view and due
to its connection to the questions of CR origin and acceleration
\citep{schlikeiser2002, Strong-etal-2007}. It is furthermore believed
that the CR component of the interstellar medium plays an important
role in its overall dynamics, for example in the generation of the
galactic magnetic field through dynamo processes
\citep[e.g.][]{Hanasz-etal-2009}, or galactic wind production
\citep[e.g.][]{Fichtner-etal-1991, Breitschwerdt-etal-1991,
  Everett-etal-2010}. A very recent overview on CR transport and its
historic development can, for example, be found in
\citet{Schlickeiser-2015}.

In the past, most of the detailed numerical studies of galactic CR
propagation relied on grid-based schemes to solve the
diffusion-convection transport equation. The most widely used tool in
the community, GALPROP \citep{Strong-Moskalenko-1998}, for example, is
based on a finite-difference scheme. The SDE approach offers a distinct
alternative to such grid-based schemes. In particular, with the
time-backward method, the distribution function can be calculated to
high precision for a few phase-space points of interest, without any
restriction from numerical stability and the need for a highly
resolved computational grid for the entire galaxy. However, to enable
CR modeling for the galactic transport problem, a proper treatment of
CR sources, as introduced in this section, is essential.

\subsubsection{The Galactic transport problem}
Similar to the transport equation in CR modulation
(Eq.~\ref{Eq:TPE_parker}), the propagation of CRs in the galaxy
is described by a diffusion-convection equation for the differential
galactic CR intensity $j(\vec{r},p,t)=p^2f(\vec{r},p,t)$
\citep[e.g.][]{Strong-etal-2007}:
\begin{equation}
\label{Eq:gcr}
  \displaystyle
  \frac{\partial j}{\partial t} =
  \nabla\cdot\left(\mathbf{K}\nabla\, j - \vec{u} j \right) 
  - \frac{\partial}{\partial p}\left[\dot{p} j
    - \frac{p}{3}(\nabla\cdot\vec{u}) j \right]
  + Q
\end{equation}
where we have neglected already any momentum diffusion and $\vec{u}$
denotes again the effect of combined background advection effects,
e.g. of a galactic wind. Momentum loss processes enter the coefficient
$\dot{p}$; for protons in the GeV energy range, for example, the
dominant loss processes are ionization and pion production
\citep[e.g.][]{Mannheim-Schlickeiser-1994}. Further linear terms can
appear in the equation to describe catastrophic losses or sources of
particles, through spallation and/or annihilation reactions
\citep[e.g.][]{Busching-etal-2005,Blasi-Amato-2012}. Although the
diffusion of CR particles is in general given by a diffusion tensor
$\mathbf{K}$, the majority of studies up to now have assumed only a
single scalar diffusion coefficient. $Q$ represents the combination of
all CR sources, like for example supernova remnants, pulsars or even
astrospheres of other stars
\citep{Scherer-etal-2008,Scherer-etal-2015}. A transport equation like
Eq.~\ref{Eq:gcr} can be solved for any CR species of interest, and
they are coupled by a complicated nuclear reaction network
\citep{Strong-etal-2007}.

\subsubsection{Selected results}
Instead of elaborating further about the many details that enter the
CR transport equation, we restrict ourselves here to a concise
discussion of some of the recent studies that made use of the SDE
solution method and those related to them.

One of the first studies to use the complete SDE equivalence for the
solution of the galactic transport equation was
\citet{Farahat-etal-2008}. They already employed a sophisticated
nuclear reaction network and benchmarked their model by comparing
their resulting primary to secondary ratios with previous GALPROP
calculations by \citet{Strong-Moskalenko-1998} and observations.  They
particularly emphasized the usefulness of the backward time
integration to determine the distribution at specific positions in the
galaxy and also the easy treatment of extended sources with this
method.

\begin{figure}[!t]
    \centering
    \includegraphics[width=1\textwidth]{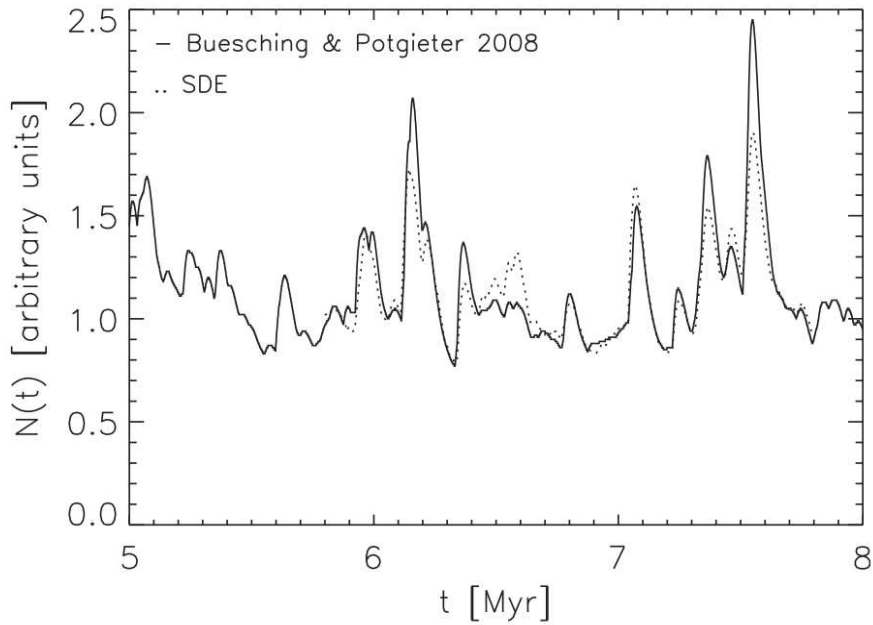}
    \caption{From \citet{Kopp-etal-2014}: Temporal variation of the
      cosmic ray flux for 1~GeV protons for an inter-arm region. The solid
      line gives the \citet{Buesching-Potgieter-2008} result, while
      the dashed line is obtained by the novel SDE code with the same
      set of point sources. Reprinted with permission from Elsevier.}
    \label{fig:kopp-benchmark}
\end{figure}

The influence of CR source distributions on the CR flux variation
throughout the galaxy has gained increased attention in recent
years. \citet{Buesching-Potgieter-2008} already considered a source
distribution consisting of more than a hundred thousand discrete
supernova remnants (SNRs) with locations following the galactic spiral
pattern \citep[e.g.][]{Vallee-1995,Vallee-2014}. In their study, they
did not employ an SDE method but rather numerics based on a
Fourier-Bessel series and a Crank-Nicholson scheme.  Their results
served as a benchmark for the study by \citet{Kopp-etal-2014}, in
which the SDE method was used to solve a similar problem (see
Fig.~\ref{fig:kopp-benchmark}). Furthermore, they considered the
influence of a spatial variation of the diffusion coefficient with the
spiral arm pattern. Both studies show a significant variation of the
CR flux along the solar orbit around the galactic center (i.e. between
arm and inter-arm regions) as well as shorter temporal variations due
to the discreetness of the sources in space and time. Different spiral
source distributions have also recently been studied by
\citet{Werner-etal-2015} and \citet{Kissmann-etal-2015} using novel
grid based elliptic solvers \citep{Kissmann-2014}.

Another study on the influence of discrete sources using the SDE
method has recently been published by \citet{Miyake-etal-2015}. They
show how the SDE method can yield path lengths and age distributions
for the arrival of CRs at Earth, thus emphasizing the strong influence
of discrete sources on the observed flux and primary/secondary
ratios. These aspects have been discussed already earlier by,
e.g. \citet{Blasi-Amato-2012} and \citet{Mertsch-2011}.

The backward SDE methods was employed in
\citet{Effenberger-etal-2012b} to study the influence of a full
diffusion tensor on the GCR propagation in a continuous spiral source
model. The background galactic magnetic field was chosen to be aligned
with the spiral arm pattern so that the diffusion could be set to a
smaller value perpendicular to the arms.
\begin{figure}[!t]
    \centering
    \includegraphics[width=1\textwidth]{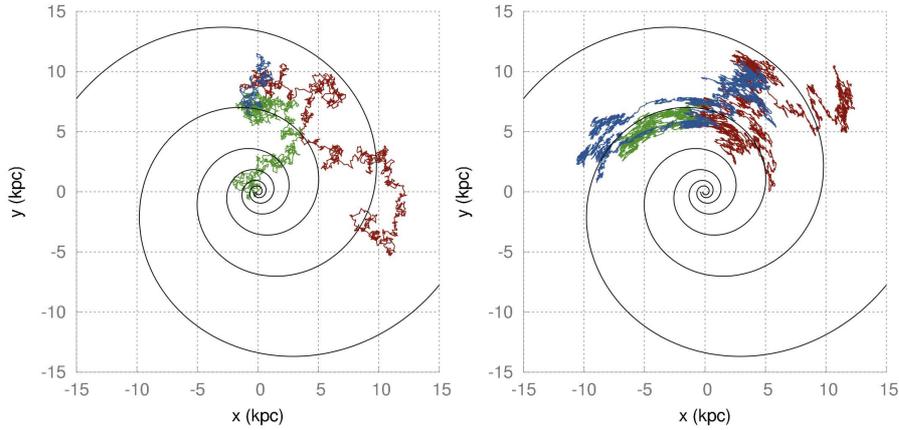}
    \caption{From \citet{Effenberger-etal-2012b}: Exemplary
      pseudo-particle trajectories in the galactic magnetic field,
      projected onto the galactic plane. The left panel is for
      isotropic diffusion while the right panel shows the effect of
      anisotropic diffusion with
      $\kappa_{\perp}/\kappa_{\parallel}=0.1$. Reproduced with
      permission from Astronomy \& Astrophysics, \copyright\ ESO.}
    \label{fig:effenberger-traj}
\end{figure}
Figure~\ref{fig:effenberger-traj} shows how the trajectories of pseudo
particles allow to visualize how the phase-space is traced differently
when anisotropic diffusion is considered, i.e. when the diffusion
along the mean magnetic field, $\kappa_{||}$, is more effective than
diffusion perpendicular to it, $\kappa_{\perp}$. Anisotropic diffusion
was found to be also of importance for the details and efficiency of
galactic magnetic dynamo action
\citep{Hanasz-etal-2009,Pakmor-etal-2016}.

\newpage
\section{Importance Sampling: Solar Energetic Particle Transport and Diffusive Shock Acceleration}
\epigraph{\emph{One of the things Ford Prefect had always found hardest to understand about humans was their habit of continually stating and repeating the very very obvious.}}{h2g2}
\label{Sec:importance_sampling}

Consider the diffusion equation with a convective part added
\begin{equation}
\frac{\partial f(x,t)}{\partial t} = V \frac{\partial f}{\partial x} + \kappa \frac{\partial^2 f}{\partial x^2}
\end{equation}
where we choose $V=-2$ and $\kappa = 0.5$ as constants. 

\begin{figure}[!ht]
    \centering
    \includegraphics[width=1\textwidth]{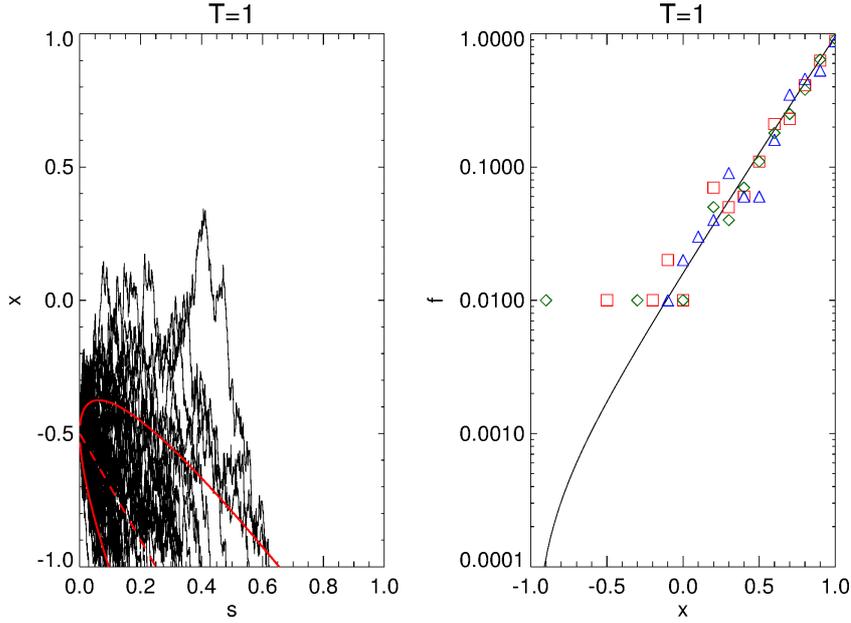}
    \caption{The left panel shows $50$ pseudo-particle traces starting
      from $x=-0.5$, while the right panel compared the finite
      difference (solid line) and SDE solutions (symbols) using three
      different sequences of Wiener processes.}
    \label{fig5_1}
\end{figure}

The equivalent SDE formulation is
\begin{equation}
dx = V dt + b  dW
\end{equation}
Using the boundary conditions, $f(x=-1,t)=0$ and $f(x=1,t)=1$,
Fig. \ref{fig5_1} shows the solution at $T=1$. The left panel shows
pseudo-particle traces for $50$ particles, starting from $x=-0.5$,
along with the average position of the particles
$y=\langle \Delta x\rangle$ (dashed red line; the first-order moment)
and the $1\sigma$ deviation from this value,
$y' = \langle \Delta x\rangle \pm \sqrt{\langle \Delta x^2\rangle}$
(solid red line; essentially the second-order moment). The right panel
shows the solution of the finite difference model (solid line) and
the SDE model (symbols). Three cases for the SDE model are shown
(denoted by the different symbols). These correspond to the same
simulation ($100$ pseudo-particles at each position) but using
different sequences of Wiener processes, i.e. each time the
integration process is initiating with a different seed.

For the set-up discussed above, the SDE model runs into problems with
the statistics of the resulting pseudo-particles. With the large
negative value of $V$, the pseudo-particles are convected towards
smaller $x$ and hence only a very limited number interact with the
non-zero boundary at $x=1$. With so few particles reaching it, the
statistics of the solution decreases, i.e. the SDE solution becomes
more noisy. This is clear from the right panel of the figure, which
illustrates that, especially for $x<0$, the SDE solution oscillates
around the actual solution. As $x \rightarrow -1$, this effect of
course increases with the SDE solution showing a very coarse discrete
nature, i.e. if only $1$ in $100$ particles interact with the
boundary, the value of $f$ becomes $f=0.01$. For two particles,
$f=0.02$, etc.

There are two ways in which to overcome this difficulty. This first is
to increase the number of pseudo-particles used. This brute-force
method is however computationally very expensive, and does not imply
much better statistics. A second, perhaps more elegant method would
be to implement an ``importance sampling" technique. We discuss here a
hands-on method to implement importance sampling for our present model
in order to enhances the statistics of the SDE solution near
$x=-1$. The basic idea is to add (in a self-consistent manner) an
additional and artificial convective term into the SDE equation,
thereby forcing the pseudo-particles to sample a spatial region we are
interested in. For this case, we want them to be convected more
towards $x=1$.

We start by converting the original $f(x,t)$ to a modified function $g(x,t)$, 
\begin{equation}
A(x)g(x,t) \equiv f(x,t)
\end{equation}
where $A(x)$ is a function chosen later on. As will be shown below,
the dependence of $A(x)$ on $x$ will lead to the desired additional
convection speed in the $x$-coordinate. The relevant PDE is now
\begin{equation}
\label{Eq:with_linear}
\frac{\partial g(x,t)}{\partial t} = \left( V + 2\frac{\kappa}{A}\frac{dA}{dx} \right) \frac{\partial g}{\partial x} + \kappa \frac{\partial^2 g}{\partial x^2} + \left( \frac{V}{A} \frac{dA}{dx} + \frac{\kappa}{A} \frac{d^2A}{dx^2} \right)g
\end{equation}
which includes an additional convective term, so that the total convective term becomes
\begin{equation}
V' =  \underbrace{V}_\text{original convection} + \underbrace{2\frac{\kappa}{A}\frac{dA}{dx}}_\text{artificial convection}
\end{equation}
where the additional convective speed can be chosen arbitrarily in
order to increase the statistics of the SDE solver. Note that this
equation also has a linear gain/loss term with a coefficient,
\begin{equation}
 \mathcal{L} = \frac{V}{A} \frac{dA}{dx} + \frac{\kappa}{A} \frac{d^2A}{dx^2}
\end{equation}
which needs to be taken into account and is discussed below.

The equivalent SDE formulation is then
\begin{equation}
dx = V' dt + b  dW,
\end{equation}
with the addition that, because linear terms are present in the PDE,
the particle weight for each pseudo-particle also needs to be
evaluated at each time step,
\begin{equation}
\label{Eq:solve_weight}
\alpha_i(s + ds) = \alpha_i(s) \exp \left\{ \mathcal{L}(x_i(s)) ds  \right\}
\end{equation}
where, initially $\alpha_i(s=0) = 1$ and the index $i$ refers to a
single pseudo-particle. 

Similarly to the particle amplitude discussed in the previous section,
the particle weight is a correction introduced when linear terms are
present in the PDE. Casting Eq. \ref{Eq:with_linear} into the form
\begin{equation}
\frac{\partial g(x,t)}{\partial t} = V' \frac{\partial g}{\partial x} + \kappa \frac{\partial^2 g}{\partial x^2} + \frac{d\alpha}{dt}
\end{equation}
with, once again, an axillary term
\begin{equation}
\label{Eq:weight_aux}
\frac{d\alpha}{dt} = \mathcal{L}\alpha \Rightarrow \Delta \alpha = \exp \left\{ \mathcal{L} \Delta t  \right\}.
\end{equation}

The standard SDE methodology is used to solve the diffusion equation,
while the particle weight accounts for the axillary equation,
Eq. \ref{Eq:weight_aux}, by solving it, via Eq. \ref{Eq:solve_weight},
along the trajectory of each pseudo-particle.

Say, for our example, we require that
\begin{equation}
V' =  V + 2\frac{\kappa}{A}\frac{dA}{dx} = 1,
\end{equation}
i.e that the artificially introduced convection speed counteracts the
original convection speed. The required value of $A(x)$ is readily
evaluated (once again we choose $V=-2$ for the actual convection
speed),
\begin{equation}
\frac{1}{A}\frac{dA}{dx} = \frac{d\ln A(x)}{dx} = 3 \Rightarrow A(x) = C e^{3x}.
\end{equation}
Note that the value of the integration constant, $C$, cancels in all
calculations. These values will be used in the following simulations
to illustrate the effectiveness of the importance sampling technique
in enhancing the statistics of the normal SDE solver.

\begin{figure}[!ht]
    \centering
    \includegraphics[width=1\textwidth]{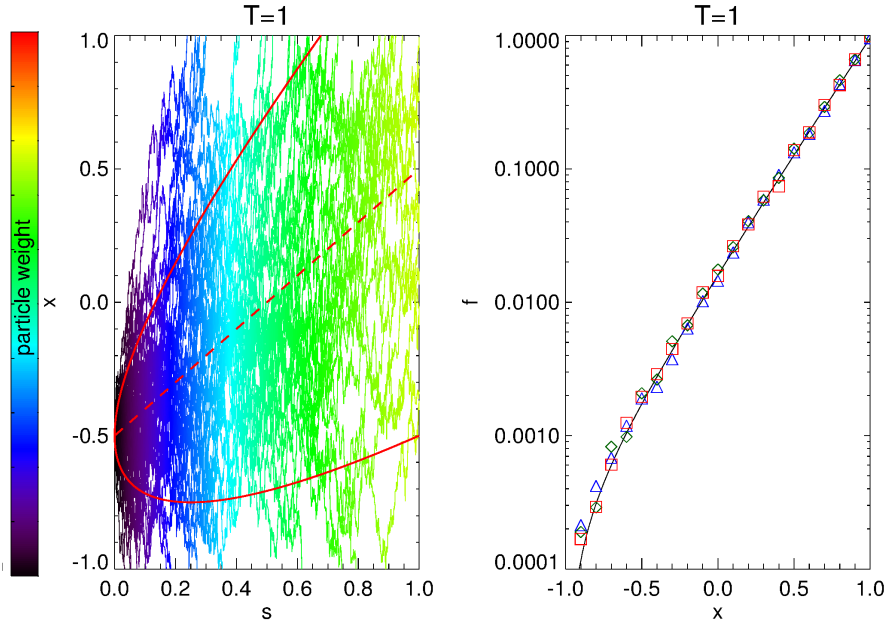}
    \caption{Similar to Fig. \ref{fig5_1}, but now with the inclusion of importance sampling.}
    \label{fig5_2}
\end{figure}

Similar to the particle amplitude, the value of an individual
pseudo-particle's weight also enter the calculation of $f$, but this
time, changing $G(x,T)$ directly. When a particle reaches a temporal
or spatial integration boundary, it is not counted (in the discretized
version) as a single particle, but as a fraction $\alpha_i(s=T)$ of a
particle (depending on the sign of $\mathcal{L}$ this fraction may be
either smaller or larger than unity). In Eq. \ref{Eq:calculate_G_1}
the binning procedure was explained, for the case of no linear terms,
as counting the number of pseudo-particles in each bin $(N_j)$ and
normalizing by the total number of particles $(N)$, i.e.
\begin{equation}
G(x,T)dx \approx \frac{N_j}{N} = \frac{1}{N} \sum_{i=1}^{N_j}1
\end{equation}
when $\alpha \neq 1$ (i.e. when linear terms are introduced) this now
becomes the addition of non-integers e.g.
\begin{equation}
G(x,T)dx \approx \frac{1}{N} \sum_{i=1}^{N_j} \alpha^{i}(s=T) .
\end{equation}
The convolution with the boundary condition remains unchanged,
although the normalization condition is no longer true
\begin{equation}
\int_x \int_t G(x,t)dtxt \neq 1.
\end{equation}

The results of this modified PDE are shown in Fig. \ref{fig5_2}. The
left panel shows again the pseudo-particle traces (technically now
used to calculate $g(x,t)$), colored according to the particle
weight, while the right compares the calculated $f(x,t)$ to the finite
difference model. Now, all regions of the computational region have
adequate sampling and more accurate solutions are obtained. The number
of solved pseudo-particle trajectories are the same as in
Fig. \ref{fig5_1}, but the statistics of the solution is much better,
especially near the $x=-1$ boundary.

\subsection{Literature Review}

The importance sampling technique appears to be applicable in any
modeling scenario where problems related to pseudo-particle statistics
are present, i.e. pseudo-particles, due to the processes they represent,
are unable to reach model boundaries, cannot contribute to the
intensity, and hence, become {\it redundant} in some sense. This
technique has not been implemented extensively in astrophysical SDE
applications, although a limited number, discussed below, of such
implementations do exist.

\subsubsection{Solar Energetic Particles}

Solar energetic particles (SEPs) are energetic particles accelerated
near the Sun during transient solar events. As reviewed by
\citet{reames_2013}, SEPs events can be divided into two classes:
So-called {\it impulsive events}, where SEP acceleration is believed
to occur at solar flares and {\it gradual events}, where the
acceleration is though to occur at the shock associated with a
propagating coronal mass ejection. Primarily due to very efficient
focussing near the Sun (the diverging HMF tends to focus the SEPs into
a narrow beam), any SEP distribution is usually highly anisotropic (in
terms of pitch-angle), so that the \citet{parker1965} TPE cannot be
used to model their propagation. A more general, focused transport
approach is needed
\citep[e.g.][]{roelof1969,skilling_1971,ruffolo1995,Litvinenko-2012,Litvinenko-Noble-2013,fred_2014,Litvinenko-etal-2015,ek_en_horst}. We
also mention that some models of pick-up ion transport in the
heliosphere, like for example in \citet{Chalov-etal-1995} and
\citet{Fichtner-etal-1996}, employ SDE methods which are related to
SEP transport models in many aspects. These applications are not
discussed further in this review.

The transport of SEPs has been modelled extensively, using the time
forward SDE formalism, by e.g. \citet{drogeetal2010},
\citet{dresing2012}, \citet{droge2014} and \citet{timo}. When
simulating SEP events in the time-backward SDE framework, a problem
related to pseudo-particle statistics is, however, present: Starting
from, e.g., Earth, the pseudo-particles are traced, as always, back
towards their source, which, for SEPs, is a small spatial region close
to the Sun. The problem is that only an extremely small fraction of
the pseudo-particles will reach this small source region; this is
clear by simply comparing the volume of the source region (usually
placed at $r=0.005$ AU) to that of the simulation volume (with an
outer boundary placed at a minimum distance of $r=3$ AU), to give the
expected probability of SEPs reaching the source region as
$\sim 5 \times 10^{-7}$ \%. To overcome this difficulty,
\citet{qin_etal_2005} and \citet{2zhang_etal_2009} implemented an
importance sampling technique, along the lines discussed in this work,
very successfully. In essence, an artificial convective velocity is
added to advect the pseudo-particles towards the inner boundary. If
such an additional convective speed is included in a self-consistent
manner, as illustrated in the previous paragraph, the results are both
valid, and have much increased statistics. See also the more recent
modelling of e.g. \citet{QinWang_2015}.

We further want to note here that also in the context of solar flare
physics SDEs have been employed successfully. The Fokker-Planck
equation used to describe the particle acceleration in a flare is very
similar to the transport equations discussed previously. It describes
the energy gains through a second order process in momentum, which is
due to interactions with the turbulence background created by the
reconnection events during a flare \citep[e.g.][]{Petrosian-2012}.
\citet{Mackinnon-Craig-1991} noted already early that the evolution of
the electron distribution in space and pitch-angle due to binary
collisions can be described by an equivalent set of ordinary
stochastic differential equations. \citet{Fletcher-1995},
\citet{Fletcher-1997} and \citet{Jeffrey-etal-2014} applied the SDE
method to the impulsive loop-top emission of solar flares and their
trapping due to magnetic fields \citep{Fletcher-1997} and
\citet{park_petrosian_1996} compared different solution methods for
the flare acceleration problem. More generally, the stochastic
acceleration of solar flare protons and electrons has been discussed
by e.g.\
\citet{Barbosa-1979,Schlickeiser-Steinacker-1989,Miller-Roberts-1995}
and \citet{Petrosian-Liu-2004}.

\subsubsection{Shock Acceleration}

A problem with pseudo-particle statistics may also be present
when simulating DSA in a time-backward fashion: The pseudo-particles
are initially released, e.g., at the acceleration region (shock) with
an energy of $\sim 100$ MeV. These particles, in the time-backward
scenario, constantly loose energy while being in the acceleration region,
until some of them might reach the injection energy of the source
population - this injection energy is usually much lower than the
energies of interest, e.g., $\sim 100$ keV. For an ensemble of
pseudo-particles, at any higher energy, to have any contribution to
the total intensity, they therefore have to sample the acceleration region
long enough to reach the injection energy. Because spatial diffusion
is also present, the likelihood of this happening is extremely low,
and an importance sampling technique can be applied to remedy the lack
of pseudo-particle statistics, and interestingly, this can be
accomplished by two very different approaches.

The first approach would be to increase the amount of time a
pseudo-particle spends in the acceleration region. For example, to
include an artificial spatial advective speed to force the
pseudo-particles back towards the shock region. A second, and perhaps
more elegant approach, is to include an artificial convection speed in
energy space, so that the pseudo-particles reach the injection
energies faster. The latter approach was followed by Zhang et
al. (private communication, 2015), who concluded that such a
modification in an SDE model leads to an increase in pseudo-particle
statistics, while reproducing the standard SDE results.

\newpage
\section{Summary}
\epigraph{\emph{Forty-two}}{h2g2}

The transport equations describing the propagation of non-thermal
particles through turbulent plasmas have, in recent years, become
increasingly complex and must therefore be integrated numerically. An
increasingly popular numerical method is the use of SDEs. In this
review paper we have tried to give a practical introduction to SDEs
and how these equations are solved numerically for a variety of
Fokker-Planck type transport equations arising in different fields of
space- and astrophysics. We have focussed on possible numerical
pit-falls that may arise, e.g., choosing an appropriate time-step,
and, demonstrated how to overcome these issues by solving simpler 1D
``toy models". These models also included discussions regarding the
handling of sources/sinks and linear gain/loss terms in the SDE
formulation by keeping track of the so-called particle amplitude and
weight, and introduced a hands-on technique of incorporating
importance sampling in an SDE model. After each technical section, we
gave a brief overview of the current literature that have applied
these techniques, and presented selected results.

It is noteworthy to point to the possibility of further extensions of
the SDE scheme to non-diffusive, in particular sub- and superdiffusive
processes \citep[e.g.][]{Magdziarz-Weron-2007, Effenberger-2014,
  Stern-etal-2014}. The basic idea is to generalize the Wiener process
in the SDE formulation to a wider class of stochastic processes, for
example so called L\'evy or $\alpha$-stable processes, which describe
the deviations from Gaussian-like diffusion behavior. This in turn can
lead to more ``heavy-tailed" distributions and power-laws instead of
exponentials in particle configuration space or energy. We recommend
the reviews by \citet{Perrone-etal-2013} and
\citet{Zimbardo-etal-2015} to the interested reader, who wants to
learn more about the physics background and applications of these
concepts.

Although, from our experience in writing this review, the number of
studies using the SDE method in space physics and astrophysics is
still limited, the trend in the last years is certainly
encouraging. The advantages of SDE formulations and numerics for a
large class of particle kinetics problems due to the opportunities for
parallel computing are significant.  With this review, it is our hope
that we could provide a contribution in enabling future studies
using this powerful and yet simple method.

\begin{acknowledgements}
  We thank Horst Fichtner for his careful reading of the draft
  manuscript and Marius Potgieter, Ingo B{\"u}sching, Andreas Kopp,
  Yuri Litvinenko and Phillip Dunzlaff for collaborating on different
  aspects of SDE modelling. RDS is funded through the National Research Foundation (NRF) of South Africa. Opinions
  expressed and conclusions arrived at are those of the authors and
  are not necessarily to be attributed to the NRF. Partial financial
  support from the Alexander von Humboldt Foundation and the Fulbright
  Visiting Scholar Program are acknowledged. FE is supported by NASA
  grant NNX14AG03G and part of the work was completed during a
  fellowship at the University of Waikato.
\end{acknowledgements}

\bibliographystyle{aps-nameyear}      
\bibliography{References}                

\begin{thebibliography}{162}
\ifx \bisbn   \undefined \def \bisbn  #1{ISBN #1}\fi
\ifx \binits  \undefined \def \binits#1{#1} \fi
\ifx \bauthor  \undefined \def \bauthor#1{#1} \fi
\ifx \bjtitle  \undefined \def \bjtitle#1{\textrm{#1}}\fi
\ifx \batitle  \undefined \def \batitle#1{#1} \fi
\ifx \bctitle  \undefined \def \bctitle#1{#1} \fi
\ifx \bvolume  \undefined \def \bvolume#1{\textbf{#1}}\fi
\ifx \byear  \undefined \def \byear#1{#1} \fi
\ifx \bissue  \undefined \def \bissue#1{#1} \fi
\ifx \bfpage  \undefined \def \bfpage#1{#1} \fi
\ifx \blpage  \undefined \def \blpage #1{#1} \fi
\ifx \burl  \undefined \def \burl#1{#1} \fi
\ifx \doiurl  \undefined \def \doiurl#1{#1} \fi
\ifx \betal  \undefined \def \betal{et al.} \fi
\ifx \binstitute  \undefined \def \binstitute#1{#1} \fi
\ifx \beditor  \undefined \def \beditor#1{#1} \fi
\ifx \bpublisher  \undefined \def \bpublisher#1{#1} \fi
\ifx \bbtitle  \undefined \def \bbtitle#1{\textit{#1}} \fi
\ifx \bedition  \undefined \def \bedition#1{#1} \fi
\ifx \bseriesno  \undefined \def \bseriesno#1{#1} \fi
\ifx \blocation  \undefined \def \blocation#1{#1} \fi
\ifx \bsertitle  \undefined \def \bsertitle#1{#1} \fi
\ifx \bsnm \undefined \def \bsnm#1{#1} \fi
\ifx \bsuffix \undefined \def \bsuffix#1{#1} \fi
\ifx \bparticle \undefined \def \bparticle#1{#1} \fi
\ifx \barticle \undefined \def \barticle#1{#1} \fi
\ifx \botherref \undefined \def \botherref #1{#1} \fi
\ifx \url \undefined \def \url#1{#1} \fi
\ifx \bchapter \undefined \def \bchapter#1{#1} \fi
\ifx \bbook \undefined \def \bbook#1{#1} \fi
\ifx \bcomment \undefined \def \bcomment#1{#1} \fi
\ifx \oauthor \undefined \def \oauthor#1{#1} \fi
\ifx \citeauthoryear \undefined \def \citeauthoryear#1{#1} \fi
\ifx \texttildelow  \undefined \def \texttildelow{\symbol{126}} \fi
\def \endbibitem {}
\ifx \bconflocation  \undefined \def \bconflocation#1{#1} \fi

\bibitem[\protect\citeauthoryear{{Achterberg} and
  {Kr{\"u}lls}}{1992}]{achterberg_krulls_1992}
\begin{barticle}
\bauthor{\binits{A.} \bsnm{{Achterberg}}},
\bauthor{\binits{W.M.} \bsnm{{Kr{\"u}lls}}},
\batitle{{A fast simulation method for particle acceleration}}.
\bjtitle{Astron. Astrophys.}
\bvolume{265},
\bfpage{13}--\blpage{16}
(\byear{1992})
\end{barticle}
\endbibitem

\bibitem[\protect\citeauthoryear{{Achterberg} and
  {Schure}}{2011}]{schterberg_schure_2011}
\begin{barticle}
\bauthor{\binits{A.} \bsnm{{Achterberg}}},
\bauthor{\binits{K.M.} \bsnm{{Schure}}},
\batitle{{A more accurate numerical scheme for diffusive shock acceleration}}.
\bjtitle{MNRAS}
\bvolume{411},
\bfpage{2628}--\blpage{2636}
(\byear{2011}).
doi:\doiurl{10.1111/j.1365-2966.2010.17868.x}
\end{barticle}
\endbibitem

\bibitem[\protect\citeauthoryear{Adams}{1979}]{h2g2}
\begin{bbook}
\bauthor{\binits{D.} \bsnm{Adams}},
\bbtitle{{The Hitchhiker's Guide to the Galaxy}}
(\bpublisher{Pan Books},
\blocation{London}, \byear{1979})
\end{bbook}
\endbibitem

\bibitem[\protect\citeauthoryear{Alanko-Huotari et~al.}{2007}]{ososkin}
\begin{barticle}
\bauthor{\binits{K.} \bsnm{Alanko-Huotari}},
\bauthor{\binits{I.G.} \bsnm{Usoskin}},
\bauthor{\binits{K.} \bsnm{Mursula}},
\bauthor{\binits{G.A.} \bsnm{Kovaltsov}},
\batitle{{Stochastic simulation of cosmic ray modulation including a wavy
  heliospheric current sheet}}.
\bjtitle{J. Geophys. Res.}
\bvolume{112},
\bfpage{08101}
(\byear{2007})
\end{barticle}
\endbibitem

\bibitem[\protect\citeauthoryear{{Armstrong}
  et~al.}{2012}]{armstrong_etal_2012}
\begin{barticle}
\bauthor{\binits{C.K.} \bsnm{{Armstrong}}},
\bauthor{\binits{Y.E.} \bsnm{{Litvinenko}}},
\bauthor{\binits{I.J.D.} \bsnm{{Craig}}},
\batitle{{Modeling Focused Acceleration of Cosmic-Ray Particles by Stochastic
  Methods}}.
\bjtitle{Astrophys. J.}
\bvolume{757},
\bfpage{165}
(\byear{2012}).
doi:\doiurl{10.1088/0004-637X/757/2/165}
\end{barticle}
\endbibitem

\bibitem[\protect\citeauthoryear{{Ball} et~al.}{2005}]{balletal2005}
\begin{barticle}
\bauthor{\binits{B.} \bsnm{{Ball}}},
\bauthor{\binits{M.} \bsnm{{Zhang}}},
\bauthor{\binits{H.} \bsnm{{Rassoul}}},
\bauthor{\binits{T.} \bsnm{{Linde}}},
\batitle{{Galactic cosmic-ray modulation using a solar minimum MHD heliosphere:
  A stochastic particle approach}}.
\bjtitle{Astrophys. J.}
\bvolume{634},
\bfpage{1116}--\blpage{1125}
(\byear{2005})
\end{barticle}
\endbibitem

\bibitem[\protect\citeauthoryear{{Barbosa}}{1979}]{Barbosa-1979}
\begin{barticle}
\bauthor{\binits{D.D.} \bsnm{{Barbosa}}},
\batitle{{Stochastic acceleration of solar flare protons}}.
\bjtitle{\apj}
\bvolume{233},
\bfpage{383}--\blpage{394}
(\byear{1979}).
doi:\doiurl{10.1086/157399}
\end{barticle}
\endbibitem

\bibitem[\protect\citeauthoryear{{Blasi} and {Amato}}{2012}]{Blasi-Amato-2012}
\begin{barticle}
\bauthor{\binits{P.} \bsnm{{Blasi}}},
\bauthor{\binits{E.} \bsnm{{Amato}}},
\batitle{{Diffusive propagation of cosmic rays from supernova remnants in the
  Galaxy. I: spectrum and chemical composition}}.
\bjtitle{\jcap}
\bvolume{1},
\bfpage{10}
(\byear{2012}).
doi:\doiurl{10.1088/1475-7516/2012/01/010}
\end{barticle}
\endbibitem

\bibitem[\protect\citeauthoryear{{Bobik} et~al.}{2008}]{bobik_etal_2008}
\begin{barticle}
\bauthor{\binits{P.} \bsnm{{Bobik}}},
\bauthor{\binits{K.} \bsnm{{Kudela}}},
\bauthor{\binits{M.} \bsnm{{Boschini}}},
\bauthor{\binits{D.} \bsnm{{Grandi}}},
\bauthor{\binits{M.} \bsnm{{Gervasi}}},
\bauthor{\binits{P.G.} \bsnm{{Rancoita}}},
\batitle{{Solar modulation model with reentrant particles}}.
\bjtitle{Advances in Space Research}
\bvolume{41},
\bfpage{339}--\blpage{342}
(\byear{2008}).
doi:\doiurl{10.1016/j.asr.2007.02.085}
\end{barticle}
\endbibitem

\bibitem[\protect\citeauthoryear{{Bobik} et~al.}{2011}]{bobik_et_al_2011}
\begin{bchapter}
\bauthor{\binits{P.} \bsnm{{Bobik}}},
\bauthor{\binits{G.} \bsnm{{Boella}}},
\bauthor{\binits{M.J.} \bsnm{{Boschini}}},
\bauthor{\binits{C.} \bsnm{{Consolandi}}},
\bauthor{\binits{S.} \bsnm{{Della Torre}}},
\bauthor{\binits{M.} \bsnm{{Gervasi}}},
\bauthor{\binits{D.} \bsnm{{Grandi}}},
\bauthor{\binits{M.} \bsnm{{Elmo}}},
\bauthor{\binits{K.} \bsnm{{Kudela}}},
\bauthor{\binits{E.} \bsnm{{Memola}}},
\bauthor{\binits{S.} \bsnm{{Pensotti}}},
\bauthor{\binits{P.G.} \bsnm{{Rancoita}}},
\bauthor{\binits{D.} \bsnm{{Rozza}}},
\bauthor{\binits{M.} \bsnm{{Tacconi}}},
\bctitle{{Energy Loss for Electrons in the Heliosphere and Local Interstellar
  Spectrum for Solar Modulation}},
in \bbtitle{{Cosmic Rays for Particle and Astroparticle Physics}},
ed. by \beditor{\binits{S.} \bsnm{{Giani}}},
\beditor{\binits{C.} \bsnm{{Leroy}}},
\beditor{\binits{P.G.} \bsnm{{Rancoita}}},
\byear{2011},
pp. \bfpage{482}--\blpage{489}
\end{bchapter}
\endbibitem

\bibitem[\protect\citeauthoryear{{Bobik} et~al.}{2012}]{bobik_etal_2012}
\begin{barticle}
\bauthor{\binits{P.} \bsnm{{Bobik}}},
\bauthor{\binits{G.} \bsnm{{Boella}}},
\bauthor{\binits{M.J.} \bsnm{{Boschini}}},
\bauthor{\binits{C.} \bsnm{{Consolandi}}},
\bauthor{\binits{S.} \bsnm{{Della Torre}}},
\bauthor{\binits{M.} \bsnm{{Gervasi}}},
\bauthor{\binits{D.} \bsnm{{Grandi}}},
\bauthor{\binits{K.} \bsnm{{Kudela}}},
\bauthor{\binits{S.} \bsnm{{Pensotti}}},
\bauthor{\binits{P.G.} \bsnm{{Rancoita}}},
\bauthor{\binits{M.} \bsnm{{Tacconi}}},
\batitle{{Systematic Investigation of Solar Modulation of Galactic Protons for
  Solar Cycle 23 Using a Monte Carlo Approach with Particle Drift Effects and
  Latitudinal Dependence}}.
\bjtitle{Astrophys. J.}
\bvolume{745},
\bfpage{132}
(\byear{2012}).
doi:\doiurl{10.1088/0004-637X/745/2/132}
\end{barticle}
\endbibitem

\bibitem[\protect\citeauthoryear{Bobik et~al.}{2016}]{Bobik_etal_2016}
\begin{botherref}
\oauthor{\binits{P.} \bsnm{Bobik}},
\oauthor{\binits{M.J.} \bsnm{Boschini}},
\oauthor{\binits{S.D.} \bsnm{Torre}},
\oauthor{\binits{M.} \bsnm{Gervasi}},
\oauthor{\binits{D.} \bsnm{Grandi}},
\oauthor{\binits{G.L.} \bsnm{Vacca}},
\oauthor{\binits{S.} \bsnm{Pensotti}},
\oauthor{\binits{M.} \bsnm{Putis}},
\oauthor{\binits{P.G.} \bsnm{Rancoita}},
\oauthor{\binits{D.} \bsnm{Rozza}},
\oauthor{\binits{M.} \bsnm{Tacconi}},
\oauthor{\binits{M.} \bsnm{Zannoni}},
{On the Forward-Backward-in-time approach for Monte Carlo solution of Parker's
  transport equation: 1-dimensional case}.
J. Geophys. Res.,
(2016).
2015JA022237.
doi:\doiurl{10.1002/2015JA022237}.
\url{http://dx.doi.org/10.1002/2015JA022237}
\end{botherref}
\endbibitem

\bibitem[\protect\citeauthoryear{{Breitschwerdt}
  et~al.}{1991}]{Breitschwerdt-etal-1991}
\begin{barticle}
\bauthor{\binits{D.} \bsnm{{Breitschwerdt}}},
\bauthor{\binits{J.F.} \bsnm{{McKenzie}}},
\bauthor{\binits{H.J.} \bsnm{{Voelk}}},
\batitle{{Galactic winds. I - Cosmic ray and wave-driven winds from the
  Galaxy}}.
\bjtitle{\aap}
\bvolume{245},
\bfpage{79}--\blpage{98}
(\byear{1991})
\end{barticle}
\endbibitem

\bibitem[\protect\citeauthoryear{{Burger} and
  {Potgieter}}{1989}]{burgerpotgieter1989}
\begin{barticle}
\bauthor{\binits{R.A.} \bsnm{{Burger}}},
\bauthor{\binits{M.S.} \bsnm{{Potgieter}}},
\batitle{{The calculation of neutral sheet drift in two-dimensional cosmic-ray
  modulation models}}.
\bjtitle{Astrophys. J.}
\bvolume{339},
\bfpage{501}--\blpage{511}
(\byear{1989})
\end{barticle}
\endbibitem

\bibitem[\protect\citeauthoryear{Burger et~al.}{1985}]{burgeretal1985}
\begin{barticle}
\bauthor{\binits{R.A.} \bsnm{Burger}},
\bauthor{\binits{H.} \bsnm{Moraal}},
\bauthor{\binits{G.M.} \bsnm{Webb}},
\batitle{{Drift theory of charged particles in electric and magnetic fields}}.
\bjtitle{Astrophys. Space Sci.}
\bvolume{116},
\bfpage{107}--\blpage{129}
(\byear{1985})
\end{barticle}
\endbibitem

\bibitem[\protect\citeauthoryear{{B{\"u}sching} and
  {Potgieter}}{2008}]{Buesching-Potgieter-2008}
\begin{barticle}
\bauthor{\binits{I.} \bsnm{{B{\"u}sching}}},
\bauthor{\binits{M.S.} \bsnm{{Potgieter}}},
\batitle{{The variability of the proton cosmic ray flux on the Sun's way around
  the galactic center}}.
\bjtitle{Advances in Space Research}
\bvolume{42},
\bfpage{504}--\blpage{509}
(\byear{2008}).
doi:\doiurl{10.1016/j.asr.2007.05.051}
\end{barticle}
\endbibitem

\bibitem[\protect\citeauthoryear{{B{\"u}sching}
  et~al.}{2005}]{Busching-etal-2005}
\begin{barticle}
\bauthor{\binits{I.} \bsnm{{B{\"u}sching}}},
\bauthor{\binits{A.} \bsnm{{Kopp}}},
\bauthor{\binits{M.} \bsnm{{Pohl}}},
\bauthor{\binits{R.} \bsnm{{Schlickeiser}}},
\bauthor{\binits{C.} \bsnm{{Perrot}}},
\bauthor{\binits{I.} \bsnm{{Grenier}}},
\batitle{{Cosmic-Ray Propagation Properties for an Origin in Supernova
  Remnants}}.
\bjtitle{\apj}
\bvolume{619},
\bfpage{314}--\blpage{326}
(\byear{2005}).
doi:\doiurl{10.1086/426537}
\end{barticle}
\endbibitem

\bibitem[\protect\citeauthoryear{{B{\"u}sching}
  et~al.}{2008}]{buschingetaL2008}
\begin{barticle}
\bauthor{\binits{I.} \bsnm{{B{\"u}sching}}},
\bauthor{\binits{O.C.} \bsnm{{de Jager}}},
\bauthor{\binits{M.S.} \bsnm{{Potgieter}}},
\bauthor{\binits{C.} \bsnm{{Venter}}},
\batitle{{A Cosmic-Ray Positron Anisotropy due to Two Middle-Aged, Nearby
  Pulsars?}}
\bjtitle{\apjl}
\bvolume{678},
\bfpage{39}--\blpage{42}
(\byear{2008}).
doi:\doiurl{10.1086/588465}
\end{barticle}
\endbibitem

\bibitem[\protect\citeauthoryear{{Chalov} et~al.}{1995}]{Chalov-etal-1995}
\begin{barticle}
\bauthor{\binits{S.V.} \bsnm{{Chalov}}},
\bauthor{\binits{H.J.} \bsnm{{Fahr}}},
\bauthor{\binits{V.} \bsnm{{Izmodenov}}},
\batitle{{Spectra of energized pick-up ions upstream. of the heliospheric
  termination shock I. The role of Alfvenic turbulences.}}
\bjtitle{\aap}
\bvolume{304},
\bfpage{609}
(\byear{1995})
\end{barticle}
\endbibitem

\bibitem[\protect\citeauthoryear{{Chandrasekhar}}{1943}]{Chandrasekhar-1943}
\begin{barticle}
\bauthor{\binits{S.} \bsnm{{Chandrasekhar}}},
\batitle{{Stochastic Problems in Physics and Astronomy}}.
\bjtitle{Reviews of Modern Physics}
\bvolume{15},
\bfpage{1}--\blpage{89}
(\byear{1943}).
doi:\doiurl{10.1103/RevModPhys.15.1}
\end{barticle}
\endbibitem

\bibitem[\protect\citeauthoryear{Chenette}{1980}]{chenette1980}
\begin{barticle}
\bauthor{\binits{D.L.} \bsnm{Chenette}},
\batitle{{The propagation of Jovian electrons to Earth}}.
\bjtitle{J. Geophys. Res.}
\bvolume{85},
\bfpage{2243}--\blpage{2256}
(\byear{1980})
\end{barticle}
\endbibitem

\bibitem[\protect\citeauthoryear{{Corti} et~al.}{2016}]{Corti_et_al_2016}
\begin{barticle}
\bauthor{\binits{C.} \bsnm{{Corti}}},
\bauthor{\binits{V.} \bsnm{{Bindi}}},
\bauthor{\binits{C.} \bsnm{{Consolandi}}},
\bauthor{\binits{K.} \bsnm{{Whitman}}},
\batitle{{Solar Modulation of the Local Interstellar Spectrum with Voyager 1,
  AMS-02, PAMELA, and BESS}}.
\bjtitle{\apj}
\bvolume{829},
\bfpage{8}
(\byear{2016}).
doi:\doiurl{10.3847/0004-637X/829/1/8}
\end{barticle}
\endbibitem

\bibitem[\protect\citeauthoryear{Courant et~al.}{1928}]{CFL}
\begin{barticle}
\bauthor{\binits{R.} \bsnm{Courant}},
\bauthor{\binits{K.} \bsnm{Friedrichs}},
\bauthor{\binits{H.} \bsnm{Lewy}},
\batitle{{{\"U}ber die partiellen differenzengleichungen der mathematischen
  physik}}.
\bjtitle{Mathematische Annalen}
\bvolume{100},
\bfpage{32}--\blpage{74}
(\byear{1928})
\end{barticle}
\endbibitem

\bibitem[\protect\citeauthoryear{{Decker} et~al.}{2005}]{deckeretal2005}
\begin{barticle}
\bauthor{\binits{R.B.} \bsnm{{Decker}}},
\bauthor{\binits{S.M.} \bsnm{{Krimigis}}},
\bauthor{\binits{E.C.} \bsnm{{Roelof}}},
\bauthor{\binits{M.E.} \bsnm{{Hill}}},
\bauthor{\binits{T.P.} \bsnm{{Armstrong}}},
\bauthor{\binits{G.} \bsnm{{Gloeckler}}},
\bauthor{\binits{D.C.} \bsnm{{Hamilton}}},
\bauthor{\binits{L.J.} \bsnm{{Lanzerotti}}},
\batitle{{Voyager 1 in the foreshock, termination shock, and heliosheath}}.
\bjtitle{Science}
\bvolume{309},
\bfpage{2020}--\blpage{2024}
(\byear{2005})
\end{barticle}
\endbibitem

\bibitem[\protect\citeauthoryear{{Della Torre}
  et~al.}{2012}]{Della_torre_etal_2012}
\begin{barticle}
\bauthor{\binits{S.} \bsnm{{Della Torre}}},
\bauthor{\binits{P.} \bsnm{{Bobik}}},
\bauthor{\binits{M.J.} \bsnm{{Boschini}}},
\bauthor{\binits{C.} \bsnm{{Consolandi}}},
\bauthor{\binits{M.} \bsnm{{Gervasi}}},
\bauthor{\binits{D.} \bsnm{{Grandi}}},
\bauthor{\binits{K.} \bsnm{{Kudela}}},
\bauthor{\binits{S.} \bsnm{{Pensotti}}},
\bauthor{\binits{P.G.} \bsnm{{Rancoita}}},
\bauthor{\binits{D.} \bsnm{{Rozza}}},
\bauthor{\binits{M.} \bsnm{{Tacconi}}},
\batitle{{Effects of solar modulation on the cosmic ray positron fraction}}.
\bjtitle{Advances in Space Research}
\bvolume{49},
\bfpage{1587}--\blpage{1592}
(\byear{2012}).
doi:\doiurl{10.1016/j.asr.2012.02.017}
\end{barticle}
\endbibitem

\bibitem[\protect\citeauthoryear{{Dresing} et~al.}{2012}]{dresing2012}
\begin{barticle}
\bauthor{\binits{N.} \bsnm{{Dresing}}},
\bauthor{\binits{R.} \bsnm{{G{\'o}mez-Herrero}}},
\bauthor{\binits{A.} \bsnm{{Klassen}}},
\bauthor{\binits{B.} \bsnm{{Heber}}},
\bauthor{\binits{Y.} \bsnm{{Kartavykh}}},
\bauthor{\binits{W.} \bsnm{{Dr{\"o}ge}}},
\batitle{{The Large Longitudinal Spread of Solar Energetic Particles During the
  17 January 2010 Solar Event}}.
\bjtitle{\solphys}
\bvolume{281},
\bfpage{281}--\blpage{300}
(\byear{2012}).
doi:\doiurl{10.1007/s11207-012-0049-y}
\end{barticle}
\endbibitem

\bibitem[\protect\citeauthoryear{{Dr{\"o}ge} et~al.}{2010}]{drogeetal2010}
\begin{barticle}
\bauthor{\binits{W.} \bsnm{{Dr{\"o}ge}}},
\bauthor{\binits{Y.Y.} \bsnm{{Kartavykh}}},
\bauthor{\binits{B.} \bsnm{{Klecker}}},
\bauthor{\binits{G.A.} \bsnm{{Kovaltsov}}},
\batitle{{Anisotropic Three-Dimensional Focused Transport of Solar Energetic
  Particles in the Inner Heliosphere}}.
\bjtitle{\apj}
\bvolume{709},
\bfpage{912}--\blpage{919}
(\byear{2010}).
doi:\doiurl{10.1088/0004-637X/709/2/912}
\end{barticle}
\endbibitem

\bibitem[\protect\citeauthoryear{{Dr{\"o}ge} et~al.}{2014}]{droge2014}
\begin{barticle}
\bauthor{\binits{W.} \bsnm{{Dr{\"o}ge}}},
\bauthor{\binits{Y.Y.} \bsnm{{Kartavykh}}},
\bauthor{\binits{N.} \bsnm{{Dresing}}},
\bauthor{\binits{B.} \bsnm{{Heber}}},
\bauthor{\binits{A.} \bsnm{{Klassen}}},
\batitle{{Wide longitudinal distribution of interplanetary electrons following
  the 7 February 2010 solar event: Observations and transport modeling}}.
\bjtitle{Journal of Geophysical Research (Space Physics)}
\bvolume{119},
\bfpage{6074}--\blpage{6094}
(\byear{2014}).
doi:\doiurl{10.1002/2014JA019933}
\end{barticle}
\endbibitem

\bibitem[\protect\citeauthoryear{{Drury}}{1983}]{drury1983}
\begin{barticle}
\bauthor{\binits{L.} \bsnm{{Drury}}},
\batitle{{On particle acceleration in supernova remnants}}.
\bjtitle{Space Sci. Rev.}
\bvolume{36},
\bfpage{57}--\blpage{60}
(\byear{1983})
\end{barticle}
\endbibitem

\bibitem[\protect\citeauthoryear{{Dunzlaff} et~al.}{2015}]{flippie}
\begin{barticle}
\bauthor{\binits{P.} \bsnm{{Dunzlaff}}},
\bauthor{\binits{R.D.} \bsnm{{Strauss}}},
\bauthor{\binits{M.S.} \bsnm{{Potgieter}}},
\batitle{{A stochastic differential equation code for multidimensional
  Fokker-Planck type problems}}.
\bjtitle{Comp. Phys. Comm.}
\bvolume{192},
\bfpage{156}--\blpage{165}
(\byear{2015})
\end{barticle}
\endbibitem

\bibitem[\protect\citeauthoryear{{Effenberger}}{2014}]{Effenberger-2014}
\begin{bchapter}
\bauthor{\binits{F.} \bsnm{{Effenberger}}},
\bctitle{{Energetic Particle Transport with Stochastic Differential Equations:
  General Methods and the Extension to Anomalous Diffusion Regimes}},
in \bbtitle{{8th International Conference of Numerical Modeling of Space Plasma
  Flows (ASTRONUM 2013)}},
ed. by \beditor{\binits{N.V.} \bsnm{{Pogorelov}}},
\beditor{\binits{E.} \bsnm{{Audit}}},
\beditor{\binits{G.P.} \bsnm{{Zank}}}
\bsertitle{{Astronomical Society of the Pacific Conference Series}},
vol. \bseriesno{488},
\byear{2014},
p. \bfpage{201}
\end{bchapter}
\endbibitem

\bibitem[\protect\citeauthoryear{{Effenberger} and
  {Litvinenko}}{2014}]{fred_2014}
\begin{barticle}
\bauthor{\binits{F.} \bsnm{{Effenberger}}},
\bauthor{\binits{Y.E.} \bsnm{{Litvinenko}}},
\batitle{{The Diffusion Approximation versus the Telegraph Equation for
  Modeling Solar Energetic Particle Transport with Adiabatic Focusing. I.
  Isotropic Pitch-angle Scattering}}.
\bjtitle{Astrophys. J.}
\bvolume{783},
\bfpage{15}
(\byear{2014}).
doi:\doiurl{10.1088/0004-637X/783/1/15}
\end{barticle}
\endbibitem

\bibitem[\protect\citeauthoryear{{Effenberger}
  et~al.}{2012a}]{effenbergeretal2012}
\begin{barticle}
\bauthor{\binits{F.} \bsnm{{Effenberger}}},
\bauthor{\binits{H.} \bsnm{{Fichtner}}},
\bauthor{\binits{K.} \bsnm{{Scherer}}},
\bauthor{\binits{S.} \bsnm{{Barra}}},
\bauthor{\binits{J.} \bsnm{{Kleimann}}},
\bauthor{\binits{R.D.} \bsnm{{Strauss}}},
\batitle{{A generalized diffusion tensor for fully anisotropic diffusion of
  energetic particles in the heliospheric magnetic field}}.
\bjtitle{Astrophys. J.}
\bvolume{750},
\bfpage{108}
(\byear{2012}a)
\end{barticle}
\endbibitem

\bibitem[\protect\citeauthoryear{{Effenberger}
  et~al.}{2012b}]{Effenberger-etal-2012b}
\begin{barticle}
\bauthor{\binits{F.} \bsnm{{Effenberger}}},
\bauthor{\binits{H.} \bsnm{{Fichtner}}},
\bauthor{\binits{K.} \bsnm{{Scherer}}},
\bauthor{\binits{I.} \bsnm{{B{\"u}sching}}},
\batitle{{Anisotropic diffusion of Galactic cosmic ray protons and their
  steady-state azimuthal distribution}}.
\bjtitle{\aap}
\bvolume{547},
\bfpage{120}
(\byear{2012}b).
doi:\doiurl{10.1051/0004-6361/201220203}
\end{barticle}
\endbibitem

\bibitem[\protect\citeauthoryear{{Engelbrecht} and
  {Burger}}{2013}]{engelbrechtburger2013}
\begin{barticle}
\bauthor{\binits{N.E.} \bsnm{{Engelbrecht}}},
\bauthor{\binits{R.A.} \bsnm{{Burger}}},
\batitle{{An Ab Initio Model for Cosmic-ray Modulation}}.
\bjtitle{\apj}
\bvolume{772},
\bfpage{46}
(\byear{2013}).
doi:\doiurl{10.1088/0004-637X/772/1/46}
\end{barticle}
\endbibitem

\bibitem[\protect\citeauthoryear{{Everett} et~al.}{2010}]{Everett-etal-2010}
\begin{barticle}
\bauthor{\binits{J.E.} \bsnm{{Everett}}},
\bauthor{\binits{Q.G.} \bsnm{{Schiller}}},
\bauthor{\binits{E.G.} \bsnm{{Zweibel}}},
\batitle{{Synchrotron Constraints on a Hybrid Cosmic-ray and Thermally Driven
  Galactic Wind}}.
\bjtitle{\apj}
\bvolume{711},
\bfpage{13}--\blpage{24}
(\byear{2010}).
doi:\doiurl{10.1088/0004-637X/711/1/13}
\end{barticle}
\endbibitem

\bibitem[\protect\citeauthoryear{{Farahat} et~al.}{2008}]{Farahat-etal-2008}
\begin{barticle}
\bauthor{\binits{A.} \bsnm{{Farahat}}},
\bauthor{\binits{M.} \bsnm{{Zhang}}},
\bauthor{\binits{H.} \bsnm{{Rassoul}}},
\bauthor{\binits{J.J.} \bsnm{{Connell}}},
\batitle{{Cosmic Ray Transport and Production in the Galaxy: A Stochastic
  Propagation Simulation Approach}}.
\bjtitle{\apj}
\bvolume{681},
\bfpage{1334}--\blpage{1340}
(\byear{2008}).
doi:\doiurl{10.1086/588374}
\end{barticle}
\endbibitem

\bibitem[\protect\citeauthoryear{{Fermi}}{1949}]{fermi1949}
\begin{barticle}
\bauthor{\binits{E.} \bsnm{{Fermi}}},
\batitle{{On the Origin of the Cosmic Radiation}}.
\bjtitle{Physical Review}
\bvolume{75},
\bfpage{1169}--\blpage{1174}
(\byear{1949}).
doi:\doiurl{10.1103/PhysRev.75.1169}
\end{barticle}
\endbibitem

\bibitem[\protect\citeauthoryear{{Fichtner}}{2001}]{Fichtner2001}
\begin{barticle}
\bauthor{\binits{H.} \bsnm{{Fichtner}}},
\batitle{{Anomalous cosmic rays: Messengers from the outer heliosphere}}.
\bjtitle{Space Sci. Rev.}
\bvolume{95},
\bfpage{639}--\blpage{754}
(\byear{2001})
\end{barticle}
\endbibitem

\bibitem[\protect\citeauthoryear{{Fichtner} et~al.}{1991}]{Fichtner-etal-1991}
\begin{barticle}
\bauthor{\binits{H.} \bsnm{{Fichtner}}},
\bauthor{\binits{H.J.} \bsnm{{Fahr}}},
\bauthor{\binits{W.} \bsnm{{Neutsch}}},
\bauthor{\binits{R.} \bsnm{{Schlickeiser}}},
\bauthor{\binits{A.} \bsnm{{Crusius-W{\"a}tzel}}},
\bauthor{\binits{H.} \bsnm{{Lesch}}},
\batitle{{Cosmic-ray-driven galactic wind.}}
\bjtitle{Nuovo Cimento B Serie}
\bvolume{106},
\bfpage{909}--\blpage{925}
(\byear{1991}).
doi:\doiurl{10.1007/BF02723185}
\end{barticle}
\endbibitem

\bibitem[\protect\citeauthoryear{{Fichtner} et~al.}{1996}]{Fichtner-etal-1996}
\begin{barticle}
\bauthor{\binits{H.} \bsnm{{Fichtner}}},
\bauthor{\binits{J.A.} \bsnm{{Le Roux}}},
\bauthor{\binits{U.} \bsnm{{Mall}}},
\bauthor{\binits{D.} \bsnm{{Rucinski}}},
\batitle{{On the transport of pick-up ions in the heliosphere.}}
\bjtitle{\aap}
\bvolume{314},
\bfpage{650}--\blpage{662}
(\byear{1996})
\end{barticle}
\endbibitem

\bibitem[\protect\citeauthoryear{{Fletcher}}{1995}]{Fletcher-1995}
\begin{barticle}
\bauthor{\binits{L.} \bsnm{{Fletcher}}},
\batitle{{On the generation of loop-top impulsive hard X-ray sources.}}
\bjtitle{\aap}
\bvolume{303},
\bfpage{9}
(\byear{1995})
\end{barticle}
\endbibitem

\bibitem[\protect\citeauthoryear{{Fletcher}}{1997}]{Fletcher-1997}
\begin{barticle}
\bauthor{\binits{L.} \bsnm{{Fletcher}}},
\batitle{{Numerical simulations of coronal particle trapping.}}
\bjtitle{\aap}
\bvolume{326},
\bfpage{1259}--\blpage{1267}
(\byear{1997})
\end{barticle}
\endbibitem

\bibitem[\protect\citeauthoryear{{Florinski} and
  {Pogorelov}}{2009}]{florinksipogorelov2009}
\begin{barticle}
\bauthor{\binits{V.} \bsnm{{Florinski}}},
\bauthor{\binits{N.V.} \bsnm{{Pogorelov}}},
\batitle{{Four-dimensional transport of galactic cosmic rays in the outer
  heliosphere and heliosheath}}.
\bjtitle{Astrophys. J.}
\bvolume{701},
\bfpage{642}--\blpage{651}
(\byear{2009})
\end{barticle}
\endbibitem

\bibitem[\protect\citeauthoryear{{Florinski}
  et~al.}{2013}]{florinski_etaL_2013}
\begin{barticle}
\bauthor{\binits{V.} \bsnm{{Florinski}}},
\bauthor{\binits{J.R.} \bsnm{{Jokipii}}},
\bauthor{\binits{F.} \bsnm{{Alouani-Bibi}}},
\bauthor{\binits{J.A.} \bsnm{{le Roux}}},
\batitle{{Energetic Particle Anisotropies at the Heliospheric Boundary}}.
\bjtitle{Astrophys. J. Lett.}
\bvolume{776},
\bfpage{37}
(\byear{2013}).
doi:\doiurl{10.1088/2041-8205/776/2/L37}
\end{barticle}
\endbibitem

\bibitem[\protect\citeauthoryear{Gardiner}{1983}]{Gardiner1983}
\begin{bbook}
\bauthor{\binits{C.W.} \bsnm{Gardiner}},
\bbtitle{{Handbook of Stochastic Methods for Physics, Chemistry and the Natural
  Sciences}}
(\bpublisher{Springer},
\blocation{Berlin}, \byear{1983})
\end{bbook}
\endbibitem

\bibitem[\protect\citeauthoryear{Gardiner}{2009}]{Gardiner}
\begin{bbook}
\bauthor{\binits{C.W.} \bsnm{Gardiner}},
\bbtitle{{Stochastic Methods: A Handbook for the Natural and Social Sciences}}
(\bpublisher{Springer},
\blocation{Berlin}, \byear{2009})
\end{bbook}
\endbibitem

\bibitem[\protect\citeauthoryear{{Gervasi} et~al.}{1999}]{gervasietal1999}
\begin{barticle}
\bauthor{\binits{M.} \bsnm{{Gervasi}}},
\bauthor{\binits{P.G.} \bsnm{{Rancoita}}},
\bauthor{\binits{I.G.} \bsnm{{Usoskin}}},
\bauthor{\binits{G.A.} \bsnm{{Kovaltsov}}},
\batitle{{Monte-Carlo approach to galactic cosmic ray propagation in the
  heliosphere}}.
\bjtitle{Nuclear Physics}
\bvolume{78},
\bfpage{26}--\blpage{31}
(\byear{1999})
\end{barticle}
\endbibitem

\bibitem[\protect\citeauthoryear{{Guo} and
  {Florinski}}{2014a}]{guo_florinski_2014_b}
\begin{barticle}
\bauthor{\binits{X.} \bsnm{{Guo}}},
\bauthor{\binits{V.} \bsnm{{Florinski}}},
\batitle{{Corotating interaction regions and the 27 day variation of galactic
  cosmic rays intensity at 1 AU during the cycle 23/24 solar minimum}}.
\bjtitle{Journal of Geophysical Research (Space Physics)}
\bvolume{119},
\bfpage{2411}--\blpage{2429}
(\byear{2014}a).
doi:\doiurl{10.1002/2013JA019546}
\end{barticle}
\endbibitem

\bibitem[\protect\citeauthoryear{{Guo} and
  {Florinski}}{2014b}]{guo_florinski_2014}
\begin{barticle}
\bauthor{\binits{X.} \bsnm{{Guo}}},
\bauthor{\binits{V.} \bsnm{{Florinski}}},
\batitle{{Galactic Cosmic-Ray Modulation near the Heliopause}}.
\bjtitle{Astrophys. J.}
\bvolume{793},
\bfpage{18}
(\byear{2014}b).
doi:\doiurl{10.1088/0004-637X/793/1/18}
\end{barticle}
\endbibitem

\bibitem[\protect\citeauthoryear{{Gurnett} et~al.}{2013}]{gurnett_etal_2013}
\begin{barticle}
\bauthor{\binits{D.A.} \bsnm{{Gurnett}}},
\bauthor{\binits{W.S.} \bsnm{{Kurth}}},
\bauthor{\binits{L.F.} \bsnm{{Burlaga}}},
\bauthor{\binits{N.F.} \bsnm{{Ness}}},
\batitle{{In Situ Observations of Interstellar Plasma with Voyager 1}}.
\bjtitle{Science}
\bvolume{341},
\bfpage{1489}--\blpage{1492}
(\byear{2013})
\end{barticle}
\endbibitem

\bibitem[\protect\citeauthoryear{{Hanasz} et~al.}{2009}]{Hanasz-etal-2009}
\begin{barticle}
\bauthor{\binits{M.} \bsnm{{Hanasz}}},
\bauthor{\binits{K.} \bsnm{{Otmianowska-Mazur}}},
\bauthor{\binits{G.} \bsnm{{Kowal}}},
\bauthor{\binits{H.} \bsnm{{Lesch}}},
\batitle{{Cosmic-ray-driven dynamo in galactic disks. A parameter study}}.
\bjtitle{Astron. Astrophys.}
\bvolume{498},
\bfpage{335}--\blpage{346}
(\byear{2009}).
doi:\doiurl{10.1051/0004-6361/200810279}
\end{barticle}
\endbibitem

\bibitem[\protect\citeauthoryear{{Herbst} et~al.}{2012}]{herbstetal2012}
\begin{barticle}
\bauthor{\binits{K.} \bsnm{{Herbst}}},
\bauthor{\binits{B.} \bsnm{{Heber}}},
\bauthor{\binits{A.} \bsnm{{Kopp}}},
\bauthor{\binits{O.} \bsnm{{Sternal}}},
\bauthor{\binits{F.} \bsnm{{Steinhilber}}},
\batitle{{The local interstellar spectrum beyond the heliopause: What can be
  learned from Voyager in the inner heliosheath?}}
\bjtitle{Astrophys. J.}
\bvolume{761},
\bfpage{17}
(\byear{2012})
\end{barticle}
\endbibitem

\bibitem[\protect\citeauthoryear{{Hillas}}{2006}]{hillas2006}
\begin{botherref}
\oauthor{\binits{A.M.} \bsnm{{Hillas}}},
{Cosmic Rays: Recent Progress and some Current Questions}.
ArXiv Astrophysics e-prints
(2006)
\end{botherref}
\endbibitem

\bibitem[\protect\citeauthoryear{{Jeffrey} et~al.}{2014}]{Jeffrey-etal-2014}
\begin{barticle}
\bauthor{\binits{N.L.S.} \bsnm{{Jeffrey}}},
\bauthor{\binits{E.P.} \bsnm{{Kontar}}},
\bauthor{\binits{N.H.} \bsnm{{Bian}}},
\bauthor{\binits{A.G.} \bsnm{{Emslie}}},
\batitle{{On the Variation of Solar Flare Coronal X-Ray Source Sizes with
  Energy}}.
\bjtitle{\apj}
\bvolume{787},
\bfpage{86}
(\byear{2014}).
doi:\doiurl{10.1088/0004-637X/787/1/86}
\end{barticle}
\endbibitem

\bibitem[\protect\citeauthoryear{{Johnson} et~al.}{2002}]{johnsonetal2002}
\begin{bbook}
\bauthor{\binits{L.W.} \bsnm{{Johnson}}},
\bauthor{\binits{R.D.} \bsnm{{Riess}}},
\bauthor{\binits{J.T.} \bsnm{{Arnold}}},
\bbtitle{{Introduction to linear algebra, $5^{th}$ ed.}}
(\bpublisher{Addison Wesley},
\blocation{New York, USA}, \byear{2002})
\end{bbook}
\endbibitem

\bibitem[\protect\citeauthoryear{{Jokipii}}{1966}]{jokipii1966}
\begin{barticle}
\bauthor{\binits{J.R.} \bsnm{{Jokipii}}},
\batitle{{Cosmic-ray propagation. I. Charged particles in a random magnetic
  field}}.
\bjtitle{Astrophys. J.}
\bvolume{146},
\bfpage{480}--\blpage{487}
(\byear{1966})
\end{barticle}
\endbibitem

\bibitem[\protect\citeauthoryear{{Jokipii}}{2001}]{jokipii2001}
\begin{bchapter}
\bauthor{\binits{J.R.} \bsnm{{Jokipii}}},
\bctitle{{Acceleration of galactic and anomalous cosmic rays in the
  heliosheath}},
in \bbtitle{{The outer heliosphere: The next frontiers}},
ed. by \beditor{\binits{K.} \bsnm{{Scherer}}},
\beditor{\binits{H.} \bsnm{{Fichtner}}},
\beditor{\binits{H.-J.} \bsnm{{Fahr}}},
\beditor{\binits{E.} \bsnm{{Marsch}}}
\bsertitle{{COSPAR colloquia series}},
vol. \bseriesno{11},
\byear{2001},
p. \bfpage{227}
\end{bchapter}
\endbibitem

\bibitem[\protect\citeauthoryear{{Jokipii} and
  {Kopriva}}{1979}]{jokipiikopriva1979}
\begin{barticle}
\bauthor{\binits{J.R.} \bsnm{{Jokipii}}},
\bauthor{\binits{D.A.} \bsnm{{Kopriva}}},
\batitle{{Effects of particle drift on the transport of cosmic rays. III -
  Numerical models of galactic cosmic-ray modulation}}.
\bjtitle{Astrophys. J.}
\bvolume{234},
\bfpage{384}--\blpage{392}
(\byear{1979})
\end{barticle}
\endbibitem

\bibitem[\protect\citeauthoryear{{Jokipii} and
  {Levy}}{1977}]{jokipii_levy_1997}
\begin{barticle}
\bauthor{\binits{J.R.} \bsnm{{Jokipii}}},
\bauthor{\binits{E.H.} \bsnm{{Levy}}},
\batitle{{Effects of particle drifts on the solar modulation of galactic cosmic
  rays}}.
\bjtitle{Astrophys. K. Lett.}
\bvolume{213},
\bfpage{85}--\blpage{88}
(\byear{1977}).
doi:\doiurl{10.1086/182415}
\end{barticle}
\endbibitem

\bibitem[\protect\citeauthoryear{{Jokipii} et~al.}{1977}]{jokipiietal1977}
\begin{barticle}
\bauthor{\binits{J.R.} \bsnm{{Jokipii}}},
\bauthor{\binits{E.H.} \bsnm{{Levy}}},
\bauthor{\binits{W.B.} \bsnm{{Hubbard}}},
\batitle{{Effects of particle drift on cosmic-ray transport. I - General
  properties, application to solar modulation}}.
\bjtitle{Astrophys. J.}
\bvolume{213},
\bfpage{861}--\blpage{868}
(\byear{1977})
\end{barticle}
\endbibitem

\bibitem[\protect\citeauthoryear{{Kartavykh}
  et~al.}{2016}]{Kartavykh-etal-2016}
\begin{barticle}
\bauthor{\binits{Y.Y.} \bsnm{{Kartavykh}}},
\bauthor{\binits{W.} \bsnm{{Dr{\"o}ge}}},
\bauthor{\binits{M.} \bsnm{{Gedalin}}},
\batitle{{Simulation of Energetic Particle Transport and Acceleration at Shock
  Waves in a Focused Transport Model: Implications for Mixed Solar Particle
  Events}}.
\bjtitle{\apj}
\bvolume{820},
\bfpage{24}
(\byear{2016}).
doi:\doiurl{10.3847/0004-637X/820/1/24}
\end{barticle}
\endbibitem

\bibitem[\protect\citeauthoryear{{Kissmann}}{2014}]{Kissmann-2014}
\begin{barticle}
\bauthor{\binits{R.} \bsnm{{Kissmann}}},
\batitle{{PICARD: A novel code for the Galactic Cosmic Ray propagation
  problem}}.
\bjtitle{Astroparticle Physics}
\bvolume{55},
\bfpage{37}--\blpage{50}
(\byear{2014}).
doi:\doiurl{10.1016/j.astropartphys.2014.02.002}
\end{barticle}
\endbibitem

\bibitem[\protect\citeauthoryear{{Kissmann} et~al.}{2015}]{Kissmann-etal-2015}
\begin{barticle}
\bauthor{\binits{R.} \bsnm{{Kissmann}}},
\bauthor{\binits{M.} \bsnm{{Werner}}},
\bauthor{\binits{O.} \bsnm{{Reimer}}},
\bauthor{\binits{A.W.} \bsnm{{Strong}}},
\batitle{{Propagation in 3D spiral-arm cosmic-ray source distribution models
  and secondary particle production using PICARD}}.
\bjtitle{Astroparticle Physics}
\bvolume{70},
\bfpage{39}--\blpage{53}
(\byear{2015}).
doi:\doiurl{10.1016/j.astropartphys.2015.04.003}
\end{barticle}
\endbibitem

\bibitem[\protect\citeauthoryear{{Kloeden} and {Platen}}{1999}]{Kloeden}
\begin{bbook}
\bauthor{\binits{P.E.} \bsnm{{Kloeden}}},
\bauthor{\binits{E.} \bsnm{{Platen}}},
\bbtitle{{Numerical solution of stochastic differential equations}}
(\bpublisher{Springer},
\blocation{Berlin}, \byear{1999})
\end{bbook}
\endbibitem

\bibitem[\protect\citeauthoryear{{Kloeden} et~al.}{1994}]{Kloeden2}
\begin{bbook}
\bauthor{\binits{P.E.} \bsnm{{Kloeden}}},
\bauthor{\binits{E.} \bsnm{{Platen}}},
\bauthor{\binits{H.} \bsnm{{Schurz}}},
\bbtitle{{Numerical Solution of SDE Through Computer Experiments}}
(\bpublisher{Springer},
\blocation{Berlin}, \byear{1994})
\end{bbook}
\endbibitem

\bibitem[\protect\citeauthoryear{{Kopp} et~al.}{2012}]{koppetal2012}
\begin{barticle}
\bauthor{\binits{A.} \bsnm{{Kopp}}},
\bauthor{\binits{I.} \bsnm{{B{\"u}sching}}},
\bauthor{\binits{R.D.} \bsnm{{Strauss}}},
\bauthor{\binits{M.S.} \bsnm{{Potgieter}}},
\batitle{{A stochastic differential equation code for multidimensional
  Fokker-Planck type problems}}.
\bjtitle{Comp. Phys. Comm.}
\bvolume{183},
\bfpage{530}--\blpage{542}
(\byear{2012})
\end{barticle}
\endbibitem

\bibitem[\protect\citeauthoryear{{Kopp} et~al.}{2014}]{Kopp-etal-2014}
\begin{barticle}
\bauthor{\binits{A.} \bsnm{{Kopp}}},
\bauthor{\binits{I.} \bsnm{{B{\"u}sching}}},
\bauthor{\binits{M.S.} \bsnm{{Potgieter}}},
\bauthor{\binits{R.D.} \bsnm{{Strauss}}},
\batitle{{A stochastic approach to Galactic proton propagation: Influence of
  the spiral arm structure}}.
\bjtitle{\na}
\bvolume{30},
\bfpage{32}--\blpage{37}
(\byear{2014}).
doi:\doiurl{10.1016/j.newast.2014.01.006}
\end{barticle}
\endbibitem

\bibitem[\protect\citeauthoryear{{K{\'o}ta} and
  {Jokipii}}{2014}]{kota_jokipii_2014}
\begin{barticle}
\bauthor{\binits{J.} \bsnm{{K{\'o}ta}}},
\bauthor{\binits{J.R.} \bsnm{{Jokipii}}},
\batitle{{Are Cosmic Rays Modulated beyond the Heliopause?}}
\bjtitle{Astrophys. J.}
\bvolume{782},
\bfpage{24}
(\byear{2014}).
doi:\doiurl{10.1088/0004-637X/782/1/24}
\end{barticle}
\endbibitem

\bibitem[\protect\citeauthoryear{{Krimigis} et~al.}{2013}]{krimigis_etal_2013}
\begin{barticle}
\bauthor{\binits{S.M.} \bsnm{{Krimigis}}},
\bauthor{\binits{R.B.} \bsnm{{Decker}}},
\bauthor{\binits{E.C.} \bsnm{{Roelof}}},
\bauthor{\binits{M.E.} \bsnm{{Hill}}},
\bauthor{\binits{T.P.} \bsnm{{Armstrong}}},
\bauthor{\binits{G.} \bsnm{{Gloeckler}}},
\bauthor{\binits{D.C.} \bsnm{{Hamilton}}},
\bauthor{\binits{L.J.} \bsnm{{Lanzerotti}}},
\batitle{{Search for the Exit: Voyager 1 at Heliosphere's Border with the
  Galaxy}}.
\bjtitle{Science}
\bvolume{341},
\bfpage{144}--\blpage{147}
(\byear{2013}).
doi:\doiurl{10.1126/science.1235721}
\end{barticle}
\endbibitem

\bibitem[\protect\citeauthoryear{{Kr{\"u}lls} and
  {Achterberg}}{1994}]{krullsachterberg1994}
\begin{barticle}
\bauthor{\binits{W.M.} \bsnm{{Kr{\"u}lls}}},
\bauthor{\binits{A.} \bsnm{{Achterberg}}},
\batitle{{Computation of cosmic-ray acceleration by Ito's stochastic
  differential equations}}.
\bjtitle{Astron. Astrophys.}
\bvolume{286},
\bfpage{314}--\blpage{327}
(\byear{1994})
\end{barticle}
\endbibitem

\bibitem[\protect\citeauthoryear{{Laitinen} et~al.}{2015}]{timo}
\begin{botherref}
\oauthor{\binits{T.} \bsnm{{Laitinen}}},
\oauthor{\binits{A.} \bsnm{{Kopp}}},
\oauthor{\binits{F.} \bsnm{{Effenberger}}},
\oauthor{\binits{S.} \bsnm{{Dalla}}},
\oauthor{\binits{M.S.} \bsnm{{Marsh}}},
{Solar Energetic Particle Access to Distant Longitudes via Turbulent Field-Line
  Meandering}.
ArXiv e-prints
(2015)
\end{botherref}
\endbibitem

\bibitem[\protect\citeauthoryear{{le Roux} et~al.}{1996}]{leroux_etal_1996}
\begin{barticle}
\bauthor{\binits{J.A.} \bsnm{{le Roux}}},
\bauthor{\binits{M.S.} \bsnm{{Potgieter}}},
\bauthor{\binits{V.S.} \bsnm{{Ptuskin}}},
\batitle{{A transport model for the diffusive shock acceleration and modulation
  of anomalous cosmic rays in the heliosphere}}.
\bjtitle{J. Geophys. Res.}
\bvolume{101},
\bfpage{4791}--\blpage{4804}
(\byear{1996}).
doi:\doiurl{10.1029/95JA03472}
\end{barticle}
\endbibitem

\bibitem[\protect\citeauthoryear{Lemons}{2002}]{Lemons}
\begin{bbook}
\bauthor{\binits{D.S.} \bsnm{Lemons}},
\bbtitle{{An Introduction to Stochastic Processes in Physics}}.
\bsertitle{{Johns Hopkins Paperback}}
(\bpublisher{Johns Hopkins University Press},
\blocation{Baltimore}, \byear{2002})
\end{bbook}
\endbibitem

\bibitem[\protect\citeauthoryear{{Litvinenko}}{2012}]{Litvinenko-2012}
\begin{barticle}
\bauthor{\binits{Y.E.} \bsnm{{Litvinenko}}},
\batitle{{Effects of Non-isotropic Scattering, Magnetic Helicity, and Adiabatic
  Focusing on Diffusive Transport of Solar Energetic Particles}}.
\bjtitle{\apj}
\bvolume{752},
\bfpage{16}
(\byear{2012}).
doi:\doiurl{10.1088/0004-637X/752/1/16}
\end{barticle}
\endbibitem

\bibitem[\protect\citeauthoryear{{Litvinenko} and
  {Noble}}{2013}]{Litvinenko-Noble-2013}
\begin{barticle}
\bauthor{\binits{Y.E.} \bsnm{{Litvinenko}}},
\bauthor{\binits{P.L.} \bsnm{{Noble}}},
\batitle{{A Numerical Study of Diffusive Cosmic-Ray Transport with Adiabatic
  Focusing}}.
\bjtitle{\apj}
\bvolume{765},
\bfpage{31}
(\byear{2013}).
doi:\doiurl{10.1088/0004-637X/765/1/31}
\end{barticle}
\endbibitem

\bibitem[\protect\citeauthoryear{{Litvinenko}
  et~al.}{2015}]{Litvinenko-etal-2015}
\begin{barticle}
\bauthor{\binits{Y.E.} \bsnm{{Litvinenko}}},
\bauthor{\binits{F.} \bsnm{{Effenberger}}},
\bauthor{\binits{R.} \bsnm{{Schlickeiser}}},
\batitle{{The Telegraph Approximation for Focused Cosmic-Ray Transport in the
  Presence of Boundaries}}.
\bjtitle{\apj}
\bvolume{806},
\bfpage{217}
(\byear{2015}).
doi:\doiurl{10.1088/0004-637X/806/2/217}
\end{barticle}
\endbibitem

\bibitem[\protect\citeauthoryear{{Luo} et~al.}{2011}]{luo_etal_2011}
\begin{barticle}
\bauthor{\binits{X.} \bsnm{{Luo}}},
\bauthor{\binits{M.} \bsnm{{Zhang}}},
\bauthor{\binits{H.K.} \bsnm{{Rassoul}}},
\bauthor{\binits{N.V.} \bsnm{{Pogorelov}}},
\batitle{{Cosmic-ray Modulation by the Global Merged Interaction Region in the
  Heliosheath}}.
\bjtitle{Astrophys. J.}
\bvolume{730},
\bfpage{13}
(\byear{2011}).
doi:\doiurl{10.1088/0004-637X/730/1/13}
\end{barticle}
\endbibitem

\bibitem[\protect\citeauthoryear{{Luo} et~al.}{2013}]{luoetal2013}
\begin{barticle}
\bauthor{\binits{X.} \bsnm{{Luo}}},
\bauthor{\binits{M.} \bsnm{{Zhang}}},
\bauthor{\binits{H.K.} \bsnm{{Rassoul}}},
\bauthor{\binits{N.V.} \bsnm{{Pogorelov}}},
\bauthor{\binits{J.} \bsnm{{Heerikhuisen}}},
\batitle{{Galactic cosmic-ray modulation in a realistic global
  magnetohydrodynamic heliosphere}}.
\bjtitle{Astrophys. J.}
\bvolume{764},
\bfpage{85}
(\byear{2013})
\end{barticle}
\endbibitem

\bibitem[\protect\citeauthoryear{{Luo} et~al.}{2015}]{luo_etaL_2016}
\begin{barticle}
\bauthor{\binits{X.} \bsnm{{Luo}}},
\bauthor{\binits{M.} \bsnm{{Zhang}}},
\bauthor{\binits{M.} \bsnm{{Potgieter}}},
\bauthor{\binits{X.} \bsnm{{Feng}}},
\bauthor{\binits{N.V.} \bsnm{{Pogorelov}}},
\batitle{{A Numerical Simulation of Cosmic-Ray Modulation Near the
  Heliopause}}.
\bjtitle{\apj}
\bvolume{808},
\bfpage{82}
(\byear{2015}).
doi:\doiurl{10.1088/0004-637X/808/1/82}
\end{barticle}
\endbibitem

\bibitem[\protect\citeauthoryear{{Maccione}}{2013}]{maccione_2013}
\begin{barticle}
\bauthor{\binits{L.} \bsnm{{Maccione}}},
\batitle{{Low Energy Cosmic Ray Positron Fraction Explained by Charge-Sign
  Dependent Solar Modulation}}.
\bjtitle{Physical Review Letters}
\bvolume{110}(\bissue{8}),
\bfpage{081101}
(\byear{2013}).
doi:\doiurl{10.1103/PhysRevLett.110.081101}
\end{barticle}
\endbibitem

\bibitem[\protect\citeauthoryear{{MacKinnon} and
  {Craig}}{1991}]{Mackinnon-Craig-1991}
\begin{barticle}
\bauthor{\binits{A.L.} \bsnm{{MacKinnon}}},
\bauthor{\binits{I.J.D.} \bsnm{{Craig}}},
\batitle{{Stochastic simulation of fast particle diffusive transport}}.
\bjtitle{\aap}
\bvolume{251},
\bfpage{693}--\blpage{699}
(\byear{1991})
\end{barticle}
\endbibitem

\bibitem[\protect\citeauthoryear{{Magdziarz} and
  {Weron}}{2007}]{Magdziarz-Weron-2007}
\begin{barticle}
\bauthor{\binits{M.} \bsnm{{Magdziarz}}},
\bauthor{\binits{A.} \bsnm{{Weron}}},
\batitle{{Competition between subdiffusion and L{\'e}vy flights: A Monte Carlo
  approach}}.
\bjtitle{\pre}
\bvolume{75}(\bissue{5}),
\bfpage{056702}
(\byear{2007}).
doi:\doiurl{10.1103/PhysRevE.75.056702}
\end{barticle}
\endbibitem

\bibitem[\protect\citeauthoryear{{Mannheim} and
  {Schlickeiser}}{1994}]{Mannheim-Schlickeiser-1994}
\begin{barticle}
\bauthor{\binits{K.} \bsnm{{Mannheim}}},
\bauthor{\binits{R.} \bsnm{{Schlickeiser}}},
\batitle{{Interactions of cosmic ray nuclei}}.
\bjtitle{\aap}
\bvolume{286},
\bfpage{983}--\blpage{996}
(\byear{1994})
\end{barticle}
\endbibitem

\bibitem[\protect\citeauthoryear{{Marcowith} and
  {Kirk}}{1999}]{Marcowith-Kirk-1999}
\begin{barticle}
\bauthor{\binits{A.} \bsnm{{Marcowith}}},
\bauthor{\binits{J.G.} \bsnm{{Kirk}}},
\batitle{{Computation of diffusive shock acceleration using stochastic
  differential equations}}.
\bjtitle{\aap}
\bvolume{347},
\bfpage{391}--\blpage{400}
(\byear{1999})
\end{barticle}
\endbibitem

\bibitem[\protect\citeauthoryear{Maruyama}{1955}]{maruyama}
\begin{barticle}
\bauthor{\binits{G.} \bsnm{Maruyama}},
\batitle{{Continuous Markov processes and stochastic equations}}.
\bjtitle{Rend. Circ. Mat. Palermo}
\bvolume{4},
\bfpage{48}--\blpage{90}
(\byear{1955})
\end{barticle}
\endbibitem

\bibitem[\protect\citeauthoryear{Matsumoto and Nishimura}{1988}]{twister}
\begin{barticle}
\bauthor{\binits{M.} \bsnm{Matsumoto}},
\bauthor{\binits{T.} \bsnm{Nishimura}},
\batitle{{Mersenne twister: A 623-dimensionally equidistributed uniform
  pseudorandom number generator}}.
\bjtitle{ACM Trans. Model. and Comp. Sims.}
\bvolume{31},
\bfpage{1192}--\blpage{1201}
(\byear{1988})
\end{barticle}
\endbibitem

\bibitem[\protect\citeauthoryear{{Matthaeus} et~al.}{2003}]{Matthaeusetal2003}
\begin{barticle}
\bauthor{\binits{W.H.} \bsnm{{Matthaeus}}},
\bauthor{\binits{G.} \bsnm{{Qin}}},
\bauthor{\binits{J.W.} \bsnm{{Bieber}}},
\bauthor{\binits{G.P.} \bsnm{{Zank}}},
\batitle{{Nonlinear collisionless perpendicular diffusion of charged
  particles}}.
\bjtitle{Astrophys. J. Lett.}
\bvolume{590},
\bfpage{53}--\blpage{56}
(\byear{2003})
\end{barticle}
\endbibitem

\bibitem[\protect\citeauthoryear{{Mertsch}}{2011}]{Mertsch-2011}
\begin{barticle}
\bauthor{\binits{P.} \bsnm{{Mertsch}}},
\batitle{{Cosmic ray electrons and positrons from discrete stochastic
  sources}}.
\bjtitle{\jcap}
\bvolume{2},
\bfpage{31}
(\byear{2011}).
doi:\doiurl{10.1088/1475-7516/2011/02/031}
\end{barticle}
\endbibitem

\bibitem[\protect\citeauthoryear{{Miller} and
  {Roberts}}{1995}]{Miller-Roberts-1995}
\begin{barticle}
\bauthor{\binits{J.A.} \bsnm{{Miller}}},
\bauthor{\binits{D.A.} \bsnm{{Roberts}}},
\batitle{{Stochastic Proton Acceleration by Cascading Alfven Waves in Impulsive
  Solar Flares}}.
\bjtitle{\apj}
\bvolume{452},
\bfpage{912}
(\byear{1995}).
doi:\doiurl{10.1086/176359}
\end{barticle}
\endbibitem

\bibitem[\protect\citeauthoryear{{Milstein} et~al.}{2004}]{milsteinetal_2004}
\begin{barticle}
\bauthor{\binits{G.N.} \bsnm{{Milstein}}},
\bauthor{\binits{J.G.M.} \bsnm{{Schoenmakers}}},
\bauthor{\binits{V.} \bsnm{{Spokoiny}}},
\batitle{{Transition density estimation for stochastic differential equations
  via forward-reverse representations}}.
\bjtitle{Bernoulli}
\bvolume{10},
\bfpage{281}--\blpage{312}
(\byear{2004})
\end{barticle}
\endbibitem

\bibitem[\protect\citeauthoryear{Miyake and Yanagita}{2005}]{miyake2005}
\begin{bchapter}
\bauthor{\binits{S.} \bsnm{Miyake}},
\bauthor{\binits{S.} \bsnm{Yanagita}},
\bctitle{{Effects of the tilted and wavy current sheet on the solar modulation
  of galactic cosmic rays}},
in \bbtitle{{International Cosmic Ray Conference Proceedings}},
vol. \bseriesno{2},
\byear{2005},
pp. \bfpage{203}--\blpage{207}
\end{bchapter}
\endbibitem

\bibitem[\protect\citeauthoryear{{Miyake} et~al.}{2015}]{Miyake-etal-2015}
\begin{barticle}
\bauthor{\binits{S.} \bsnm{{Miyake}}},
\bauthor{\binits{H.} \bsnm{{Muraishi}}},
\bauthor{\binits{S.} \bsnm{{Yanagita}}},
\batitle{{A stochastic simulation of the propagation of Galactic cosmic rays
  reflecting the discreteness of cosmic ray sources Age and path length
  distribution}}.
\bjtitle{\aap}
\bvolume{573},
\bfpage{134}
(\byear{2015}).
doi:\doiurl{10.1051/0004-6361/201424442}
\end{barticle}
\endbibitem

\bibitem[\protect\citeauthoryear{{Moses}}{1987}]{moses1987}
\begin{barticle}
\bauthor{\binits{D.} \bsnm{{Moses}}},
\batitle{{Jovian electrons at 1 AU - 1978-1984}}.
\bjtitle{Astrophys. J.}
\bvolume{313},
\bfpage{471}--\blpage{486}
(\byear{1987})
\end{barticle}
\endbibitem

\bibitem[\protect\citeauthoryear{{Moskalenko}
  et~al.}{2002}]{moskalenkoetal2002}
\begin{barticle}
\bauthor{\binits{I.V.} \bsnm{{Moskalenko}}},
\bauthor{\binits{A.W.} \bsnm{{Strong}}},
\bauthor{\binits{J.F.} \bsnm{{Ormes}}},
\bauthor{\binits{M.S.} \bsnm{{Potgieter}}},
\batitle{{Secondary antiprotons and propagation of cosmic rays in the galaxy
  and heliosphere}}.
\bjtitle{Astrophys. J.}
\bvolume{565},
\bfpage{280}--\blpage{296}
(\byear{2002})
\end{barticle}
\endbibitem

\bibitem[\protect\citeauthoryear{{\O}ksendal}{2003}]{Oksendal}
\begin{bbook}
\bauthor{\binits{B.} \bsnm{{\O}ksendal}},
\bbtitle{{Stochastic Differential Equations: An Introduction with
  Applications}}
(\bpublisher{Springer},
\blocation{Berlin}, \byear{2003})
\end{bbook}
\endbibitem

\bibitem[\protect\citeauthoryear{Oughton et~al.}{2011}]{oughtonetal2011}
\begin{barticle}
\bauthor{\binits{S.} \bsnm{Oughton}},
\bauthor{\binits{W.H.} \bsnm{Matthaeus}},
\bauthor{\binits{C.W.} \bsnm{Smith}},
\bauthor{\binits{B.} \bsnm{Breech}},
\bauthor{\binits{P.A.} \bsnm{Isenberg}},
\batitle{{Transport of solar wind fluctuations: A two-component model}}.
\bjtitle{J. Geophys. Res.}
\bvolume{116},
\bfpage{08105}
(\byear{2011})
\end{barticle}
\endbibitem

\bibitem[\protect\citeauthoryear{{Pakmor} et~al.}{2016}]{Pakmor-etal-2016}
\begin{botherref}
\oauthor{\binits{R.} \bsnm{{Pakmor}}},
\oauthor{\binits{C.} \bsnm{{Pfrommer}}},
\oauthor{\binits{C.M.} \bsnm{{Simpson}}},
\oauthor{\binits{V.} \bsnm{{Springel}}},
{Galactic winds driven by isotropic and anisotropic cosmic ray diffusion in
  disk galaxies}.
ArXiv e-prints
(2016)
\end{botherref}
\endbibitem

\bibitem[\protect\citeauthoryear{{Park} and
  {Petrosian}}{1996}]{park_petrosian_1996}
\begin{barticle}
\bauthor{\binits{B.T.} \bsnm{{Park}}},
\bauthor{\binits{V.} \bsnm{{Petrosian}}},
\batitle{{Fokker-Planck Equations of Stochastic Acceleration: A Study of
  Numerical Methods}}.
\bjtitle{Astrophys. J. Suppl. Series}
\bvolume{103},
\bfpage{255}
(\byear{1996}).
doi:\doiurl{10.1086/192278}
\end{barticle}
\endbibitem

\bibitem[\protect\citeauthoryear{{Parker}}{1958}]{parker1958}
\begin{barticle}
\bauthor{\binits{E.N.} \bsnm{{Parker}}},
\batitle{{Dynamics of the interplanetary gas and magnetic fields.}}
\bjtitle{Astrophys. J.}
\bvolume{128},
\bfpage{664}--\blpage{676}
(\byear{1958})
\end{barticle}
\endbibitem

\bibitem[\protect\citeauthoryear{{Parker}}{1965}]{parker1965}
\begin{barticle}
\bauthor{\binits{E.N.} \bsnm{{Parker}}},
\batitle{{The passage of energetic charged particles through interplanetary
  space}}.
\bjtitle{Planet Space Sci.}
\bvolume{13},
\bfpage{9}--\blpage{49}
(\byear{1965})
\end{barticle}
\endbibitem

\bibitem[\protect\citeauthoryear{Parker}{1966}]{parker1966}
\begin{barticle}
\bauthor{\binits{E.N.} \bsnm{Parker}},
\batitle{{The effect of adiabatic deceleration on the cosmic ray spectrum in
  the solar system}}.
\bjtitle{Planet. Space Sci.}
\bvolume{14},
\bfpage{371}--\blpage{380}
(\byear{1966})
\end{barticle}
\endbibitem

\bibitem[\protect\citeauthoryear{Pei et~al.}{2012}]{pei2011}
\begin{botherref}
\oauthor{\binits{C.} \bsnm{Pei}},
\oauthor{\binits{J.W.} \bsnm{Bieber}},
\oauthor{\binits{R.A.} \bsnm{Burger}},
{Three-dimensional wavy current sheet drifts}.
Astrophys. J.,
170--175
(2012)
\end{botherref}
\endbibitem

\bibitem[\protect\citeauthoryear{Pei et~al.}{2010}]{peietal2010}
\begin{barticle}
\bauthor{\binits{C.} \bsnm{Pei}},
\bauthor{\binits{J.} \bsnm{Bieber}},
\bauthor{\binits{R.A.} \bsnm{Burger}},
\bauthor{\binits{J.} \bsnm{Clem}},
\batitle{{A general time dependent stochastic method for solving Parker's
  equation in spherical coordinates}}.
\bjtitle{J. Geophys. Res.}
\bvolume{115},
\bfpage{12107}
(\byear{2010})
\end{barticle}
\endbibitem

\bibitem[\protect\citeauthoryear{{Perrone} et~al.}{2013}]{Perrone-etal-2013}
\begin{barticle}
\bauthor{\binits{D.} \bsnm{{Perrone}}},
\bauthor{\binits{R.O.} \bsnm{{Dendy}}},
\bauthor{\binits{I.} \bsnm{{Furno}}},
\bauthor{\binits{R.} \bsnm{{Sanchez}}},
\bauthor{\binits{G.} \bsnm{{Zimbardo}}},
\bauthor{\binits{A.} \bsnm{{Bovet}}},
\bauthor{\binits{A.} \bsnm{{Fasoli}}},
\bauthor{\binits{K.} \bsnm{{Gustafson}}},
\bauthor{\binits{S.} \bsnm{{Perri}}},
\bauthor{\binits{P.} \bsnm{{Ricci}}},
\bauthor{\binits{F.} \bsnm{{Valentini}}},
\batitle{{Nonclassical Transport and Particle-Field Coupling: from Laboratory
  Plasmas to the Solar Wind}}.
\bjtitle{\ssr}
\bvolume{178},
\bfpage{233}--\blpage{270}
(\byear{2013}).
doi:\doiurl{10.1007/s11214-013-9966-9}
\end{barticle}
\endbibitem

\bibitem[\protect\citeauthoryear{{Petrosian}}{2012}]{Petrosian-2012}
\begin{barticle}
\bauthor{\binits{V.} \bsnm{{Petrosian}}},
\batitle{{Stochastic Acceleration by Turbulence}}.
\bjtitle{\ssr}
\bvolume{173},
\bfpage{535}--\blpage{556}
(\byear{2012}).
doi:\doiurl{10.1007/s11214-012-9900-6}
\end{barticle}
\endbibitem

\bibitem[\protect\citeauthoryear{{Petrosian} and
  {Liu}}{2004}]{Petrosian-Liu-2004}
\begin{barticle}
\bauthor{\binits{V.} \bsnm{{Petrosian}}},
\bauthor{\binits{S.} \bsnm{{Liu}}},
\batitle{{Stochastic Acceleration of Electrons and Protons. I. Acceleration by
  Parallel-Propagating Waves}}.
\bjtitle{\apj}
\bvolume{610},
\bfpage{550}--\blpage{571}
(\byear{2004}).
doi:\doiurl{10.1086/421486}
\end{barticle}
\endbibitem

\bibitem[\protect\citeauthoryear{{Potgieter}}{2013}]{potgieter2013}
\begin{barticle}
\bauthor{\binits{M.} \bsnm{{Potgieter}}},
\batitle{{Solar Modulation of Cosmic Rays}}.
\bjtitle{Living Reviews in Solar Physics}
\bvolume{10},
\bfpage{3}
(\byear{2013})
\end{barticle}
\endbibitem

\bibitem[\protect\citeauthoryear{{Potgieter} and
  {Moraal}}{1985}]{potgietermoraal1985}
\begin{barticle}
\bauthor{\binits{M.S.} \bsnm{{Potgieter}}},
\bauthor{\binits{H.} \bsnm{{Moraal}}},
\batitle{{A drift model for the modulation of galactic cosmic rays}}.
\bjtitle{Astrophys. J.}
\bvolume{294},
\bfpage{425}--\blpage{440}
(\byear{1985})
\end{barticle}
\endbibitem

\bibitem[\protect\citeauthoryear{{Qin} and {Wang}}{2015}]{QinWang_2015}
\begin{barticle}
\bauthor{\binits{G.} \bsnm{{Qin}}},
\bauthor{\binits{Y.} \bsnm{{Wang}}},
\batitle{{Simulations of a Gradual Solar Energetic Particle Event Observed by
  Helios 1, Helios 2, and IMP 8}}.
\bjtitle{\apj}
\bvolume{809},
\bfpage{177}
(\byear{2015}).
doi:\doiurl{10.1088/0004-637X/809/2/177}
\end{barticle}
\endbibitem

\bibitem[\protect\citeauthoryear{{Qin} et~al.}{2005}]{qin_etal_2005}
\begin{barticle}
\bauthor{\binits{G.} \bsnm{{Qin}}},
\bauthor{\binits{M.} \bsnm{{Zhang}}},
\bauthor{\binits{J.R.} \bsnm{{Dwyer}}},
\bauthor{\binits{H.K.} \bsnm{{Rassoul}}},
\bauthor{\binits{G.M.} \bsnm{{Mason}}},
\batitle{{The Model Dependence of Solar Energetic Particle Mean Free Paths
  under Weak Scattering}}.
\bjtitle{Astrophys. J.}
\bvolume{627},
\bfpage{562}--\blpage{566}
(\byear{2005}).
doi:\doiurl{10.1086/430136}
\end{barticle}
\endbibitem

\bibitem[\protect\citeauthoryear{{Raath} et~al.}{2015}]{Raath-etal-2015}
\begin{barticle}
\bauthor{\binits{J.L.} \bsnm{{Raath}}},
\bauthor{\binits{R.D.} \bsnm{{Strauss}}},
\bauthor{\binits{M.S.} \bsnm{{Potgieter}}},
\batitle{{New insights from modeling the neutral heliospheric current sheet}}.
\bjtitle{\apss}
\bvolume{360},
\bfpage{24}
(\byear{2015}).
doi:\doiurl{10.1007/s10509-015-2556-4}
\end{barticle}
\endbibitem

\bibitem[\protect\citeauthoryear{{Raath} et~al.}{2016}]{Raath-etal-2016}
\begin{barticle}
\bauthor{\binits{J.L.} \bsnm{{Raath}}},
\bauthor{\binits{M.S.} \bsnm{{Potgieter}}},
\bauthor{\binits{R.D.} \bsnm{{Strauss}}},
\bauthor{\binits{A.} \bsnm{{Kopp}}},
\batitle{{The effects of magnetic field modifications on the solar modulation
  of cosmic rays with a SDE-based model}}.
\bjtitle{Advances in Space Research}
\bvolume{57},
\bfpage{1965}--\blpage{1977}
(\byear{2016}).
doi:\doiurl{10.1016/j.asr.2016.01.017}
\end{barticle}
\endbibitem

\bibitem[\protect\citeauthoryear{{Reames}}{2013}]{reames_2013}
\begin{barticle}
\bauthor{\binits{D.V.} \bsnm{{Reames}}},
\batitle{{The Two Sources of Solar Energetic Particles}}.
\bjtitle{Space Sci. Rev.}
\bvolume{175},
\bfpage{53}--\blpage{92}
(\byear{2013}).
doi:\doiurl{10.1007/s11214-013-9958-9}
\end{barticle}
\endbibitem

\bibitem[\protect\citeauthoryear{{Roelof}}{1969}]{roelof1969}
\begin{bchapter}
\bauthor{\binits{E.C.} \bsnm{{Roelof}}},
\bctitle{{Propagation of Solar Cosmic Rays in the Interplanetary Magnetic
  Field}},
in \bbtitle{{Lectures in High-Energy Astrophysics}},
ed. by \beditor{\binits{H.} \bsnm{{{\"O}gelman}}},
\beditor{\binits{J.R.} \bsnm{{Wayland}}},
\byear{1969},
p. \bfpage{111}
\end{bchapter}
\endbibitem

\bibitem[\protect\citeauthoryear{{Ruffolo}}{1995}]{ruffolo1995}
\begin{barticle}
\bauthor{\binits{D.} \bsnm{{Ruffolo}}},
\batitle{{Effect of adiabatic deceleration on the focused transport of solar
  cosmic rays}}.
\bjtitle{Astrophys. J.}
\bvolume{442},
\bfpage{861}--\blpage{874}
(\byear{1995}).
doi:\doiurl{10.1086/175489}
\end{barticle}
\endbibitem

\bibitem[\protect\citeauthoryear{{Scherer} et~al.}{2008}]{Scherer-etal-2008}
\begin{barticle}
\bauthor{\binits{K.} \bsnm{{Scherer}}},
\bauthor{\binits{H.} \bsnm{{Fichtner}}},
\bauthor{\binits{S.E.S.} \bsnm{{Ferreira}}},
\bauthor{\binits{I.} \bsnm{{B{\"u}sching}}},
\bauthor{\binits{M.S.} \bsnm{{Potgieter}}},
\batitle{{Are Anomalous Cosmic Rays the Main Contribution to the Low-Energy
  Galactic Cosmic Ray Spectrum?}}
\bjtitle{\apjl}
\bvolume{680},
\bfpage{105}
(\byear{2008}).
doi:\doiurl{10.1086/589969}
\end{barticle}
\endbibitem

\bibitem[\protect\citeauthoryear{{Scherer} et~al.}{2011}]{schereretal2011}
\begin{barticle}
\bauthor{\binits{K.} \bsnm{{Scherer}}},
\bauthor{\binits{H.} \bsnm{{Fichtner}}},
\bauthor{\binits{R.D.} \bsnm{{Strauss}}},
\bauthor{\binits{S.E.S.} \bsnm{{Ferreira}}},
\bauthor{\binits{M.S.} \bsnm{{Potgieter}}},
\bauthor{\binits{H.-J.} \bsnm{{Fahr}}},
\batitle{{On cosmic ray modulation beyond the heliopause: Where is the
  modulation boundary?}}
\bjtitle{Astrophys. J.}
\bvolume{735},
\bfpage{128}
(\byear{2011})
\end{barticle}
\endbibitem

\bibitem[\protect\citeauthoryear{{Scherer} et~al.}{2015}]{Scherer-etal-2015}
\begin{barticle}
\bauthor{\binits{K.} \bsnm{{Scherer}}},
\bauthor{\binits{A.} \bsnm{{van der Schyff}}},
\bauthor{\binits{D.J.} \bsnm{{Bomans}}},
\bauthor{\binits{S.E.S.} \bsnm{{Ferreira}}},
\bauthor{\binits{H.} \bsnm{{Fichtner}}},
\bauthor{\binits{J.} \bsnm{{Kleimann}}},
\bauthor{\binits{R.D.} \bsnm{{Strauss}}},
\bauthor{\binits{K.} \bsnm{{Weis}}},
\bauthor{\binits{T.} \bsnm{{Wiengarten}}},
\bauthor{\binits{T.} \bsnm{{Wodzinski}}},
\batitle{{Cosmic rays in astrospheres}}.
\bjtitle{\aap}
\bvolume{576},
\bfpage{97}
(\byear{2015}).
doi:\doiurl{10.1051/0004-6361/201425091}
\end{barticle}
\endbibitem

\bibitem[\protect\citeauthoryear{{Schlickeiser}}{1989}]{Schlickeiser-1989}
\begin{barticle}
\bauthor{\binits{R.} \bsnm{{Schlickeiser}}},
\batitle{{Cosmic-ray transport and acceleration. I - Derivation of the kinetic
  equation and application to cosmic rays in static cold media. II - Cosmic
  rays in moving cold media with application to diffusive shock wave
  acceleration}}.
\bjtitle{\apj}
\bvolume{336},
\bfpage{243}--\blpage{293}
(\byear{1989}).
doi:\doiurl{10.1086/167009}
\end{barticle}
\endbibitem

\bibitem[\protect\citeauthoryear{{Schlickeiser}}{2002}]{schlikeiser2002}
\begin{bbook}
\bauthor{\binits{R.} \bsnm{{Schlickeiser}}},
\bbtitle{{Cosmic Ray Astrophysics}}
(\bpublisher{Springer},
\blocation{Germany}, \byear{2002})
\end{bbook}
\endbibitem

\bibitem[\protect\citeauthoryear{{Schlickeiser}}{2015}]{Schlickeiser-2015}
\begin{barticle}
\bauthor{\binits{R.} \bsnm{{Schlickeiser}}},
\batitle{{Cosmic ray transport in astrophysical plasmas}}.
\bjtitle{Physics of Plasmas}
\bvolume{22}(\bissue{9}),
\bfpage{091502}
(\byear{2015}).
doi:\doiurl{10.1063/1.4928940}
\end{barticle}
\endbibitem

\bibitem[\protect\citeauthoryear{{Schlickeiser} and
  {Steinacker}}{1989}]{Schlickeiser-Steinacker-1989}
\begin{barticle}
\bauthor{\binits{R.} \bsnm{{Schlickeiser}}},
\bauthor{\binits{J.} \bsnm{{Steinacker}}},
\batitle{{Particle acceleration in impulsive solar flares. II - Nonrelativistic
  protons and ions}}.
\bjtitle{\solphys}
\bvolume{122},
\bfpage{29}--\blpage{52}
(\byear{1989}).
doi:\doiurl{10.1007/BF00162827}
\end{barticle}
\endbibitem

\bibitem[\protect\citeauthoryear{{Senanayake} and
  {Florinski}}{2013}]{senanayake_florinski_2013}
\begin{barticle}
\bauthor{\binits{U.K.} \bsnm{{Senanayake}}},
\bauthor{\binits{V.} \bsnm{{Florinski}}},
\batitle{{Is the Acceleration of Anomalous Cosmic Rays affected by the Geometry
  of the Termination Shock?}}
\bjtitle{Astrophys. J.}
\bvolume{778},
\bfpage{122}
(\byear{2013}).
doi:\doiurl{10.1088/0004-637X/778/2/122}
\end{barticle}
\endbibitem

\bibitem[\protect\citeauthoryear{{Shalchi}}{2009}]{Shalchi-2009}
\begin{bbook}
\bauthor{\binits{A.} \bsnm{{Shalchi}}},
\bbtitle{{Nonlinear Cosmic Ray Diffusion Theories}}.
\bsertitle{Astrophysics and Space Science Library},
vol. \bseriesno{362}
\byear{2009}.
doi:\doiurl{10.1007/978-3-642-00309-7}
\end{bbook}
\endbibitem

\bibitem[\protect\citeauthoryear{{Skilling}}{1971}]{skilling_1971}
\begin{barticle}
\bauthor{\binits{J.} \bsnm{{Skilling}}},
\batitle{{Cosmic Rays in the Galaxy: Convection or Diffusion?}}
\bjtitle{Astrophys. J.}
\bvolume{170},
\bfpage{265}
(\byear{1971}).
doi:\doiurl{10.1086/151210}
\end{barticle}
\endbibitem

\bibitem[\protect\citeauthoryear{{Stawicki}}{2003}]{stawicki2003}
\begin{botherref}
\oauthor{\binits{O.} \bsnm{{Stawicki}}},
{On solar wind magnetic fluctuations on their influence on the transport of
  charged particles in the heliosphere},
PhD thesis,
Ruhr Universit{\"a}t, Bochum, Germany,
2003
\end{botherref}
\endbibitem

\bibitem[\protect\citeauthoryear{Stern et~al.}{2014}]{Stern-etal-2014}
\begin{barticle}
\bauthor{\binits{R.} \bsnm{Stern}},
\bauthor{\binits{F.} \bsnm{Effenberger}},
\bauthor{\binits{H.} \bsnm{Fichtner}},
\bauthor{\binits{T.} \bsnm{Sch{\"a}fer}},
\batitle{{The space-fractional diffusion-advection equation: Analytical
  solutions and critical assessment of numerical solutions}}.
\bjtitle{Fractional Calculus and Applied Analysis}
\bvolume{17}(\bissue{1}),
\bfpage{171}--\blpage{190}
(\byear{2014})
\end{barticle}
\endbibitem

\bibitem[\protect\citeauthoryear{{Stone} et~al.}{2005}]{stoneetal2005}
\begin{barticle}
\bauthor{\binits{E.C.} \bsnm{{Stone}}},
\bauthor{\binits{A.C.} \bsnm{{Cummings}}},
\bauthor{\binits{F.B.} \bsnm{{McDonald}}},
\bauthor{\binits{B.C.} \bsnm{{Heikkila}}},
\bauthor{\binits{N.} \bsnm{{Lal}}},
\bauthor{\binits{W.R.} \bsnm{{Webber}}},
\batitle{{Voyager 1 explores the termination shock region and the heliosheath
  beyond}}.
\bjtitle{Science}
\bvolume{309},
\bfpage{2017}--\blpage{2020}
(\byear{2005})
\end{barticle}
\endbibitem

\bibitem[\protect\citeauthoryear{{Stone} et~al.}{2013}]{Stone_etal_2013}
\begin{barticle}
\bauthor{\binits{E.C.} \bsnm{{Stone}}},
\bauthor{\binits{A.C.} \bsnm{{Cummings}}},
\bauthor{\binits{F.B.} \bsnm{{McDonald}}},
\bauthor{\binits{B.C.} \bsnm{{Heikkila}}},
\bauthor{\binits{N.} \bsnm{{Lal}}},
\bauthor{\binits{W.R.} \bsnm{{Webber}}},
\batitle{{Voyager 1 Observes Low-Energy Galactic Cosmic Rays in a Region
  Depleted of Heliospheric Ions}}.
\bjtitle{Science}
\bvolume{341},
\bfpage{150}--\blpage{153}
(\byear{2013}).
doi:\doiurl{10.1126/science.1236408}
\end{barticle}
\endbibitem

\bibitem[\protect\citeauthoryear{{Strauss} and
  {Fichtner}}{2014}]{strauss_fichtner_2015}
\begin{barticle}
\bauthor{\binits{R.D.} \bsnm{{Strauss}}},
\bauthor{\binits{H.} \bsnm{{Fichtner}}},
\batitle{{Cosmic ray anisotropies near the heliopause}}.
\bjtitle{Astron. Astrophys. Lett.}
\bvolume{572},
\bfpage{3}
(\byear{2014}).
doi:\doiurl{10.1051/0004-6361/201424842}
\end{barticle}
\endbibitem

\bibitem[\protect\citeauthoryear{{Strauss} and {Fichtner}}{2015}]{ek_en_horst}
\begin{barticle}
\bauthor{\binits{R.D.} \bsnm{{Strauss}}},
\bauthor{\binits{H.} \bsnm{{Fichtner}}},
\batitle{{On Aspects Pertaining to the Perpendicular Diffusion of Solar
  Energetic Particles}}.
\bjtitle{Astrophys. J.}
\bvolume{801},
\bfpage{29}
(\byear{2015}).
doi:\doiurl{10.1088/0004-637X/801/1/29}
\end{barticle}
\endbibitem

\bibitem[\protect\citeauthoryear{Strauss et~al.}{2012}]{straussetal2012c}
\begin{barticle}
\bauthor{\binits{R.D.} \bsnm{Strauss}},
\bauthor{\binits{M.S.} \bsnm{Potgieter}},
\bauthor{\binits{S.E.S.} \bsnm{Ferreira}},
\batitle{{Modeling ground and space based cosmic ray observations}}.
\bjtitle{Adv. Space Res.}
\bvolume{49},
\bfpage{392}
(\byear{2012})
\end{barticle}
\endbibitem

\bibitem[\protect\citeauthoryear{Strauss et~al.}{2013}]{straussetal2013a}
\begin{barticle}
\bauthor{\binits{R.D.} \bsnm{Strauss}},
\bauthor{\binits{M.S.} \bsnm{Potgieter}},
\bauthor{\binits{S.E.S.} \bsnm{Ferreira}},
\batitle{{Modelling and observing Jovian electron propagation times in the
  inner heliosphere}}.
\bjtitle{Adv. Space Res.}
\bvolume{51},
\bfpage{339}--\blpage{349}
(\byear{2013})
\end{barticle}
\endbibitem

\bibitem[\protect\citeauthoryear{{Strauss} et~al.}{2010}]{strauss_etal_2010}
\begin{barticle}
\bauthor{\binits{R.D.} \bsnm{{Strauss}}},
\bauthor{\binits{M.S.} \bsnm{{Potgieter}}},
\bauthor{\binits{S.E.S.} \bsnm{{Ferreira}}},
\bauthor{\binits{M.E.} \bsnm{{Hill}}},
\batitle{{Modelling anomalous cosmic ray oxygen in the heliosheath}}.
\bjtitle{Astron. Astrophys.}
\bvolume{522},
\bfpage{35}
(\byear{2010}).
doi:\doiurl{10.1051/0004-6361/201014528}
\end{barticle}
\endbibitem

\bibitem[\protect\citeauthoryear{Strauss et~al.}{2011a}]{straussetal2011}
\begin{barticle}
\bauthor{\binits{R.D.} \bsnm{Strauss}},
\bauthor{\binits{M.S.} \bsnm{Potgieter}},
\bauthor{\binits{I.} \bsnm{B{\"u}sching}},
\bauthor{\binits{A.} \bsnm{Kopp}},
\batitle{{Modeling the modulation of galactic and Jovian electrons by
  stochastic processes}}.
\bjtitle{Astrophys. J.}
\bvolume{735},
\bfpage{83}--\blpage{96}
(\byear{2011}a)
\end{barticle}
\endbibitem

\bibitem[\protect\citeauthoryear{Strauss et~al.}{2011b}]{straussetal2011b}
\begin{barticle}
\bauthor{\binits{R.D.} \bsnm{Strauss}},
\bauthor{\binits{M.S.} \bsnm{Potgieter}},
\bauthor{\binits{A.} \bsnm{Kopp}},
\bauthor{\binits{I.} \bsnm{B{\"u}sching}},
\batitle{{On the propagation times and energy losses of cosmic rays in the
  heliosphere}}.
\bjtitle{J. Geophys. Res.}
\bvolume{116},
\bfpage{12105}
(\byear{2011}b)
\end{barticle}
\endbibitem

\bibitem[\protect\citeauthoryear{Strauss et~al.}{2012}]{straussetal2012}
\begin{barticle}
\bauthor{\binits{R.D.} \bsnm{Strauss}},
\bauthor{\binits{M.S.} \bsnm{Potgieter}},
\bauthor{\binits{I.} \bsnm{B{\"u}sching}},
\bauthor{\binits{A.} \bsnm{Kopp}},
\batitle{{Modelling heliospheric current sheet drift in stochastic transport
  models}}.
\bjtitle{Astrophys. Space Sci.}
\bvolume{339},
\bfpage{223}--\blpage{236}
(\byear{2012})
\end{barticle}
\endbibitem

\bibitem[\protect\citeauthoryear{Strauss et~al.}{2013}]{straussetal2013b}
\begin{barticle}
\bauthor{\binits{R.D.} \bsnm{Strauss}},
\bauthor{\binits{M.S.} \bsnm{Potgieter}},
\bauthor{\binits{S.E.S.} \bsnm{Ferreira}},
\bauthor{\binits{H.} \bsnm{Fichtner}},
\bauthor{\binits{K.} \bsnm{Scherer}},
\batitle{{Cosmic ray modulation beyond the heliopause: A hybrid modelling
  approach}}.
\bjtitle{Astrophys. J. Lett.}
\bvolume{765},
\bfpage{18}
(\byear{2013})
\end{barticle}
\endbibitem

\bibitem[\protect\citeauthoryear{{Strong} and
  {Moskalenko}}{1998}]{Strong-Moskalenko-1998}
\begin{barticle}
\bauthor{\binits{A.W.} \bsnm{{Strong}}},
\bauthor{\binits{I.V.} \bsnm{{Moskalenko}}},
\batitle{{Propagation of Cosmic-Ray Nucleons in the Galaxy}}.
\bjtitle{\apj}
\bvolume{509},
\bfpage{212}--\blpage{228}
(\byear{1998}).
doi:\doiurl{10.1086/306470}
\end{barticle}
\endbibitem

\bibitem[\protect\citeauthoryear{{Strong} et~al.}{2007}]{Strong-etal-2007}
\begin{barticle}
\bauthor{\binits{A.W.} \bsnm{{Strong}}},
\bauthor{\binits{I.V.} \bsnm{{Moskalenko}}},
\bauthor{\binits{V.S.} \bsnm{{Ptuskin}}},
\batitle{{Cosmic-Ray Propagation and Interactions in the Galaxy}}.
\bjtitle{Annual Review of Nuclear and Particle Science}
\bvolume{57},
\bfpage{285}--\blpage{327}
(\byear{2007}).
doi:\doiurl{10.1146/annurev.nucl.57.090506.123011}
\end{barticle}
\endbibitem

\bibitem[\protect\citeauthoryear{{Strong} et~al.}{2011}]{strongetal2011}
\begin{barticle}
\bauthor{\binits{A.W.} \bsnm{{Strong}}},
\bauthor{\binits{E.} \bsnm{{Orlando}}},
\bauthor{\binits{T.R.} \bsnm{{Jaffe}}},
\batitle{{The interstellar cosmic-ray electron spectrum from synchrotron
  radiation and direct measurements}}.
\bjtitle{Astron. Astrophys.}
\bvolume{534},
\bfpage{54}
(\byear{2011})
\end{barticle}
\endbibitem

\bibitem[\protect\citeauthoryear{{Usmanov} et~al.}{2014}]{usmanovetal2014}
\begin{barticle}
\bauthor{\binits{A.V.} \bsnm{{Usmanov}}},
\bauthor{\binits{M.L.} \bsnm{{Goldstein}}},
\bauthor{\binits{W.H.} \bsnm{{Matthaeus}}},
\batitle{{Three-fluid, Three-dimensional Magnetohydrodynamic Solar Wind Model
  with Eddy Viscosity and Turbulent Resistivity}}.
\bjtitle{\apj}
\bvolume{788},
\bfpage{43}
(\byear{2014}).
doi:\doiurl{10.1088/0004-637X/788/1/43}
\end{barticle}
\endbibitem

\bibitem[\protect\citeauthoryear{{Vall{\'e}e}}{1995}]{Vallee-1995}
\begin{barticle}
\bauthor{\binits{J.P.} \bsnm{{Vall{\'e}e}}},
\batitle{{The Milky Way's Spiral Arms Traced by Magnetic Fields, Dust, Gas, and
  Stars}}.
\bjtitle{\apj}
\bvolume{454},
\bfpage{119}
(\byear{1995}).
doi:\doiurl{10.1086/176470}
\end{barticle}
\endbibitem

\bibitem[\protect\citeauthoryear{{Vall{\'e}e}}{2014}]{Vallee-2014}
\begin{barticle}
\bauthor{\binits{J.P.} \bsnm{{Vall{\'e}e}}},
\batitle{{The Spiral Arms of the Milky Way: The Relative Location of Each
  Different Arm Tracer within a Typical Spiral Arm Width}}.
\bjtitle{\aj}
\bvolume{148},
\bfpage{5}
(\byear{2014}).
doi:\doiurl{10.1088/0004-6256/148/1/5}
\end{barticle}
\endbibitem

\bibitem[\protect\citeauthoryear{{Vogt} et~al.}{2015}]{adrian}
\begin{bchapter}
\bauthor{\binits{A.} \bsnm{{Vogt}}},
\bauthor{\binits{P.} \bsnm{{Dunzlaff}}},
\bauthor{\binits{B.} \bsnm{{Heber}}},
\bauthor{\binits{A.} \bsnm{{Kopp}}},
\bauthor{\binits{P.} \bsnm{{K{\"u}hl}}},
\bauthor{\binits{R.D.} \bsnm{{Strauss}}},
\bctitle{{Jovian Electrons In The Inner Heliosphere: A Parameter Study On
  Intensity Profiles Near Earth}},
in \bbtitle{{Proceedings, 34th International Cosmic Ray Conference (ICRC
  2015)}},
\byear{2015}
\end{bchapter}
\endbibitem

\bibitem[\protect\citeauthoryear{{Vos} and
  {Potgieter}}{2015}]{vos_potgieter_2015}
\begin{barticle}
\bauthor{\binits{E.E.} \bsnm{{Vos}}},
\bauthor{\binits{M.S.} \bsnm{{Potgieter}}},
\batitle{{New Modeling of Galactic Proton Modulation during the Minimum of
  Solar Cycle 23/24}}.
\bjtitle{\apj}
\bvolume{815},
\bfpage{119}
(\byear{2015}).
doi:\doiurl{10.1088/0004-637X/815/2/119}
\end{barticle}
\endbibitem

\bibitem[\protect\citeauthoryear{{Wawrzynczak} et~al.}{2015a}]{polish2014}
\begin{barticle}
\bauthor{\binits{A.} \bsnm{{Wawrzynczak}}},
\bauthor{\binits{R.} \bsnm{{Modzelewska}}},
\bauthor{\binits{A.} \bsnm{{Gil}}},
\batitle{{Stochastic approach to the numerical solution of the non-stationary
  Parker's transport equation}}.
\bjtitle{Journal of Physics Conference Series}
\bvolume{574}(\bissue{1}),
\bfpage{012078}
(\byear{2015}a).
doi:\doiurl{10.1088/1742-6596/574/1/012078}
\end{barticle}
\endbibitem

\bibitem[\protect\citeauthoryear{{Wawrzynczak} et~al.}{2015b}]{anna2015}
\begin{barticle}
\bauthor{\binits{A.} \bsnm{{Wawrzynczak}}},
\bauthor{\binits{R.} \bsnm{{Modzelewska}}},
\bauthor{\binits{M.} \bsnm{{Kluczek}}},
\batitle{{Numerical methods for solution of the stochastic differential
  equations equivalent to the non-stationary Parkers transport equation}}.
\bjtitle{Journal of Physics Conference Series}
\bvolume{633}(\bissue{1}),
\bfpage{012058}
(\byear{2015}b).
doi:\doiurl{10.1088/1742-6596/633/1/012058}
\end{barticle}
\endbibitem

\bibitem[\protect\citeauthoryear{Webb and Gleeson}{1977}]{webbgleeson1977}
\begin{barticle}
\bauthor{\binits{G.M.} \bsnm{Webb}},
\bauthor{\binits{L.J.} \bsnm{Gleeson}},
\batitle{{Green's theorem and Green's function for the steady-state cosmic-ray
  equation of transport}}.
\bjtitle{Astrophys. Space Sci.}
\bvolume{50},
\bfpage{205}--\blpage{223}
(\byear{1977})
\end{barticle}
\endbibitem

\bibitem[\protect\citeauthoryear{Webb and Gleeson}{1979}]{webbgleeson1979}
\begin{barticle}
\bauthor{\binits{G.M.} \bsnm{Webb}},
\bauthor{\binits{L.J.} \bsnm{Gleeson}},
\batitle{{On the equation of transport for cosmic-ray particles in the
  interplanetary region}}.
\bjtitle{Astrophys. Space Sci.}
\bvolume{60},
\bfpage{335}--\blpage{351}
(\byear{1979})
\end{barticle}
\endbibitem

\bibitem[\protect\citeauthoryear{{Werner} et~al.}{2015}]{Werner-etal-2015}
\begin{barticle}
\bauthor{\binits{M.} \bsnm{{Werner}}},
\bauthor{\binits{R.} \bsnm{{Kissmann}}},
\bauthor{\binits{A.W.} \bsnm{{Strong}}},
\bauthor{\binits{O.} \bsnm{{Reimer}}},
\batitle{{Spiral arms as cosmic ray source distributions}}.
\bjtitle{Astroparticle Physics}
\bvolume{64},
\bfpage{18}--\blpage{33}
(\byear{2015}).
doi:\doiurl{10.1016/j.astropartphys.2014.10.005}
\end{barticle}
\endbibitem

\bibitem[\protect\citeauthoryear{Yamada et~al.}{1998}]{yamadaetal1998}
\begin{barticle}
\bauthor{\binits{Y.} \bsnm{Yamada}},
\bauthor{\binits{S.} \bsnm{Yanagita}},
\bauthor{\binits{T.} \bsnm{Yoshida}},
\batitle{{A stochastic view of the solar modulation phenomena of cosmic rays}}.
\bjtitle{Geophys. Res. Lett.}
\bvolume{25},
\bfpage{2353}--\blpage{2356}
(\byear{1998})
\end{barticle}
\endbibitem

\bibitem[\protect\citeauthoryear{{Yamada} et~al.}{1999}]{yamada_etal_1999}
\begin{barticle}
\bauthor{\binits{Y.} \bsnm{{Yamada}}},
\bauthor{\binits{S.} \bsnm{{Yanagita}}},
\bauthor{\binits{T.} \bsnm{{Yoshida}}},
\batitle{{A stochastic simulation method for the solar cycle modulation of
  cosmic rays}}.
\bjtitle{Advances in Space Research}
\bvolume{23},
\bfpage{505}--\blpage{508}
(\byear{1999}).
doi:\doiurl{10.1016/S0273-1177(99)00114-3}
\end{barticle}
\endbibitem

\bibitem[\protect\citeauthoryear{{Zank} et~al.}{2012}]{zanketal2012}
\begin{barticle}
\bauthor{\binits{G.P.} \bsnm{{Zank}}},
\bauthor{\binits{A.} \bsnm{{Dosch}}},
\bauthor{\binits{P.} \bsnm{{Hunana}}},
\bauthor{\binits{V.} \bsnm{{Florinski}}},
\bauthor{\binits{W.H.} \bsnm{{Matthaeus}}},
\bauthor{\binits{G.M.} \bsnm{{Webb}}},
\batitle{{The transport of low-frequency turbulence in astrophysical flows. I.
  Governing equations}}.
\bjtitle{Astrophys. J.}
\bvolume{745},
\bfpage{35}
(\byear{2012})
\end{barticle}
\endbibitem

\bibitem[\protect\citeauthoryear{Zhang}{1999}]{zhang1999}
\begin{barticle}
\bauthor{\binits{M.} \bsnm{Zhang}},
\batitle{{A Markov stochastic process theory of cosmic-ray modulation}}.
\bjtitle{Astrophys. J.}
\bvolume{513},
\bfpage{409}--\blpage{420}
(\byear{1999})
\end{barticle}
\endbibitem

\bibitem[\protect\citeauthoryear{{Zhang}}{2000}]{zhang_2000}
\begin{barticle}
\bauthor{\binits{M.} \bsnm{{Zhang}}},
\batitle{{Calculation of Diffusive Shock Acceleration of Charged Particles by
  Skew Brownian Motion}}.
\bjtitle{Astrophys. J.}
\bvolume{541},
\bfpage{428}--\blpage{435}
(\byear{2000}).
doi:\doiurl{10.1086/309429}
\end{barticle}
\endbibitem

\bibitem[\protect\citeauthoryear{{Zhang} et~al.}{2015}]{zhang2015}
\begin{barticle}
\bauthor{\binits{M.} \bsnm{{Zhang}}},
\bauthor{\binits{X.} \bsnm{{Luo}}},
\bauthor{\binits{N.} \bsnm{{Pogorelov}}},
\batitle{{Where is the cosmic-ray modulation boundary of the heliosphere?}}
\bjtitle{Physics of Plasmas}
\bvolume{22}(\bissue{9}),
\bfpage{091501}
(\byear{2015}).
doi:\doiurl{10.1063/1.4928945}
\end{barticle}
\endbibitem

\bibitem[\protect\citeauthoryear{{Zhang} et~al.}{2009}]{2zhang_etal_2009}
\begin{barticle}
\bauthor{\binits{M.} \bsnm{{Zhang}}},
\bauthor{\binits{G.} \bsnm{{Qin}}},
\bauthor{\binits{H.} \bsnm{{Rassoul}}},
\batitle{{Propagation of Solar Energetic Particles in Three-Dimensional
  Interplanetary Magnetic Fields}}.
\bjtitle{Astrophys. J.}
\bvolume{692},
\bfpage{109}--\blpage{132}
(\byear{2009}).
doi:\doiurl{10.1088/0004-637X/692/1/109}
\end{barticle}
\endbibitem

\bibitem[\protect\citeauthoryear{{Zimbardo} et~al.}{2015}]{Zimbardo-etal-2015}
\begin{barticle}
\bauthor{\binits{G.} \bsnm{{Zimbardo}}},
\bauthor{\binits{E.} \bsnm{{Amato}}},
\bauthor{\binits{A.} \bsnm{{Bovet}}},
\bauthor{\binits{F.} \bsnm{{Effenberger}}},
\bauthor{\binits{A.} \bsnm{{Fasoli}}},
\bauthor{\binits{H.} \bsnm{{Fichtner}}},
\bauthor{\binits{I.} \bsnm{{Furno}}},
\bauthor{\binits{K.} \bsnm{{Gustafson}}},
\bauthor{\binits{P.} \bsnm{{Ricci}}},
\bauthor{\binits{S.} \bsnm{{Perri}}},
\batitle{{Superdiffusive transport in laboratory and astrophysical plasmas}}.
\bjtitle{Journal of Plasma Physics}
\bvolume{81}(\bissue{6}),
\bfpage{495810601}
(\byear{2015}).
doi:\doiurl{10.1017/S0022377815001117}
\end{barticle}
\endbibitem

\bibitem[\protect\citeauthoryear{{Zuo} et~al.}{2013}]{zuo_etaL_2013}
\begin{barticle}
\bauthor{\binits{P.} \bsnm{{Zuo}}},
\bauthor{\binits{M.} \bsnm{{Zhang}}},
\bauthor{\binits{H.K.} \bsnm{{Rassoul}}},
\batitle{{The Role of Cross-shock Potential on Pickup Ion Shock Acceleration in
  the Framework of Focused Transport Theory}}.
\bjtitle{Astrophys. J.}
\bvolume{776},
\bfpage{93}
(\byear{2013}).
doi:\doiurl{10.1088/0004-637X/776/2/93}
\end{barticle}
\endbibitem

\bibitem[\protect\citeauthoryear{{Zuo} et~al.}{2011}]{Zuo_etal_2011}
\begin{barticle}
\bauthor{\binits{P.} \bsnm{{Zuo}}},
\bauthor{\binits{M.} \bsnm{{Zhang}}},
\bauthor{\binits{K.} \bsnm{{Gamayunov}}},
\bauthor{\binits{H.} \bsnm{{Rassoul}}},
\bauthor{\binits{X.} \bsnm{{Luo}}},
\batitle{{Energy Spectrum of Energetic Particles Accelerated by Shock Waves:
  From Focused Transport to Diffusive Acceleration}}.
\bjtitle{Astrophys. J.}
\bvolume{738},
\bfpage{168}
(\byear{2011}).
doi:\doiurl{10.1088/0004-637X/738/2/168}
\end{barticle}
\endbibitem

\end{thebibliography}
\vspace{0.5cm}
\small\emph{Exactly!" said Deep Thought. "So once you do know what the
    question actually is, you'll know what the answer means.}
\end{document}